\documentclass[reprint,superscriptaddress,amsmath, amssymb,aps,prd,notitlepage,longbibliography,floatfix,nofootinbib,onecolumn]{revtex4-1}

\usepackage{tensor}     
\usepackage{graphicx}   
\usepackage{subfigure}
\usepackage[
colorlinks=true,        
citecolor=blue,         
linkcolor=blue,         
urlcolor=blue           
]{hyperref}             
\usepackage{bm}         
\usepackage{xcolor}     
\usepackage{lipsum}
\usepackage{color}      
\usepackage[utf8]{inputenc} 
\usepackage[section]{placeins} 
\usepackage{multirow}
\usepackage[title]{appendix}
\usepackage{simpler-wick}
\newcommand{\nc}{\newcommand*}

\nc{\al}{\alpha}
\nc{\s}{\sigma}
\nc{\kp}{\kappa}
\nc{\dt}{\delta}
\nc{\Dt}{\Delta}
\nc{\Ld}{\Lambda}
\nc{\p}{\partial}
\nc{\Gm}{\Gamma}
\nc{\om}{\omega}
\nc{\Om}{\Omega}
\nc{\rd}{\mathrm{d}}
\def\({\left(}
\def\){\right)}
\def\[{\left[}
\def\]{\right]}
\def\e{\begin{equation}}
\def\q{\end{equation}}
\def\m{\begin{eqnarray}}
\def\n{\end{eqnarray}}
\nc{\Eq}[1]{Eq.~\eqref{#1}}     
\nc{\Fig}[1]{Fig.~\ref{#1}}     
\nc{\Table}[1]{Table~\ref{#1}}  
\nc{\Sec}[1]{Sec.~\ref{#1}}     
\nc{\Msun}{M_\odot}             
\nc{\fpbh}{f_{\mathrm{pbh}}}    
\nc{\mpbh}{m_{\mathrm{pbh}}}    
\nc{\fpbhn}{f_{\mathrm{pbh0}}}    
\nc{\mR}{\mathcal{R}} 
\nc{\seq}{\sigma_{\mathrm{eq}}}
\nc{\ogw}{\Omega_{\mathrm{GW}}}
\nc{\gpcyr}{\mathrm{Gpc}^{-3}\,\mathrm{yr}^{-1}}
\nc{\lvc}{LIGO/Virgo} 
\nc{\SNR}{\mathrm{SNR}} 
\nc{\mmin}{{m_{\mathrm{min}}}}
\nc{\mmax}{{m_{\mathrm{max}}}}
\nc{\Mmin}{{M_{\mathrm{min}}}}
\nc{\fmin}{{f_{\mathrm{min}}}}
\nc{\VT}{\mathrm{VT}}
\nc{\rhoGW}{\rho_{\mathrm{GW}}}
\nc{\vth}{\vec{\theta}}
\nc{\vd}{\vec{d}}
\nc{\vla}{\vec{\lambda}}
\nc{\Nobs}{N_{\mathrm{obs}}}
\nc{\av}[1]{\langle #1 \rangle} 
\nc{\km}{\mathrm{km}}
\nc{\Mpc}{\mathrm{Mpc}}
\nc{\Tobs}{T_{\mathrm{obs}}}
\nc{\Ntemp}{N_{\mathrm{temp}}}
\nc{\fyr}{f_{\mathrm{yr}}}
\nc{\addref}{[\textcolor{red}{add ref}] } 
\nc{\eg}{\textit{e.g.~}}
\nc{\app}{\approx}
\nc{\hf}{\frac{1}{2}}
\nc{\discuss}{\textcolor{red}{Add discussion here!}}
\nc{\red}[1]{\textcolor{red}{#1}}
\nc{\hp}{h_+} 
\nc{\hc}{h_{\times}} 
\nc{\Oh}{\hat{\Omega}}
\nc{\vx}{\vec{x}}
\nc{\A}[1]{\mathcal{A}_{#1}}
\nc{\Ogw}[1]{\Omega_{\mathrm{#1}}}
\nc{\bn}[1]{\dt\bm{t}_{\text{#1}}}
\nc{\bC}[1]{\bm{C}_{\text{#1}}}
\nc{\BFST}{$107 \pm 7$}

\nc{\Fnl}{F_{\mathrm{NL}}}
\nc{\Gnl}{G_{\mathrm{NL}}}
\begin{document}
	
\title{Full analysis of the scalar-induced gravitational waves for the curvature perturbation with local-type non-Gaussianities}

	

\author{Chen Yuan}
\email{chenyuan@tecnico.ulisboa.pt}
\affiliation{CENTRA, Departamento de Física, Instituto Superior Técnico – IST, Universidade de Lisboa – UL, Avenida Rovisco Pais 1, 1049–001 Lisboa, Portugal}
\author{De-Shuang Meng}
\email{mengdeshuang@itp.ac.cn}
\affiliation{CAS Key Laboratory of Theoretical Physics,
Institute of Theoretical Physics, Chinese Academy of Sciences, Beijing 100190, China}
\affiliation{School of Physical Sciences,
University of Chinese Academy of Sciences,
No. 19A Yuquan Road, Beijing 100049, China}

\author{Qing-Guo Huang}
\email{Corresponding author: huangqg@itp.ac.cn}
\affiliation{CAS Key Laboratory of Theoretical Physics,
Institute of Theoretical Physics, Chinese Academy of Sciences,
Beijing 100190, China}
\affiliation{School of Physical Sciences,
University of Chinese Academy of Sciences,
No. 19A Yuquan Road, Beijing 100049, China}
\affiliation{School of Fundamental Physics and Mathematical Sciences
Hangzhou Institute for Advanced Study, UCAS, Hangzhou 310024, China}

\date{\today}

\begin{abstract}
Primordial black holes (PBHs) are supposed to form through the gravitational collapse of regions with large density fluctuations. The formation of PBHs inevitably leads to the emission of scalar-induced gravitational wave (SIGW) signals, offering a unique opportunity to test the hypothesis of PBHs as a constituent of dark matter (DM). Previous studies have calculated the energy spectrum of SIGWs in local-type non-Gaussian models, primarily considering the contributions from the $F_{\mathrm{NL}}$-order or the $G_{\mathrm{NL}}$-order while neglecting connected diagrams. In this study, we extend the previous work by (i) considering the full contribution of non-Gaussian diagrams up to the $G_{\mathrm{NL}}$-order; (ii) deriving the generic scaling of the SIGW energy spectrum in the infrared region. We derive semi-analytical results applicable to arbitrary primordial power spectra and numerically evaluate the energy spectrum of SIGWs for a log-normal power spectrum. 
\end{abstract}

\maketitle

\maketitle

\section{Introduction}
The nature of dark matter (DM) poses a fundamental enigma in astrophysics that has been puzzling for decades. Although its existence can be inferred from its gravitational effects, there remains a significant dearth of knowledge regarding its composition and properties. Among the potential DM candidates, primordial black holes (PBHs) have attracted considerable attention.
PBHs are hypothesized to have formed through the gravitational collapse of regions with over-density during the radiation-dominated epoch immediately after the corresponding perturbation mode entered the horizon \cite{Zeldovich:1967lct,Hawking:1971ei,Carr:1974nx,Carr:1975qj}. And the mass of PBHs is related to the comoving wavelength of the perturbation mode. Numerous studies have been conducted to constrain the abundance of PBHs across a wide mass range \cite{Carr:2009jm,Graham:2015apa,Niikura:2017zjd,EROS-2:2006ryy,Niikura:2019kqi,Wang:2016ana,Chen:2019irf,Brandt:2016aco,Chen:2019xse,Montero-Camacho:2019jte,Laha:2019ssq,Dasgupta:2019cae,Laha:2020ivk,Saha:2021pqf,Ray:2021mxu,Zheng:2022wqo,Liu:2022iuf,Chen:2022fda,Chen:2018rzo,Mittal:2021egv}. However, the question of whether PBHs within the mass range of $[10^{-16}, 10^{-14}]\Msun$ and $[10^{-13}, 10^{-12}]\Msun$ could account for the entirety of DM remains unresolved (see e.g., \cite{Carr:2020gox} for review of constraints on PBHs).

Non-Gaussianity, characterized by deviations from Gaussian statistics, plays a significant role in the abundance of PBHs by affecting the tail of the probability density function (PDF) of curvature perturbations  \cite{Franciolini:2018vbk,Biagetti:2018pjj,Atal:2018neu,Passaglia:2018ixg,Atal:2019cdz,Atal:2019erb,Taoso:2021uvl,Biagetti:2021eep,Davies:2021loj}. As a result, PBH formation might be significantly enhanced or suppressed by non-Gaussian effects.

The recent detection of gravitational waves (GWs) from the merger of two black holes by the LIGO-Virgo Collaboration \cite{LIGOScientific:2016dsl,LIGOScientific:2021djp} has inaugurated the era of GW astronomy and sparked renewed interest in the potential role of PBH as constituents of DM \cite{Sasaki:2016jop,Chen:2018czv,Raidal:2018bbj,DeLuca:2020qqa,Hall:2020daa,Bhagwat:2020bzh,Hutsi:2020sol,Wong:2020yig,DeLuca:2021wjr,Franciolini:2021tla,Chen:2021nxo}. When the primordial scalar power spectrum experiences amplification on small scales, the quadratic terms of linear scalar perturbations give rise to a second-order tensor mode that can overwhelm the inflationary first-order tensor mode. This second-order tensor mode is known as scalar-induced gravitational waves (SIGWs) \cite{Ananda:2006af,Baumann:2007zm}. The SIGWs generated during the formation of PBHs provide a new way to hunt for PBHs \cite{Ananda:2006af,Baumann:2007zm,Saito:2008jc,Arroja:2009sh,Assadullahi:2009jc,Bugaev:2009kq,Bugaev:2009zh,Saito:2009jt,Bugaev:2010bb,Alabidi:2013lya,Nakama:2016enz,Nakama:2016gzw,Inomata:2016rbd,Orlofsky:2016vbd,Garcia-Bellido:2017aan,Sasaki:2018dmp,Espinosa:2018eve,Kohri:2018awv,Cai:2018dig,Bartolo:2018evs,Bartolo:2018rku,Unal:2018yaa,Byrnes:2018txb,Inomata:2018epa,Clesse:2018ogk,Cai:2019amo,Inomata:2019ivs,Cai:2019elf,Yuan:2019udt,Cai:2019cdl,Lu:2019sti,Yuan:2019wwo,Tomikawa:2019tvi,DeLuca:2019ufz,Yuan:2019fwv,
Inomata:2020lmk,Yuan:2020iwf,Papanikolaou:2020qtd,Zhang:2020ptw,Kapadia:2020pnr,Zhang:2020uek,Domenech:2020ssp,Dalianis:2020gup,Atal:2021jyo,Chen:2021nxo,Franciolini:2021nvv,Witkowski:2021raz,Balaji:2022dbi,Cang:2022oia,Gehrman:2022imk,Braglia:2020taf,Papanikolaou:2022chm,Qiu:2022klm,Escriva:2022duf,Meng:2022low,Gehrman:2023esa,Dandoy:2023jot,Ferrante:2023bgz,NANOGrav:2023hvm,Antoniadis:2023zhi,Inomata:2023zup,Gu:2023mmd,Liu:2023ymk,Yi:2023mbm,You:2023rmn,Jin:2023wri,Balaji:2023ehk,Liu:2023pau,Basilakos:2023xof,Ragavendra:2021qdu,Franciolini:2023pbf,Ferrante:2022mui}. For reviews of SIGW, see \cite{Yuan:2021qgz,Domenech:2021ztg}.

Previous studies have considered the local-type non-Gaussianity to the second order (or the $F_{\mathrm{NL}}$-order) \cite{Unal:2018yaa,Adshead:2021hnm}, but it is necessary to extend to the third order (or the $G_{\mathrm{NL}}$-order) for a general analysis of GWs induced by non-Gaussian scalar curvature perturbations. While literature \cite{Yuan:2020iwf} discusses the contribution to the $G_{\mathrm{NL}}$ term, it overlooks the contribution of the connected components, which are not zero. More recently, \cite{Abe:2022xur} studied the non-Gaussian contribution up to $G_{\mathrm{NL}}^2$ order using the diagram approach. However, as we will show in this paper, the full contribution of $G_{\mathrm{NL}}$ non-Gaussianities include $G_{\mathrm{NL}}^3$ and $G_{\mathrm{NL}}^4$ terms which have not been investigated in previous studies.
In this paper, we perform an extensive analysis of GWs induced by local-type non-Gaussian curvature perturbations, encompassing all the contributions up to the $G_{\mathrm{NL}}$-order. We also study the scaling in the infrared region for a generic power spectrum.

\section{The energy spectrum of SIGWs}

Let's begin from the FRW perturbated metric in Newton gauge, namely
\e
\rd s^{2}=a^{2}\left\{-(1+2 \phi) \rd \eta^{2}+\left[(1-2 \psi) \delta_{i j}+\frac{h_{i j}}{2}\right] \rd x^{i} \rd x^{j}\right\},
\q
where $\phi$ and $\psi$ are the scalar modes, $h_{ij}$ is the transverse and traceless second-order tensor mode and $a$ is the scale factor. During radiation dominated (RD) period, the stress tensor is described by perfect fluid and $\phi=\psi$ in the absence of anisotropies. The equation of motion for $h_{ij}$ is given by the second-order perturbative Einstein equation. In momentum space, we have
\e\label{eqh}
h_{\lambda,\textbf{k}}''(\eta)+2 \mathcal{H} h_{\lambda,\textbf{k}}'(\eta)+k^2 h_{\lambda,\textbf{k}}(\eta)=4 S_{\lambda,\textbf{k}}(\eta),
\q
where $\mathcal{H}\equiv a'/a$ is the conformal Hubble parameter and the prime stands for the derivative with respect to the conformal time $\eta$. The subscript $\lambda$ 
indicates the two different polarization modes of gravitational waves, which are represented by $+$ and $\times$.
The source term $S_{\lambda,\textbf{k}}(\eta)$ in eq.~(\ref{eqh}) reads \cite{Ananda:2006af,Baumann:2007zm}
\e\label{S2}
S_{\lambda,\textbf{k}}(\eta)=\int \frac{\mathrm{d}^{3} q}{(2 \pi)^{3}} Q_{\lambda}(\mathbf{k}, \mathbf{q}) F(q,|\mathbf{k}-\mathbf{q}|,\eta) \Phi_q \Phi_{|\mathbf{k}-\mathbf{q}|},
\q
where $F(p,q, \eta)$ is given by
\e\label{F}
F(p,q, \eta)=3T(p\eta)T(q\eta)+\frac{1}{\mathcal{H}}\[T'(p\eta)T(q\eta)+T(p\eta)T'(q\eta)\]+\frac{1}{\mathcal{H}^2}T'(p\eta)T'(q\eta),
\q
and $T(k\eta)$ is the transfer function, encoding the linear evolution of the scalar mode $\phi_k$ after re-entering the horizon following the end of inflation and is given by the first-order Einstein equation, namely
\e\label{T}
\phi_{k}(\eta)\equiv\Phi_k
T(k\eta)=\Phi_k{9\over (k\eta)^2}\({\sin(k\eta/\sqrt{3})\over k\eta/\sqrt{3}}-\cos(k\eta/\sqrt{3})\),
\q
where $\Phi_k$ represents the initial value of $\phi_k$ when it enters the horizon and is also the value of $\phi_k$ at the end of inflation because scalar perturbation remains conserved on super-horizon scales. Note that $F(p,q, \eta)$ is symmetric for $p$ and $q$ and unbolded symbols represent the modulus of a vector, and the same convention applies below in this paper. The projection factor $Q_{\lambda}(\mathbf{k}, \mathbf{q})$ in \Eq{S2} is defined by
\e
Q_{\lambda}(\mathbf{k}, \mathbf{q})\equiv e_{ij}^{\lambda}(\textbf{k})q_iq_j,
\q
where the polarization tensors are defined as $e_{ij}^{+}=(e_i e_j - \bar{e}_i \bar{e}_j)/\sqrt{2}$ and $e_{ij}^{\times}=(e_i \bar{e}_j + \bar{e}_i e_j)/\sqrt{2}$ and $e(\textbf{k})$ and $\bar e(\textbf{k})$ are a pair of orthogonal basis vectors perpendicular to $\textbf{k}$. It obeys the following symmetries:
\e
Q_{\lambda}(\mathbf{k}, \mathbf{q})=Q_{\lambda}(\mathbf{k}, \mathbf{q}\pm\mathbf{k})=Q_{\lambda}(-\mathbf{k}, \mathbf{q})=Q_{\lambda}(\mathbf{k}, -\mathbf{q})=Q_{\lambda}(-\mathbf{k}, -\mathbf{q}).
\q
Here we choose $e=(1,0,0)$, $\bar{e}=(0,1,0)$ and $\bm{k}=(0,0,k)$ and we can express the vector $\textbf{q}$ explicitly as
\e
\mathbf{q}=q(\sin \theta \cos \phi, \sin \theta \sin \phi, \cos \theta).
\q
Then we can directly write out
\e\label{Q}
Q_{\lambda}(\mathbf{k}, \mathbf{q})=\frac{q^{2}}{\sqrt{2}} \sin ^{2}\theta \times\left\{\begin{array}{ll}
\cos (2 \phi),\quad \lambda=+ \\
\sin (2 \phi),\quad \lambda=\times
\end{array}\right..
\q

Eq.~(\ref{eqh}) can be solved by Green's function
\e\label{solh}
h_{\lambda,\textbf{k}}(\eta)={4\over a(\eta)}\int_0^\eta g_k(\eta;\eta')a(\eta')S_{\lambda,\textbf{k}}(\eta)\mathrm{d}\eta',
\q
where the Green's function takes the form $g_k(\eta;\eta')={1\over k}\sin(k\eta-k\eta')$ during RD. 

The power spectrum, $P_{\lambda}\left(k,\eta\right)$, and the dimensionless power spectrum of GWs, $\mathcal{P}_{\lambda}\left(k,\eta\right)$, are defined as
\e\label{PS}
\left<h_{\lambda,\textbf{k}}(\eta)h_{\lambda',\textbf{k}'}(\eta)\right>=(2\pi)^3\delta^{3}\left(\mathbf{k}+\mathbf{k}'\right) \delta^{\lambda \lambda'} P_{\lambda}\left(k,\eta\right)=(2\pi)^3\delta^{3}\left(\mathbf{k}+\mathbf{k}'\right) \delta^{\lambda \lambda'}\frac{2 \pi^{2}}{k^{3}} \mathcal{P}_{\lambda}\left(k,\eta\right).
\q
The energy density of GWs, $\Omega_{\mathrm{GW}}(k,\eta)$, is an observed quantity, defined as the energy of GWs per logarithm frequency normalized by the critical energy,  $\rho_{c}(\eta)$, takes the form
\e\label{ogw}
\ogw(k,\eta)\equiv{1 \over \rho_c}{\mathrm{d}\rho_{\mathrm{GW}} \over \mathrm{d} \ln k}
={1\over48}\(k\over\mathcal{H}\)^2\sum_{\lambda=+, \times}\overline{\mathcal{P}_{\lambda}\left(k,\eta\right)},
\q
where the overbar denotes the time average. The density parameter at the matter-radiation equality is $\Omega_{\mathrm{GW}}(k)\simeq \Omega_{\mathrm{GW}}(k, k\eta \rightarrow \infty)$ and the quality that would be observed today can be obtained by $\Omega_{\mathrm{GW}, 0}(k)=\Omega_{r} \times \Omega_{\mathrm{GW}}(k)$, where $\Omega_{r}$ is the density parameter of radiation today. 

Using \Eq{S2}, \Eq{solh}, \Eq{PS} and \Eq{ogw},  $\Omega_{\mathrm{GW}}(k)$ can be expressed by
\m\label{omegak}
\Omega_{\mathrm{GW}}(k)=\frac{k^3}{6\pi^2
}\left(\frac{k}{\mathcal{H}}\right)^2 \sum_{\lambda=+, \times} &&\int \frac{\mathrm{d}^{3}q \mathrm{d}^{3} q'} {(2 \pi)^{6}}  Q_{\lambda}\left(\mathbf{k}, \mathbf{q}\right) Q_{\lambda}\left(\mathbf{k'}, \mathbf{q}'\right)\overline{\tilde{I}\left(q,\left|\mathbf{k}-\mathbf{q}\right|,k\eta\to\infty\right) \tilde{I}\left(q', \left|\mathbf{k'}-\mathbf{q}'\right|,k\eta\to\infty\right)}\nonumber\\
&&\times  
\left\langle\left\langle\Phi_{\bm{q}}\Phi_{\bm{k-q}}\Phi_{\bm{q'}}\Phi_{\bm{k'-q'}}\right\rangle\right\rangle,
\n
where we define $\left\langle\left\langle\Phi_{\bm{q}}\Phi_{\bm{k-q}}\Phi_{\bm{q'}}\Phi_{\bm{k'-q'}}\right\rangle\right\rangle$ as the remaining part after extracting $(2\pi)^3\delta^{3}\left(\bm{k}+\bm{k'}\right)$ from the terms containing $\delta^{3}\left(\bm{k}+\bm{k'}\right)$ in the four-point function $\left\langle\Phi_{\bm{q}}\Phi_{\bm{k-q}}\Phi_{\bm{q'}}\Phi_{\bm{k'-q'}}\right\rangle$, i.e.
\e
\left\langle\Phi_{\bm{q}}\Phi_{\bm{k-q}}\Phi_{\bm{q'}}\Phi_{\bm{k'-q'}}\right\rangle\equiv(2\pi)^3\delta^{3}\left(\bm{k}+\bm{k'}\right)\left\langle\left\langle\Phi_{\bm{q}}\Phi_{\bm{k-q}}\Phi_{\bm{q'}}\Phi_{\bm{k'-q'}}\right\rangle\right\rangle,
\q
and the kernel function $\tilde{I}(p,q,\eta)$ is defined as
\e\label{tI}
\tilde{I}(p,q,\eta)\equiv\int\mathrm{d}\eta'{a(\eta')\over a(\eta)} g_k(\eta;\eta')F(p,q,\eta'),
\q
which contains all the time-dependent terms.

By substituting \Eq{F}, \Eq{T}, \Eq{Q} and \Eq{tI} into \Eq{omegak}, applying coordinate transformations $u = q / k$, $ v=|\mathbf{k}-\mathbf{q}| / k$ and $u' = q' / k$, $ v'=|\mathbf{k}-\mathbf{q'}| / k$, and then averaging over time, we obtain $\Omega_{\mathrm{GW}}(k)$:
\m\label{omega}
\Omega_{\mathrm{GW}}(k)=\frac{k^3}{6\pi^2}\int\frac{\mathrm{d}^3q\mathrm{d}^3q'}{(2\pi)^6}
\cos2(\phi-\phi') I\left(u,v\right) I\left(u',v'\right) \left\langle\left\langle\zeta_{\bm{q}}\zeta_{\bm{k-q}}\zeta_{\bm{q'}}\zeta_{\bm{k'-q'}}\right\rangle\right\rangle,
\n
where we have transformed the scalar perturbation $\Phi$ into the comoving curvature perturbation $\zeta$ using the relation $\Phi=(2/3)\zeta$ and absorbed the coefficient $16/81$ and the remaining projection term after removing $\cos2(\phi-\phi')$ into the kernel function. The resulting new kernel function can be expressed as
\m
I\left(u,v\right) I\left(u',v'\right)=&& \frac{9 \left(u^2+v^2-3\right)\left(u'^2+v'^2-3\right)}{1024 u^3 u'^3 v^3 v'^3} \left[4 u^2-\left(u^2-v^2+1\right)^2\right]  \left[4u'^2-\left(u'^2-v'^2+1\right)^2\right] \nonumber\\
&&\times\Bigg\{\left[\left(u^2+v^2-3\right) \ln \left(\left|\frac{(u-v)^2-3}{(u+v)^2-3}\right| \right)+4u v\right] \left[\left(u'^2+v'^2-3\right) \ln \left(\left|\frac{(u'-v')^2-3}{(u'+v')^2-3}\right| \right)+4u' v'\right] \nonumber\\
&&+\pi ^2 \left(u^2+v^2-3\right) \left(u'^2+v'^2-3\right) \Theta \left(u+v-\sqrt{3}\right) \Theta \left(u'+v'-\sqrt{3}\right)\Bigg\},
\n
where $\Theta$ is the Heaviside function.
It is worth noting that in \Eq{omega}, we have retained the momentum dependence in the integral variable and four-point function without making a change of variables. This is for the convenience of future calculations when dealing with non-Gaussianity. The following relations will be frequently used in subsequent calculations. The transformation relation for the integral variables is as follows:
\e
\int\mathrm{d}^3q\to\int_0^{\infty} \mathrm{d}u \int_{|1-u|}^{1+u}\mathrm{d}v\int_0^{2\pi} \mathrm{d}\phi uvk^3,
\q
and $\cos\theta$ and $\sin\theta$ can be expressed by
\e\label{sincos}
\cos\theta=\frac{1+u^2-v^2}{2u},\quad \sin\theta=\sqrt{1-\frac{(1+u^2-v^2)^2}{4u^2}}.
\q

\section{GWs induced by local-type non-Gaussian curvature perturbations}
The local-type non-Gaussian curvature perturbation $\zeta$ is expanded in terms of the Gaussian part $\zeta_g$ in real space as
\e
\zeta\left(\zeta_{g}\right)=\zeta_{g}+F_{\mathrm{NL}}\left(\zeta_{g}^{2}-\left\langle\zeta_{g}^{2}\right\rangle\right)+G_{\mathrm{NL}} \zeta_{g}^{3},
\q
where $F_{\mathrm{NL}}$ and $G_{\mathrm{NL}}$ are the dimensionless non-Gaussian parameters, related to the commonly used notations $f_{\mathrm{NL}}$ and $g_{\mathrm{NL}}$ by $F_{\mathrm{NL}} \equiv 3/5 f_{\mathrm{NL}}$ and $G_{\mathrm{NL}} \equiv 9/25 g_{\mathrm{NL}}$ respectively. In momentum space, the curvature perturbation is expanded by convolution of the Gaussian part
\e\label{zeta}
\zeta_{\bm{k}}=\zeta_g(\bm{k})+F_{\mathrm{NL}}\int\frac{\mathrm{d}^3p}{(2\pi)^3}\zeta_g(\bm{p})\zeta_g(\bm{k-p})+G_{\mathrm{NL}}\int\frac{\mathrm{d}^3p_1\mathrm{d}^3p_2}{(2\pi)^6}\zeta_g(\bm{p_1})\zeta_g(\bm{p_2})\zeta_g(\bm{k-p_1-p_2}).
\q
Note that we neglect the Fourier transform of the constant term $F_{\mathrm{NL}} \left\langle\zeta_{g}^{2}\right\rangle$ since this term leads to $\delta(\bm{k})$ and does not contribute to the SIGW power spectrum in the following calculation. 
The power spectrum $P_{g}(k)$ and the dimensionless power spectrum $\mathcal{P}_{g}(k)$ of the Gaussian part curvature perturbation  are defined as
\e
\left\langle\zeta_{g}\left(\bm{k}\right) \zeta_{g}\left(\bm{k}'\right)\right\rangle=(2\pi)^3\delta^{3}\left(\bm{k}+\bm{k}'\right) P_{g}\left(k\right)=(2\pi)^3\delta^{3}\left(\bm{k}+\bm{k}'\right) \frac{2\pi^2}{k^3} \mathcal{P} _{g}\left(k\right).
\q

The influence of non-Gaussianity in curvature perturbations on the GWs spectrum is manifested in the four-point function in \Eq{omega}. By substituting \Eq{zeta} into \Eq{omega} and employing Wick's theorem, we can obtain the total GWs energy density spectrum up to the $G_{\mathrm{NL}}$ order in local non-Gaussian expansion. Next, we will decompose the GWs spectrum into different powers of $F_{\mathrm{NL}}$ and $G_{\mathrm{NL}}$.

\subsection{Gaussian part}
The leading order is the Gaussian part, in which case we have
\m
\left\langle\zeta_{\bm{q}}\zeta_{\bm{k-q}}\zeta_{\bm{q'}}\zeta_{\bm{k'-q'}}\right\rangle_{g}=&&\left\langle\zeta_{g}\left(\mathbf{q}\right) \zeta_{g}\left(\mathbf{q}'\right)\right\rangle\left\langle\zeta_{g}\left(\mathbf{k-q}\right) \zeta_{g}\left(\mathbf{k'-q'}\right)\right\rangle+\left\langle\zeta_{g}\left(\mathbf{q}\right) \zeta_{g}\left(\mathbf{k'-q'}\right)\right\rangle\left\langle\zeta_{g}\left(\mathbf{k-q}\right) \zeta_{g}\left(\mathbf{q}'\right)\right\rangle\nonumber\\
&&+\left\langle\zeta_{g}\left(\mathbf{q}\right) \zeta_{g}\left(\mathbf{k-q}\right)\right\rangle\left\langle\zeta_{g}\left(\mathbf{q'}\right) \zeta_{g}\left(\mathbf{k'-q'}\right)\right\rangle,
\n
where the third term on the right-hand side of the above equation is zero because it corresponds to a disconnected diagram that does not contribute to the physical mechanism, and it also does not contain the $\delta^{3}\left(\bm{k}+\bm{k'}\right)$ term. Due to symmetry, the contributions of the first two terms on the right-hand side of the above equation are equal. By substituting this equation into \Eq{omega}, we can obtain the Gaussian part of the
GWs spectrum
\m
\Omega_{\mathrm{GW}}^{g}(k)=&&\frac{k^3}{6\pi^2}\int\frac{\mathrm{d}^3q}{(2\pi)^3}  I^2\left(u,v\right) P_{g}\left(q\right) P_{g}\left(|\bm{k-q}|\right) \times2\nonumber\\
=&&\frac{1}{3} \int_0^{\infty} \mathrm{d}u \int_{|1-u|}^{1+u}\mathrm{d}v I^2(u,v)\frac{1}{u^2v^2} \mathcal{P} _{g}\left(uk\right) \mathcal{P} _{g}\left(vk\right).
\n

\subsection{$F_{\mathrm{NL}}^2$ terms}
Since $\zeta_{g}$ is Gaussian curvature perturbation, the odd-order n-point correlation functions of which are zero. Therefore, the GWs spectrum containing $F_{\mathrm{NL}}$ and $F_{\mathrm{NL}}^3$ terms is also zero. Therefore, we only need to consider terms containing $F_{\mathrm{NL}}^2$ and $F_{\mathrm{NL}}^4$.

For terms containing $F_{\mathrm{NL}}^2$, considering symmetry, the GWs spectrum can be expressed in the following form:
\m
\Omega_{\mathrm{GW}}^{F_{\mathrm{NL}}^2}(k)=&&\frac{F_{\mathrm{NL}}^2k^3}{6\pi^2}\int\frac{\mathrm{d}^3q\mathrm{d}^3q'}{(2\pi)^6} \cos2(\phi-\phi') I\left(u,v\right) I\left(u',v'\right) \int\frac{\mathrm{d}^3p_1\mathrm{d}^3p_2}{(2\pi)^6}\nonumber\\
&&\times \Bigg[ 2\left\langle\left\langle\zeta_g(\bm{p_1})\zeta_g(\bm{q-p_1})\zeta_g(\bm{p_2})\zeta_g(\bm{k-q-p_2})\zeta_g(\bm{q'})\zeta_g(\bm{k'-q'})\right\rangle\right\rangle\nonumber\\
&&+4\left\langle\left\langle\zeta_g(\bm{p_1})\zeta_g(\bm{q-p_1})\zeta_g(\bm{k-q})\zeta_g(\bm{p_2})\zeta_g(\bm{q'-p_2})\zeta_g(\bm{k'-q'})\right\rangle\right\rangle\Bigg].
\n
Performing Wick contraction on the above two six-point functions, there are three distinct non-zero contractions denoted as 'hybrid' term, 'Z' term, and 'C' term as named in ref.\cite{Unal:2018yaa,Adshead:2021hnm}. Then we have
\e
\Omega_{\mathrm{GW}}^{F_{\mathrm{NL}}^2}(k)=\Omega_{\mathrm{GW}}^{hybrid}(k)+\Omega_{\mathrm{GW}}^{C}(k)+\Omega_{\mathrm{GW}}^{Z}(k).
\q
Note that the 'hybrid' term is a disconnected term, the 'C' term and the 'Z' term are connected terms. Ref.\cite{Yuan:2020iwf} omitted all disconnected items. We now demonstrate each of these three parts in detail.

For the 'hybrid' term, one example of the contraction is shown as follows:
\e
\left\langle\wick{\c1\zeta_g(\bm{p_1})\c2\zeta_g(\bm{q-p_1})\c3\zeta_g(\bm{k-q})\c1\zeta_g(\bm{p_2})\c2\zeta_g(\bm{q'-p_2}) \c3 \zeta_g(\bm{k'-q'})}\right\rangle,
\q
which is commonly referred to as a disconnected diagram and the term $\delta^3(\bm{q}+\bm{q'})$ is present. According to symmetry, there are two other contractions that yield the same result. Therefore, we need to multiply by a symmetry factor, which in this case is 2. The calculation of a disconnected diagram is relatively straightforward, because in this case $\cos2(\phi-\phi')=1$ and thus disappears in the integral. Expanding the correlation function and using the appearing delta functions to eliminate redundant integrals, we can obtain:
\m
\Omega_{\mathrm{GW}}^{hybrid}(k)=&&\frac{F_{\mathrm{NL}}^2k^3}{6\pi^2}\int\frac{\mathrm{d}^3q}{(2\pi)^3}  I^2\left(u,v\right) \int\frac{\mathrm{d}^3p_1}{(2\pi)^3}4P _{g}\left(p_1\right) P_{g}\left(|\bm{q-p_1}|\right) P_{g}\left(|\bm{k-q}|\right)\times2\nonumber\\
=&&\frac{2F_{\mathrm{NL}}^2}{3} \int_0^{\infty} \mathrm{d}u \int_{|1-u|}^{1+u}\mathrm{d}v \int_0^{\infty} \mathrm{d}u_1 \int_{|1-u_1|}^{1+u_1}\mathrm{d}v_1 I^2(u,v) \frac{1}{u^2v^2u_1^2v_1^2} \mathcal{P} _{g}\left(u_1uk\right) \mathcal{P} _{g}\left(v_1uk\right)\mathcal{P} _{g}\left(vk\right),
\n
where the second equality in the above equation is obtained by performing the coordinate transformation $u_1 = p_1 / q$ and $ v_1=|\bm{q-p_1}| / q$. 

For the ’Z’ term, one example of the contraction is shown as follows:
\e
\left\langle\wick{\c1\zeta_g(\bm{p_1})\c2\zeta_g(\bm{q-p_1})\c3\zeta_g(\bm{k-q})\c1\zeta_g(\bm{p_2})\c3\zeta_g(\bm{q'-p_2}) \c2 \zeta_g(\bm{k'-q'})}\right\rangle,
\q
which is commonly referred to as a connected diagram and the term $\delta^3(\bm{q}+\bm{q'})$ is not satisfied. The symmetry factor in this case is 4. Then the calculation will be more complicated than the disconnected diagram because $\cos2(\phi-\phi')$ will be retained in the integral. In this case, we have
\m
\Omega_{\mathrm{GW}}^{Z}(k)=&&\frac{F_{\mathrm{NL}}^2k^3}{6\pi^2}\int\frac{\mathrm{d}^3q\mathrm{d}^3q'}{(2\pi)^6} \cos2(\phi-\phi') I\left(u,v\right) I\left(u',v'\right)
4P_{g}\left(|\bm{k-q}|\right) P _{g}\left(|\bm{k-q'}|\right)P _{g}\left(|\bm{k-q-q'}|\right)\times4\nonumber\\
=&&\frac{F_{\mathrm{NL}}^2}{3\pi^2}\int_0^{\infty} \mathrm{d}u \int_{|1-u|}^{1+u}\mathrm{d}v \int_0^{\infty} \mathrm{d}u' \int_{|1-u'|}^{1+u'}\mathrm{d}v' \int_0^{2\pi}\mathrm{d}\phi\int_0^{2\pi}\mathrm{d}\phi' \cos2(\phi-\phi') I(u,v)I(u',v') \frac{uvu'v'}{v^3v'^3w_{012}^3}\nonumber\\
&&\times\mathcal{P}_{g}\left(vk\right) \mathcal{P}_{g}\left(v'k\right)\mathcal{P}_{g}\left(w_{012}k\right)\nonumber\\
=&&\frac{2F_{\mathrm{NL}}^2}{3\pi}\int_0^{\infty} \mathrm{d}u \int_{|1-u|}^{1+u}\mathrm{d}v \int_0^{\infty} \mathrm{d}u' \int_{|1-u'|}^{1+u'}\mathrm{d}v' \int_0^{2\pi}\mathrm{d}\varphi_1 \cos2\varphi_1 I(u,v)I(u',v') \frac{uvu'v'}{v^3v'^3w_{012}^3}\nonumber\\
&&\times\mathcal{P}_{g}\left(vk\right) \mathcal{P}_{g}\left(v'k\right)\mathcal{P}_{g}\left(w_{012}k\right),
\n
where the third equality is obtained by performing a coordinate transformation $\varphi_1=\phi-\phi'$ and $\varphi_2=\phi+\phi'$ and then we have
\e
\int_0^{2\pi}\mathrm{d}\phi\int_0^{2\pi}\mathrm{d}\phi'\to\frac{1}{2}\int_0^{2\pi}\mathrm{d}\varphi_1\int_0^{4\pi}\mathrm{d}\varphi_2=2\pi\int_0^{2\pi}\mathrm{d}\varphi_1,
\q
as $\varphi_2$ does not appear in the integral. 

Additionally, $w_{012}$ is defined as follows:
\e\label{w012}
w_{012}^2=\frac{|\bm{k-q-q'}|^2}{k^2}=1+u^2+u'^2+2uu'(\sin\theta\sin\theta'\cos\varphi_1+\cos\theta\cos\theta')-2u\cos\theta-2u'\cos\theta',
\q
where we have used the following relations
\m
\bm{q}\cdot\bm{q'}&&=uu'k^2\[\sin\theta\sin\theta'(\cos\phi\cos\phi'+\sin\phi\sin\phi')+\cos\theta\cos\theta'\],\\
\bm{k}\cdot\bm{q}&&=uk^2\cos\theta,\\
\bm{k}\cdot\bm{q'}&&=u'k^2\cos\theta',
\n
and \Eq{sincos} to replace $\sin$ and $\cos$.

As to the 'C' term, one example of the contraction is shown as follows:
\e
\left\langle\wick{\c1\zeta_g(\bm{p_1})\c2\zeta_g(\bm{q-p_1})\c1\zeta_g(\bm{p_2}) \c1\zeta_g(\bm{k-q-p_2}) \c2\zeta_g(\bm{q'}) \c1 \zeta_g(\bm{k'-q'})}\right\rangle,
\q
and the symmetry factor in this case is 8. Then we have
\m
\Omega_{\mathrm{GW}}^{C}(k)=&&\frac{F_{\mathrm{NL}}^2k^3}{6\pi^2}\int\frac{\mathrm{d}^3q\mathrm{d}^3q'}{(2\pi)^6} \cos2(\phi-\phi') I\left(u,v\right) I\left(u',v'\right)
2P_{g}\left(|\bm{q-q'}|\right) P_{g}\left(|\bm{k-q'}|\right)P _{g}\left(q'\right)\times8\nonumber\\
=&&\frac{2F_{\mathrm{NL}}^2}{3\pi}\int_0^{\infty} \mathrm{d}u \int_{|1-u|}^{1+u}\mathrm{d}v \int_0^{\infty} \mathrm{d}u' \int_{|1-u'|}^{1+u'}\mathrm{d}v' \int_0^{2\pi}\mathrm{d}\varphi_1 \cos2\varphi_1 I(u,v)I(u',v') \frac{uvu'v'}{u'^3v'^3w_{12}^3}\nonumber\\
&&\times\mathcal{P} _{g}\left(u'k\right) \mathcal{P} _{g}\left(v'k\right)\mathcal{P} _{g}\left(w_{12}k\right),
\n
where
\e\label{w12}
w_{12}^2=\frac{|\bm{q-q'}|^2}{k^2}=u^2+u'^2-2uu'(\sin\theta\sin\theta'\cos\varphi_1+\cos\theta\cos\theta').
\q

\subsection{$F_{\mathrm{NL}}^4$ terms}
For terms containing $F_{\mathrm{NL}}^4$, the GWs spectrum can be expressed in the following form:
\m
\Omega_{\mathrm{GW}}^{F_{\mathrm{NL}}^4}(k)=&&\frac{F_{\mathrm{NL}}^4k^3}{6\pi^2}\int\frac{\mathrm{d}^3q\mathrm{d}^3q'}{(2\pi)^6} \cos2(\phi-\phi') I\left(u,v\right) I\left(u',v'\right) \int\frac{\mathrm{d}^3p_1\mathrm{d}^3p_2\mathrm{d}^3p_3\mathrm{d}^3p_4}{(2\pi)^{12}}\nonumber\\
&& \times\left\langle\left\langle\zeta_g(\bm{p_1})\zeta_g(\bm{q-p_1})\zeta_g(\bm{p_2})\zeta_g(\bm{k-q-p_2})\zeta_g(\bm{p_3})\zeta_g(\bm{q'-p_3})\zeta_g(\bm{p_4})\zeta_g(\bm{k'-q'-p_4})\right\rangle\right\rangle.
\n
Performing Wick contraction on the above eight-point function, there are three distinct non-zero contractions denoted as 'reducible' term, 'planar' term, and 'non-planar' term as also named in ref.\cite{Garcia-Bellido:2017aan,Adshead:2021hnm}. Then we can write
\e
\Omega_{\mathrm{GW}}^{F_{\mathrm{NL}}^4}(k)=\Omega_{\mathrm{GW}}^{re}(k)+\Omega_{\mathrm{GW}}^{planar}(k)+\Omega_{\mathrm{GW}}^{np}(k).
\q
We now demonstrate each of these three parts in detail.

For the 'reducible' term, one example of the contraction is shown as follows:
\e
\left\langle\wick{\c1\zeta_g(\bm{p_1}) \c2\zeta_g(\bm{q-p_1}) \c3\zeta_g(\bm{p_2}) \c4\zeta_g(\bm{k-q-p_2}) \c1\zeta_g(\bm{p_3}) \c2\zeta_g(\bm{q'-p_3}) \c3\zeta_g(\bm{p_4}) \c4\zeta_g(\bm{k'-q'-p_4})}\right\rangle,
\q
which is a disconnected diagram and the symmetry factor is 8. Expanding the correlation function and using the appearing delta functions to eliminate redundant integrals, we can obtain:
\m
\Omega_{\mathrm{GW}}^{re}(k)=&&\frac{F_{\mathrm{NL}}^4k^3}{6\pi^2}\int\frac{\mathrm{d}^3q}{(2\pi)^3}  I^2\left(u,v\right) \int\frac{\mathrm{d}^3p_1\mathrm{d}^3p_2}{(2\pi)^6}P_{g}\left(p_1\right) P _{g}\left(|\bm{q-p_1}|\right) P_{g}\left(p_2\right)P_{g}\left(|\bm{k-q-p_2}|\right)\times8\nonumber\\
=&&\frac{F_{\mathrm{NL}}^4}{3} \int_0^{\infty} \mathrm{d}u \int_{|1-u|}^{1+u}\mathrm{d}v \int_0^{\infty} \mathrm{d}u_1 \int_{|1-u_1|}^{1+u_1}\mathrm{d}v_1 \int_0^{\infty} \mathrm{d}u_2 \int_{|1-u_2|}^{1+u_2}\mathrm{d}v_2 I^2(u,v) \frac{1}{u^2v^2u_1^2v_1^2u_2^2v_2^2}\nonumber\\
&&\times\mathcal{P} _{g}\left(u_1uk\right) \mathcal{P} _{g}\left(v_1uk\right)\mathcal{P} _{g}\left(u_2vk\right) \mathcal{P} _{g}\left(v_2vk\right),
\n
where the second equality in the above equation is obtained by performing the coordinate transformation $u_1 = p_1 / q$, $ v_1=|\bm{q-p_1}| / q$ and $u_2 = p_2 / |\bm{k-q}|$, $ v_2=|\bm{k-q-p_2}| / |\bm{k-q}|$.

For the `planar’ term, one example of the contraction is shown as follows:
\e
\left\langle\wick{\c2\zeta_g(\bm{p_1}) \c1\zeta_g(\bm{q-p_1}) \c1\zeta_g(\bm{p_2}) \c3\zeta_g(\bm{k-q-p_2}) \c2\zeta_g(\bm{p_3}) \c1\zeta_g(\bm{q'-p_3}) \c1\zeta_g(\bm{p_4}) \c3\zeta_g(\bm{k'-q'-p_4})}\right\rangle,
\q
and the symmetry factor in this case is 32. Then we have
\m
\Omega_{\mathrm{GW}}^{planar}(k)
=&&\frac{F_{\mathrm{NL}}^4k^3}{6\pi^2}\int\frac{\mathrm{d}^3q\mathrm{d}^3q'}{(2\pi)^6} \cos2(\phi-\phi') I\left(u,v\right) I\left(u',v'\right) \int\frac{\mathrm{d}^3p_1}{(2\pi)^3}P_{g}\left(p_1\right) P_{g}\left(|\bm{q-p_1}|\right) \nonumber\\
&&\times P_{g}\left(|\bm{k-p_1}|\right)P_{g}\left(|\bm{q'-p_1}|\right)\times32\nonumber\\
=&&\frac{F_{\mathrm{NL}}^4}{6\pi^3}\int_0^{\infty} \mathrm{d}u \int_{|1-u|}^{1+u}\mathrm{d}v \int_0^{\infty} \mathrm{d}u' \int_{|1-u'|}^{1+u'}\mathrm{d}v' \int_0^{\infty} \mathrm{d}u_1 \int_{|1-u_1|}^{1+u_1}\mathrm{d}v_1 \int_0^{2\pi}\mathrm{d}\phi\int_0^{2\pi}\mathrm{d}\phi' \int_0^{2\pi}\mathrm{d}\phi_1\nonumber\\
&&\cos2(\phi-\phi') I(u,v)I(u',v')  \frac{uvu'v'u_1v_1}{(u_1v_1w_{13}w_{23})^3}\mathcal{P} _{g}\left(u_1k\right) \mathcal{P}_{g}\left(v_1k\right)\mathcal{P} _{g}\left(w_{13}k\right)\mathcal{P} _{g}\left(w_{23}k\right)\nonumber\\
=&&\frac{F_{\mathrm{NL}}^4}{3\pi^2}\int_0^{\infty} \mathrm{d}u \int_{|1-u|}^{1+u}\mathrm{d}v \int_0^{\infty} \mathrm{d}u' \int_{|1-u'|}^{1+u'}\mathrm{d}v' \int_0^{\infty} \mathrm{d}u_1 \int_{|1-u_1|}^{1+u_1}\mathrm{d}v_1 \int_0^{2\pi}\mathrm{d}\varphi_1 \int_0^{2\pi}\mathrm{d}\varphi_2 \nonumber\\
&&\cos2\varphi_1 I(u,v)I(u',v')  \frac{uvu'v'u_1v_1}{(u_1v_1w_{13}w_{23})^3}\mathcal{P} _{g}\left(u_1k\right) \mathcal{P}_{g}\left(v_1k\right)\mathcal{P} _{g}\left(w_{13}k\right)\mathcal{P} _{g}\left(w_{23}k\right),
\n
where the third equality in the above equation is obtained by performing a coordinate transformation $\varphi_1=\phi-\phi'$, $\varphi_2=\phi-\phi_1$, and $\varphi_3=\phi+\phi'$ with the Jacobi 2 and then $\varphi_3$ can be integrated out. Additionally, $w_{13}$ and $w_{23}$ are defined as follows:
\m\label{w13}
w_{13}^2&&=\frac{|\bm{q-p_1}|^2}{k^2}=u^2+u_1^2-2uu_1(\sin\theta\sin\theta_1\cos\varphi_2+\cos\theta\cos\theta_1),\\
\label{w23}
w_{23}^2&&=\frac{|\bm{q'-p_1}|^2}{k^2}=u'^2+u_1^2-2u'u_1\[\sin\theta'\sin\theta_1\cos(\varphi_1-\varphi_2)+\cos\theta'\cos\theta_1\].
\n

As to the 'non-planar' term, one example of the contraction is shown as follows:
\e
\left\langle\wick{\c1\zeta_g(\bm{p_1}) \c2\zeta_g(\bm{q-p_1}) \c3\zeta_g(\bm{p_2}) \c4\zeta_g(\bm{k-q-p_2}) \c1\zeta_g(\bm{p_3}) \c3\zeta_g(\bm{q'-p_3}) \c2\zeta_g(\bm{p_4}) \c4\zeta_g(\bm{k'-q'-p_4})}\right\rangle,
\q
and the symmetry factor in this case is 16. Then we have
\m
\Omega_{\mathrm{GW}}^{np}(k)=&&\frac{F_{\mathrm{NL}}^4k^3}{6\pi^2}\int\frac{\mathrm{d}^3q\mathrm{d}^3q'}{(2\pi)^6} \cos2(\phi-\phi') I\left(u,v\right) I\left(u',v'\right) \int\frac{\mathrm{d}^3p_1}{(2\pi)^3}\nonumber\\
&&\times P_{g}\left(p_1\right) P_{g}\left(|\bm{q-p_1}|\right) P_{g}\left(|\bm{q'-p_1}|\right) P_{g}\left(|\bm{k+p_1-q-q'}|\right)\times16\nonumber\\
=&&\frac{F_{\mathrm{NL}}^4}{6\pi^2}\int_0^{\infty} \mathrm{d}u \int_{|1-u|}^{1+u}\mathrm{d}v \int_0^{\infty} \mathrm{d}u' \int_{|1-u'|}^{1+u'}\mathrm{d}v' \int_0^{\infty} \mathrm{d}u_1 \int_{|1-u_1|}^{1+u_1}\mathrm{d}v_1 \int_0^{2\pi}\mathrm{d}\varphi_1 \int_0^{2\pi}\mathrm{d}\varphi_2 \nonumber\\
&&\cos2\varphi_1 I(u,v)I(u',v')  \frac{uvu'v'u_1v_1}{(u_1w_{13}w_{23}w_{0123})^3}\mathcal{P} _{g}\left(u_1k\right) \mathcal{P} _{g}\left(w_{13}k\right)\mathcal{P} _{g}\left(w_{23}k\right)\mathcal{P}_{g}\left(w_{0123}k\right),
\n
where $w_{0123}$ is defined as
\m\label{w0123}
w_{0123}^2=&&\frac{|\bm{k+p_1-q-q'}|^2}{k^2}\nonumber\\
=&&1+u^2+u'^2+u_1^2-2u\cos\theta-2u'\cos\theta'+2u_1\cos\theta_1+2uu'\(\sin\theta\sin\theta'\cos\varphi_1+\cos\theta\cos\theta'\)\nonumber\\
&&-2uu_1\(\sin\theta\sin\theta_1\cos\varphi_2+\cos\theta\cos\theta_1\)-2u'u_1\[\sin\theta'\sin\theta_1\cos(\varphi_1-\varphi_2)+\cos\theta'\cos\theta_1\].
\n

\subsection{$G_{\mathrm{NL}}$ terms}
For terms containing $G_{\mathrm{NL}}$, we can easily obtain that
\e
\Omega_{\mathrm{GW}}^{G_{\mathrm{NL}}}(k)=12G_{\mathrm{NL}}\int\frac{\mathrm{d}^3p}{(2\pi)^3}P_g(p) \Omega_{\mathrm{GW}}^g(k)=12G_{\mathrm{NL}}\int\frac{\mathrm{d}p}{p}\mathcal{P}_g(p) \Omega_{\mathrm{GW}}^g(k).
\q
In particular, we normalize the power spectrum $\mathcal{P}_g(p)$ to be
\e\label{norm}
A=\int\frac{\mathrm{d}p}{p}\mathcal{P}_g(p),
\q
then $\Omega_{\mathrm{GW}}^{G_{\mathrm{NL}}}(k)=12AG_{\mathrm{NL}}\Omega_{\mathrm{GW}}^g(k)$ holds, where $A$ represents the variance of the Gaussian part of the dimensionless curvature perturbation spectrum $\mathcal{P}_g(p)$.

\subsection{$G_{\mathrm{NL}}^2$ terms}
For terms containing $G_{\mathrm{NL}}^2$, considering symmetry, the GWs spectrum can be expressed in the following form:
\m
\Omega_{\mathrm{GW}}^{G_{\mathrm{NL}}^2}(k)=&&\frac{G_{\mathrm{NL}}^2k^3}{6\pi^2}\int\frac{\mathrm{d}^3q\mathrm{d}^3q'}{(2\pi)^6} \cos2(\phi-\phi') I\left(u,v\right) I\left(u',v'\right) \int\frac{\mathrm{d}^3p_1\mathrm{d}^3p_2\mathrm{d}^3p_3\mathrm{d}^3p_4}{(2\pi)^{12}}\nonumber\\
&&\times\Bigg[2\left\langle\left\langle\zeta_g(\bm{p_1})\zeta_g(\bm{p_2})\zeta_g(\bm{q-p_1-p_2})\zeta_g(\bm{p_3})\zeta_g(\bm{p_4})\zeta_g(\bm{k-q-p_3-p_4})\zeta_g(\bm{q'})\zeta_g(\bm{k'-q'})\right\rangle\right\rangle\nonumber\\
&&+4\left\langle\left\langle\zeta_g(\bm{p_1})\zeta_g(\bm{p_2})\zeta_g(\bm{q-p_1-p_2})\zeta_g(\bm{k-q})\zeta_g(\bm{p_3})\zeta_g(\bm{p_4})\zeta_g(\bm{q'-p_3-p_4})\zeta_g(\bm{k'-q'})\right\rangle\right\rangle\Bigg]. 
\n
Performing Wick contraction on the above eight-point function, there are four distinct non-zero contractions and we name them as '2loop' term, 'tri' term, 'ring1' term, and 'ring2' term. Then we have
\e
\Omega_{\mathrm{GW}}^{G_{\mathrm{NL}}^2}(k)=\Omega_{\mathrm{GW}}^{2loop}(k)+\Omega_{\mathrm{GW}}^{tri}(k)+\Omega_{\mathrm{GW}}^{ring1}(k)+\Omega_{\mathrm{GW}}^{ring2}(k).
\q
We now demonstrate each of these four parts in detail.

The `2loop' term can be easily obtained as
\e
\Omega_{\mathrm{GW}}^{2loop}(k)=54G_{\mathrm{NL}}^2\int\frac{\mathrm{d}^3p_1\mathrm{d}^3p_2}{(2\pi)^6}P_g(p_1)P_g(p_2) \Omega_{\mathrm{GW}}^g(k)=54G_{\mathrm{NL}}^2\int\frac{\mathrm{d}p_1\mathrm{d}p_2}{p_1p_2}\mathcal{P}_g(p_1)\mathcal{P}_g(p_2) \Omega_{\mathrm{GW}}^g(k),
\q
and for power spectrum satisfying \Eq{norm}, we have $\Omega_{\mathrm{GW}}^{2loop}(k)=54A^2G_{\mathrm{NL}}^2\Omega_{\mathrm{GW}}^g(k)$.

For the 'tri' term, one example of the contraction is shown as follows:
\e
\left\langle\wick{\c1\zeta_g(\bm{p_1}) \c2\zeta_g(\bm{p_2}) \c3\zeta_g(\bm{q-p_1-p_2}) \c4\zeta_g(\bm{k-q}) \c1\zeta_g(\bm{p_3}) \c2\zeta_g(\bm{p_4}) \c3\zeta_g(\bm{q'-p_3-p_4}) \c4\zeta_g(\bm{k'-q'})}\right\rangle,
\q
which is a disconnected diagram and the symmetry factor in this case is 6. Expanding the correlation function and using the appearing delta functions to eliminate redundant integrals, we can obtain:
\m
\Omega_{\mathrm{GW}}^{tri}(k)=&&\frac{G_{\mathrm{NL}}^2k^3}{6\pi^2}\int\frac{\mathrm{d}^3q}{(2\pi)^3}  I^2\left(u,v\right) \int\frac{\mathrm{d}^3p_1\mathrm{d}^3p_2}{(2\pi)^6}4P_{g}\left(p_1\right) P_{g}\left(p_2\right) P_{g}\left(|\bm{q-p_1-p_2}|\right)P_{g}\left(|\bm{k-q}|\right)\times6\nonumber\\
=&&G_{\mathrm{NL}}^2 \int_0^{\infty} \mathrm{d}u \int_{|1-u|}^{1+u}\mathrm{d}v \int_0^{\infty} \mathrm{d}u_1 \int_{|1-u_1|}^{1+u_1}\mathrm{d}v_1 \int_0^{\infty} \mathrm{d}u_2 \int_{|1-u_2|}^{1+u_2}\mathrm{d}v_2 I^2(u,v) \frac{1}{u^2v^2u_1^2v_1^2u_2^2v_2^2}\nonumber\\
&&\times\mathcal{P} _{g}\left(u_1uk\right) \mathcal{P} _{g}\left(u_2v_1uk\right)\mathcal{P} _{g}\left(v_2v_1uk\right) \mathcal{P} _{g}\left(vk\right),
\n
where the second equality in the above equation is obtained by performing the coordinate transformation $u_1 = p_1 / q$, $ v_1=|\bm{q-p_1}| / q$ and $u_2 = p_2 / |\bm{q-p_1}|$, $ v_2=|\bm{q-p_1-p_2}| / |\bm{q-p_1}|$.

For the ’ring1’ term, one example of the contraction is shown as follows:
\e
\left\langle\wick{\c1\zeta_g(\bm{p_1}) \c2\zeta_g(\bm{p_2}) \c3\zeta_g(\bm{q-p_1-p_2}) \c1\zeta_g(\bm{p_3}) \c2\zeta_g(\bm{p_4}) \c4\zeta_g(\bm{k-q-p_3-p_4}) \c3\zeta_g(\bm{q'}) \c4\zeta_g(\bm{k'-q'})}\right\rangle,
\q
and the symmetry factor in this case is 36. Then we have
\m
\Omega_{\mathrm{GW}}^{ring1}(k)=&&\frac{G_{\mathrm{NL}}^2k^3}{6\pi^2}\int\frac{\mathrm{d}^3q\mathrm{d}^3q'}{(2\pi)^6} \cos2(\phi-\phi') I\left(u,v\right) I\left(u',v'\right) \int\frac{\mathrm{d}^3p}{(2\pi)^3}2 P_{g}\left(p\right) P_{g}\left(q'\right) P_{g}\left(|\bm{k-q'}|\right) P_{g}\left(|\bm{q+q'+p}|\right)\times36\nonumber\\
=&&\frac{3G_{\mathrm{NL}}^2}{4\pi^2}\int_0^{\infty} \mathrm{d}u \int_{|1-u|}^{1+u}\mathrm{d}v \int_0^{\infty} \mathrm{d}u' \int_{|1-u'|}^{1+u'}\mathrm{d}v' \int_0^{\infty} \mathrm{d}u_1 \int_{|1-u_1|}^{1+u_1}\mathrm{d}v_1 \int_0^{2\pi}\mathrm{d}\varphi_1 \int_0^{2\pi}\mathrm{d}\varphi_2 \nonumber\\
&&\times\cos2\varphi_1 I(u,v)I(u',v')  \frac{uvu'v'u_1v_1}{(u'v'u_1w_{123})^3}\mathcal{P}_{g}\left(u'k\right) \mathcal{P}_{g}\left(v'k\right)\mathcal{P}_{g}\left(u_1k\right)\mathcal{P}_{g}\left(w_{123}k\right),
\n
where we perform the coordinate transformation $\varphi_1=\phi-\phi'$, $\varphi_2=\phi-\phi_1$, and $\varphi_3=\phi+\phi'$. Additionally, $w_{123}$ is defined as follows:
\m
w_{123}^2=&&\frac{|\bm{q+q'+p}|^2}{k^2}\nonumber\\
=&&u^2+u'^2+u_1^2+2uu'\(\sin\theta\sin\theta'\cos\varphi_1+\cos\theta\cos\theta'\)+2uu_1(\sin\theta\sin\theta_1\cos\varphi_2+\cos\theta\cos\theta_1)\nonumber\\
&&+2u'u_1\[\sin\theta'\sin\theta_1\cos(\varphi_1-\varphi_2)+\cos\theta'\cos\theta_1\],
\n
where $\sin$ and $\cos$ are replaced using \Eq{sincos}.

For the `ring2’ term, one example of the contraction is shown as follows:
\e
\left\langle\wick{\c1\zeta_g(\bm{p_1}) \c2\zeta_g(\bm{p_2}) \c3\zeta_g(\bm{q-p_1-p_2}) \c4\zeta_g(\bm{k-q}) \c1\zeta_g(\bm{p_3}) \c2\zeta_g(\bm{p_4}) \c4\zeta_g(\bm{q'-p_3-p_4}) \c3\zeta_g(\bm{k'-q'})}\right\rangle,
\q
and the symmetry factor in this case is 18. Then we have
\m
\Omega_{\mathrm{GW}}^{ring2}(k)=&&\frac{G_{\mathrm{NL}}^2k^3}{6\pi^2}\int\frac{\mathrm{d}^3q\mathrm{d}^3q'}{(2\pi)^6} \cos2(\phi-\phi') I\left(u,v\right) I\left(u',v'\right) \int\frac{\mathrm{d}^3p}{(2\pi)^3}\nonumber\\
&&\times 4 P_{g}\left(p\right) P_{g}\left(|\bm{k-q}|\right) P_{g}\left(|\bm{k-q'}|\right) P_{g}\left(|\bm{k+p-q-q'}|\right)\times18\nonumber\\
=&&\frac{3G_{\mathrm{NL}}^2}{4\pi^2}\int_0^{\infty} \mathrm{d}u \int_{|1-u|}^{1+u}\mathrm{d}v \int_0^{\infty} \mathrm{d}u' \int_{|1-u'|}^{1+u'}\mathrm{d}v' \int_0^{\infty} \mathrm{d}u_1 \int_{|1-u_1|}^{1+u_1}\mathrm{d}v_1 \int_0^{2\pi}\mathrm{d}\varphi_1 \int_0^{2\pi}\mathrm{d}\varphi_2 \nonumber\\
&&\times\cos2\varphi_1 I(u,v)I(u',v')  \frac{uvu'v'u_1v_1}{(u_1 v v' w_{0123})^3}\mathcal{P}_{g}\left(u_1k\right) \mathcal{P}_{g}\left(vk\right)\mathcal{P}_{g}\left(v'k\right)\mathcal{P}_{g}\left(w_{0123}k\right),
\n
where $w_{0123}$ is defined as follows:
\m
w_{0123}^2=&&\frac{|\bm{k+p-q-q'}|^2}{k^2}\nonumber\\
=&&1+u^2+u'^2+u_1^2+2u_1\cos\theta_1-2u\cos\theta-2u'\cos\theta'+2uu'\(\sin\theta\sin\theta'\cos\varphi_1+\cos\theta\cos\theta'\)\nonumber\\
&&-2uu_1(\sin\theta\sin\theta_1\cos\varphi_2+\cos\theta\cos\theta_1)-2u'u_1\[\sin\theta'\sin\theta_1\cos(\varphi_1-\varphi_2)+\cos\theta'\cos\theta_1\].
\n

\subsection{$G_{\mathrm{NL}}^3$ terms}
For terms containing $G_{\mathrm{NL}}^3$, considering symmetry, the GWs spectrum can be expressed in the following form:
\m
\Omega_{\mathrm{GW}}^{G_{\mathrm{NL}}^3}(k)=&&\frac{G_{\mathrm{NL}}^3k^3}{6\pi^2}\int\frac{\mathrm{d}^3q\mathrm{d}^3q'}{(2\pi)^6} \cos2(\phi-\phi') I\left(u,v\right) I\left(u',v'\right) \int\frac{\mathrm{d}^3p_1\mathrm{d}^3p_2\mathrm{d}^3p_3\mathrm{d}^3p_4\mathrm{d}^3p_5\mathrm{d}^3p_6}{(2\pi)^{18}}4\langle\langle\zeta_g(\bm{p_1})\zeta_g(\bm{p_2})\nonumber\\
&&\times\zeta_g(\bm{q-p_1-p_2})\zeta_g(\bm{p_3})\zeta_g(\bm{p_4})\zeta_g(\bm{k-q-p_3-p_4})\zeta_g(\bm{p_5})\zeta_g(\bm{p_6})\zeta_g(\bm{q'-p_5-p_6})\zeta_g(\bm{k'-q'})\rangle\rangle.\nonumber\\
\n
Performing Wick contraction on the above ten-point function, there are four distinct non-zero contractions and we name them as the `1loop' term, the `3loop' term and the `ring3' term. Then we have
\e
\Omega_{\mathrm{GW}}^{G_{\mathrm{NL}}^3}(k)=\Omega_{\mathrm{GW}}^{1loop}(k)+\Omega_{\mathrm{GW}}^{3loop}(k)+\Omega_{\mathrm{GW}}^{ring3}(k),
\q
and we can easily obtain
\m
\Omega_{\mathrm{GW}}^{1loop}(k)=&&6G_{\mathrm{NL}}\int\frac{\mathrm{d}^3p}{(2\pi)^3}P_g(p) \(\Omega_{\mathrm{GW}}^{tri}(k)+\Omega_{\mathrm{GW}}^{ring1}(k)+\Omega_{\mathrm{GW}}^{ring2}(k)\)\nonumber\\
=&&6G_{\mathrm{NL}}\int\frac{\mathrm{d}p}{p}\mathcal{P}_g(p) \(\Omega_{\mathrm{GW}}^{tri}(k)+\Omega_{\mathrm{GW}}^{ring1}(k)+\Omega_{\mathrm{GW}}^{ring2}(k)\),
\n
and for power spectrum satisfying \Eq{norm}, we have $\Omega_{\mathrm{GW}}^{1loop}(k)=6AG_{\mathrm{NL}}\(\Omega_{\mathrm{GW}}^{tri}(k)+\Omega_{\mathrm{GW}}^{ring1}(k)+\Omega_{\mathrm{GW}}^{ring2}(k)\)$. 

While the '3loop' term can be expressed as
\m
\Omega_{\mathrm{GW}}^{3loop}(k)=&&108G_{\mathrm{NL}}^3\int\frac{\mathrm{d}^3p_1\mathrm{d}^3p_2\mathrm{d}^3p_3}{(2\pi)^9}P_g(p_1)P_g(p_2)P_g(p_3) \Omega_{\mathrm{GW}}^g(k)\nonumber\\
=&&108G_{\mathrm{NL}}^3\int\frac{\mathrm{d}p_1\mathrm{d}p_2\mathrm{d}p_3}{p_1p_2p_3}\mathcal{P}_g(p_1)\mathcal{P}_g(p_2)\mathcal{P}_g(p_3) \Omega_{\mathrm{GW}}^g(k),
\n
and for power spectrum satisfying \Eq{norm}, we have $\Omega_{\mathrm{GW}}^{3loop}(k)=108A^3G_{\mathrm{NL}}^3\Omega_{\mathrm{GW}}^g(k)$.

The `ring3' term is a disconnected diagram and one example of the contraction is shown as follows:
\e
\left\langle\wick{
\c1\zeta_g(\bm{p_1}) \c2\zeta_g(\bm{p_2}) \c3\zeta_g(\bm{q-p_1-p_2}) \c3\zeta_g(\bm{p_3}) \c4\zeta_g(\bm{p_4}) \c5\zeta_g(\bm{k-q-p_3-p_4})) \c1\zeta_g(\bm{p_5}) \c2\zeta_g(\bm{p_6}) \c4\zeta_g(\bm{q'-p_5-p_6})) \c5\zeta_g(\bm{k'-q'})}\right\rangle,
\q
and the symmetry factor in this contraction is 216. Then we have
\m
\Omega_{\mathrm{GW}}^{ring3}(k)=&&\frac{G_{\mathrm{NL}}^3k^3}{6\pi^2}\int\frac{\mathrm{d}^3q\mathrm{d}^3q'}{(2\pi)^6} \cos2(\phi-\phi') I\left(u,v\right) I\left(u',v'\right) \int\frac{\mathrm{d}^3p_1\mathrm{d}^3p_2}{(2\pi)^6}4\times 216
\nonumber\\
\times&& P_{g}\left(p_1\right) P_{g}\left(p_2\right) P_{g}\left(|\bm{q-p_1-p_2}|\right) P_{g}\left(|\bm{q'-p_1-p_2}|\right)
P_{g}\left(|\bm{k-q'}|\right)\nonumber\\
=&&\frac{9G_{\mathrm{NL}}^3}{4\pi^2}\int_0^{\infty} \mathrm{d}u \int_{|1-u|}^{1+u}\mathrm{d}v \int_0^{\infty} \mathrm{d}u' \int_{|1-u'|}^{1+u'}\mathrm{d}v' \int_0^{\infty} \mathrm{d}u_1 \int_{|1-u_1|}^{1+u_1}\mathrm{d}v_1 
\int_0^{\infty} \mathrm{d}u_2 \int_{|1-u_2|}^{1+u_2}\mathrm{d}v_2 \int_0^{2\pi}\mathrm{d}\varphi_1 
\int_0^{2\pi}\mathrm{d}\varphi_2\int_0^{2\pi}\mathrm{d}\varphi_3 \nonumber\\
&&\times\cos2\varphi_1 I(u,v)I(u',v')  \frac{uvu'v'u_1v_1u_2v_2}{(u_1u_2w_{134}w_{234}v')^3}\mathcal{P}_{g}\left(u_1k\right) \mathcal{P}_{g}\left(u_2k\right)\mathcal{P}_{g}\left(w_{134}k\right)\mathcal{P}_{g}\left(w_{234}k\right)\mathcal{P}_{g}\left(v'k\right),
\n
where we perform the coordinate transformation $\varphi_1=\phi-\phi'$, $\varphi_2=\phi-\phi_1$, $\varphi_3=\phi-\phi_2$, $\varphi_4=\phi+\phi_2$ and $\varphi_4$ can be integrated out to get $2\pi$. 
Additionally, $w_{134}$ and $w_{234}$ are defined as
\m
w_{134}^2=&&\frac{|\bm{q-p_1-p_2}|^2}{k^2}\nonumber\\
=&&u^2+u_1^2+u_2^2-2uu_1\[\sin\theta\sin\theta_1\cos\varphi_2+\cos\theta\cos\theta_1\]-2uu_2(\sin\theta\sin\theta_2\cos\varphi_3+\cos\theta\cos\theta_2)\nonumber\\
&&+2u_1u_2(\sin\theta_1\sin\theta_2\cos(\varphi_2-\varphi_3)+\cos\theta_1\cos\theta_2),\\
w_{234}^2=&&\frac{|\bm{q'-p_1-p_2}|^2}{k^2}\nonumber\\
=&&u'^2+u_1^2+u_2^2-2u'u_1\[\sin\theta'\sin\theta_1\cos(\varphi_1-\varphi_2)+\cos\theta'\cos\theta_1\]-2u'u_2(\sin\theta'\sin\theta_2\cos(\varphi_1-\varphi_3)+\cos\theta'\cos\theta_2)\nonumber\\
&&+2u_1u_2(\sin\theta_1\sin\theta_2\cos(\varphi_2-\varphi_3)+\cos\theta_1\cos\theta_2).
\n


\subsection{$G_{\mathrm{NL}}^4$ terms}
For terms containing $G_{\mathrm{NL}}^4$, considering symmetry, the GWs spectrum can be expressed in the following form:
\begin{equation}
\begin{aligned}
\Omega_{\mathrm{GW}}^{G_{\mathrm{NL}}^4}(k)= & \frac{G_{\mathrm{NL}}^4 k^3}{6 \pi^2} \int \frac{\mathrm{d}^3 q \mathrm{~d}^3 q^{\prime}}{(2 \pi)^6} \cos 2\left(\phi-\phi^{\prime}\right) I(u, v) I\left(u^{\prime}, v^{\prime}\right) \int \frac{\mathrm{d}^3 p_1 \mathrm{~d}^3 p_2 \mathrm{~d}^3 p_3 \mathrm{~d}^3 p_4 \mathrm{~d}^3 p_5 \mathrm{~d}^3 p_6 \mathrm{~d}^3 p_7 \mathrm{~d}^3 p_8}{(2 \pi)^{24}} \\
& \times\left\langle\left\langle\zeta_g\left(\boldsymbol{p}_{\mathbf{1}}\right) \zeta_g\left(\boldsymbol{p}_{\mathbf{2}}\right) \zeta_g\left(\boldsymbol{q}-\boldsymbol{p}_{\mathbf{1}}-\boldsymbol{p}_{\mathbf{2}}\right) \zeta_g\left(\boldsymbol{p}_3\right) \zeta_g\left(\boldsymbol{p}_{\mathbf{4}}\right) \zeta_g\left(\boldsymbol{k}-\boldsymbol{q}-\boldsymbol{p}_{\mathbf{3}}-\boldsymbol{p}_{\mathbf{4}}\right) \zeta_g\left(\boldsymbol{p}_{\mathbf{5}}\right) \zeta_g\left(\boldsymbol{p}_{\mathbf{6}}\right) \zeta_g\left(\boldsymbol{q}^{\prime}-\boldsymbol{p}_{\mathbf{5}}-\boldsymbol{p}_{\mathbf{6}}\right)\right.\right. \\
& \left.\left.\zeta_g\left(\boldsymbol{p}_{\boldsymbol{7}}\right) \zeta_g\left(\boldsymbol{p}_{\boldsymbol{8}}\right) \zeta_g\left(\boldsymbol{k}^{\prime}-\boldsymbol{q}^{\prime}-\boldsymbol{p}_{\boldsymbol{7}}-\boldsymbol{p}_{\boldsymbol{8}}\right)\right\rangle\right\rangle
\end{aligned}
\end{equation}
Performing Wick contraction on the above twelve-point function, there are 8 distinct non-zero contractions and we name them as the `2loops' term, the `4loop' term, the `ring3loop' term, the `double' term, the `bubble' term, the `sand clock'(sc) term, the `2rings' term and the `net' term. Then we have
\e
\Omega_{\mathrm{GW}}^{G_{\mathrm{NL}}^4}(k)=\Omega_{\mathrm{GW}}^{2loops}(k)+\Omega_{\mathrm{GW}}^{4loop}(k)+\Omega_{\mathrm{GW}}^{ring3loop}(k)+\Omega_{\mathrm{GW}}^{double}(k)+\Omega_{\mathrm{GW}}^{bubble}(k)+\Omega_{\mathrm{GW}}^{sc}(k)+\Omega_{\mathrm{GW}}^{2rings}(k)+\Omega_{\mathrm{GW}}^{net}(k),
\q
and we now demonstrate each of these 8 parts in detail.

We can easily obtain
\m
\Omega_{\mathrm{GW}}^{2loops}(k)=&&9G_{\mathrm{NL}}^2\int\frac{\mathrm{d}^3p_1\mathrm{d}^3p_2}{(2\pi)^6}P_g(p_1)P_g(p_2) \(\Omega_{\mathrm{GW}}^{tri}(k)+\Omega_{\mathrm{GW}}^{ring1}(k)+\Omega_{\mathrm{GW}}^{ring2}(k)\)\nonumber\\
=&&9G_{\mathrm{NL}}^2\int\frac{\mathrm{d}p_1\mathrm{d}p_2}{p_1p_2}\mathcal{P}_g(p_1)\mathcal{P}_g(p_2) \(\Omega_{\mathrm{GW}}^{tri}(k)+\Omega_{\mathrm{GW}}^{ring1}(k)+\Omega_{\mathrm{GW}}^{ring2}(k)\),
\n
and for power spectrum satisfying \Eq{norm}, we have $\Omega_{\mathrm{GW}}^{2loops}(k)=9A^2G_{\mathrm{NL}}^2\(\Omega_{\mathrm{GW}}^{tri}(k)+\Omega_{\mathrm{GW}}^{ring1}(k)+\Omega_{\mathrm{GW}}^{ring2}(k)\)$.

The `4loop' term can also be easily obtained as
\m
\Omega_{\mathrm{GW}}^{4loop}(k)=81G_{\mathrm{NL}}^4\int\frac{\mathrm{d}p_1\mathrm{d}p_2\mathrm{d}p_3\mathrm{d}p_4}{p_1p_2p_3p_4}\mathcal{P}_g(p_1)\mathcal{P}_g(p_2)\mathcal{P}_g(p_3)\mathcal{P}_g(p_4) \Omega_{\mathrm{GW}}^g(k),
\n
and for power spectrum satisfying \Eq{norm}, we have $\Omega_{\mathrm{GW}}^{4loop}(k)=81A^4G_{\mathrm{NL}}^4\Omega_{\mathrm{GW}}^g(k)$.

Similarly, the `ring3loop' term can be easily obtained as
\m
\Omega_{\mathrm{GW}}^{ring3loop}(k)=3AG_{\mathrm{NL}}\Omega_{\mathrm{GW}}^{ring3}(k)
\n

The `double' term is a disconnected diagram and one example of the contraction is shown as follows:
\e
\left\langle\wick{\c1\zeta_g(\bm{p_1}) \c2\zeta_g(\bm{p_2}) \c3\zeta_g(\bm{p_9}) \c4\zeta_g(\bm{p_3}) \c5\zeta_g(\bm{p_4}) \c6\zeta_g(\bm{p_{10}}) \c1\zeta_g(\bm{p_5}) \c2\zeta_g(\bm{p_6}) \c3\zeta_g(\bm{p_{11}}) \c4\zeta_g(\bm{p_7}) \c5\zeta_g(\bm{p_8}) 
\c6\zeta_g(\bm{p_{12}})}\right\rangle,
\q
where $\bm{p_9}\equiv\bm{q-p_1-p_2}$, $\bm{p_{10}}\equiv\bm{k-q-p_3-p_4}$, $\bm{p_{11}}\equiv\bm{q'-p_5-p_6}$ and $\bm{p_{12}}\equiv\bm{k'-q'-p_7-p_8}$.
The symmetry factor in this case is 72. Expanding the correlation function and using the appearing delta functions to eliminate redundant integrals, we can obtain:
\m
\Omega_{\mathrm{GW}}^{double}(k)=&&\frac{G_{\mathrm{NL}}^4k^3}{6\pi^2}\int\frac{\mathrm{d}^3q}{(2\pi)^3}  I^2\left(u,v\right) \int\frac{\mathrm{d}^3p_1\mathrm{d}^3p_2\mathrm{d}^3p_3\mathrm{d}^3p_4}{(2\pi)^{12}}P_{g}\left(p_1\right) P_{g}\left(p_2\right) P_{g}\left(|\bm{q-p_1-p_2}|\right)\nonumber\\
&&\times P_{g}\left(p_3\right) P_{g}\left(p_4\right) P_{g}\left(|\bm{k-q-p_3-p_4}|\right)\times72\nonumber\\
=&&\frac{3G_{\mathrm{NL}}^4}{4} \int_0^{\infty} \mathrm{d}u \int_{|1-u|}^{1+u}\mathrm{d}v \int_0^{\infty} \mathrm{d}u_1 \int_{|1-u_1|}^{1+u_1}\mathrm{d}v_1 \int_0^{\infty} \mathrm{d}u_2 \int_{|1-u_2|}^{1+u_2}\mathrm{d}v_2 \int_0^{\infty} \mathrm{d}u_3 \int_{|1-u_3|}^{1+u_3}\mathrm{d}v_3 \int_0^{\infty} \mathrm{d}u_4 \int_{|1-u_4|}^{1+u_4}\mathrm{d}v_4\nonumber\\
&&\times I^2(u,v) \frac{1}{u^2v^2u_1^2v_1^2u_2^2v_2^2u_3^2v_3^2u_4^2v_4^2}\mathcal{P} _{g}\left(u_1uk\right) \mathcal{P} _{g}\left(u_2v_1uk\right)\mathcal{P} _{g}\left(v_2v_1uk\right)\mathcal{P} _{g}\left(u_3vk\right) \mathcal{P} _{g}\left(u_4v_3vk\right)\mathcal{P} _{g}\left(v_4v_3vk\right),\nonumber\\
\n
where the second equality in the above equation is obtained by performing the coordinate transformation $u_1 = p_1 / q$, $ v_1=|\bm{q-p_1}| / q$, $u_2 = p_2 / |\bm{q-p_1}|$, $ v_2=|\bm{q-p_1-p_2}| / |\bm{q-p_1}|$, $u_3 = p_3 / |\bm{k-q}|$, $ v_3=|\bm{k-q-p_3}| /|\bm{k-q}|$, $u_4 = p_4 / |\bm{k-q-p_3}|$, $ v_4=|\bm{k-q-p_3-p_4}| / |\bm{k-q-p_3}|$.

For the ’bubble’ term, one example of the contraction is shown as follows:
\e
\left\langle\wick{\c1\zeta_g(\bm{p_1}) \c2\zeta_g(\bm{p_2}) \c5\zeta_g(\bm{p_9}) \c3\zeta_g(\bm{p_3}) \c4\zeta_g(\bm{p_4}) \c5\zeta_g(\bm{p_{10}}) \c1\zeta_g(\bm{p_5}) \c2\zeta_g(\bm{p_6}) \c5\zeta_g(\bm{p_{11}}) \c3\zeta_g(\bm{p_7}) \c4\zeta_g(\bm{p_8}) 
\c5\zeta_g(\bm{p_{12}})}\right\rangle,
\q
where $\bm{p_9}$, $\bm{p_{10}}$, $\bm{p_{11}}$ and $\bm{p_{12}}$ are defined the same as above and the symmetry factor in this case is 648. Then we have
\m
\Omega_{\mathrm{GW}}^{bubble}(k)=&&\frac{G_{\mathrm{NL}}^4k^3}{6\pi^2}\int\frac{\mathrm{d}^3q\mathrm{d}^3q'}{(2\pi)^6} \cos2(\phi-\phi') I\left(u,v\right) I\left(u',v'\right) \int\frac{\mathrm{d}^3p_1\mathrm{d}^3p_2\mathrm{d}^3p_3}{(2\pi)^9}\nonumber\\
&&\times P_{g}\left(p_1\right) P_{g}\left(p_2\right) P_{g}\left(p_3\right) P_{g}\left(\bm{k-p_1-p_2-p_3}\right) P_{g}\left(|\bm{q-p_1-p_2}|\right) P_{g}\left(|\bm{q'-p_1-p_2}|\right)\times648\nonumber\\
=&&\frac{27G_{\mathrm{NL}}^4}{64\pi^4}\int_0^{\infty} \mathrm{d}u \int_{|1-u|}^{1+u}\mathrm{d}v \int_0^{\infty} \mathrm{d}u' \int_{|1-u'|}^{1+u'}\mathrm{d}v' \int_0^{\infty} \mathrm{d}u_1 \int_{|1-u_1|}^{1+u_1}\mathrm{d}v_1 \int_0^{\infty} \mathrm{d}u_2 \int_{|1-u_2|}^{1+u_2}\mathrm{d}v_2\int_0^{\infty} \mathrm{d}u_3 \int_{|1-u_3|}^{1+u_3}\mathrm{d}v_3\nonumber\\
&&\times\int_0^{2\pi}\mathrm{d}\varphi_1 \int_0^{2\pi}\mathrm{d}\varphi_2 \int_0^{2\pi}\mathrm{d}\varphi_3 \int_0^{2\pi}\mathrm{d}\varphi_4  \cos2\varphi_1 I(u,v)I(u',v')  \frac{uvu'v'u_1v_1u_2v_2u_3v_3}{(u_1u_2u_3w_{0345}w_{134}w_{234})^3}\nonumber\\
&&\times\mathcal{P}_{g}\left(u_1k\right) \mathcal{P}_{g}\left(u_2k\right)\mathcal{P}_{g}\left(u_3k\right)\mathcal{P}_{g}\left(w_{0345}k\right)\mathcal{P}_{g}\left(w_{134}k\right)\mathcal{P}_{g}\left(w_{234}k\right),
\n
where we perform the coordinate transformation $\varphi_1=\phi-\phi'$, $\varphi_2=\phi-\phi_1$, $\varphi_3=\phi-\phi_2$, $\varphi_4=\phi-\phi_3$, and $\varphi_5=\phi+\phi'$. Additionally, $w_{0345}$ is defined as follows:
\m
w_{0345}^2=&&\frac{|\bm{k-p_1-p_2-p_3}|^2}{k^2}\nonumber\\
=&&1+u_1^2+u_2^2+u_3^2-2u_1\cos\theta_1-2u_2\cos\theta_2-2u_3\cos\theta_3+2u_1u_2\[\sin\theta_1\sin\theta_2\cos(\varphi_2-\varphi_3)+\cos\theta_1\cos\theta_2\]\nonumber\\
&&+2u_1u_3\[\sin\theta_1\sin\theta_3\cos(\varphi_2-\varphi_4)+\cos\theta_1\cos\theta_3\]+2u_2u_3\[\sin\theta_2\sin\theta_3\cos(\varphi_3-\varphi_4)+\cos\theta_2\cos\theta_3\]
\n

For the ’sand clock’ term, one example of the contraction is shown as follows:
\e
\left\langle\wick{\c1\zeta_g(\bm{p_1}) \c2\zeta_g(\bm{p_2}) \c5\zeta_g(\bm{p_9}) \c3\zeta_g(\bm{p_3}) \c4\zeta_g(\bm{p_4}) \c6\zeta_g(\bm{p_{10}}) \c1\zeta_g(\bm{p_5}) \c2\zeta_g(\bm{p_6}) \c6\zeta_g(\bm{p_{11}}) \c3\zeta_g(\bm{p_7}) \c4\zeta_g(\bm{p_8}) 
\c5\zeta_g(\bm{p_{12}})}\right\rangle,
\q
where $\bm{p_9}$, $\bm{p_{10}}$, $\bm{p_{11}}$ and $\bm{p_{12}}$ are defined the same as above and the symmetry factor in this case is 648. Then we have
\m
\Omega_{\mathrm{GW}}^{sc}(k)=&&\frac{G_{\mathrm{NL}}^4k^3}{6\pi^2}\int\frac{\mathrm{d}^3q\mathrm{d}^3q'}{(2\pi)^6} \cos2(\phi-\phi') I\left(u,v\right) I\left(u',v'\right) \int\frac{\mathrm{d}^3p_1\mathrm{d}^3p_2\mathrm{d}^3p_3}{(2\pi)^9}\nonumber\\
&&\times P_{g}\left(p_1\right) P_{g}\left(p_2\right) P_{g}\left(p_3\right) P_{g}\left(|\bm{q-p_1-p_2}|\right) P_{g}\left(|\bm{q'-p_1-p_2}|\right) P_{g}\left(\bm{k-q-q'+p_1+p_2-p_3}\right)\times648\nonumber\\
=&&\frac{27G_{\mathrm{NL}}^4}{64\pi^4}\int_0^{\infty} \mathrm{d}u \int_{|1-u|}^{1+u}\mathrm{d}v \int_0^{\infty} \mathrm{d}u' \int_{|1-u'|}^{1+u'}\mathrm{d}v' \int_0^{\infty} \mathrm{d}u_1 \int_{|1-u_1|}^{1+u_1}\mathrm{d}v_1 \int_0^{\infty} \mathrm{d}u_2 \int_{|1-u_2|}^{1+u_2}\mathrm{d}v_2\int_0^{\infty} \mathrm{d}u_3 \int_{|1-u_3|}^{1+u_3}\mathrm{d}v_3\nonumber\\
&&\times\int_0^{2\pi}\mathrm{d}\varphi_1 \int_0^{2\pi}\mathrm{d}\varphi_2 \int_0^{2\pi}\mathrm{d}\varphi_3 \int_0^{2\pi}\mathrm{d}\varphi_4  \cos2\varphi_1 I(u,v)I(u',v')  \frac{uvu'v'u_1v_1u_2v_2u_3v_3}{(u_1u_2u_3w_{134}w_{234}w)^3}\nonumber\\
&&\times\mathcal{P}_{g}\left(u_1k\right) \mathcal{P}_{g}\left(u_2k\right)\mathcal{P}_{g}\left(u_3k\right)\mathcal{P}_{g}\left(w_{134}k\right)\mathcal{P}_{g}\left(w_{234}k\right)\mathcal{P}_{g}\left(wk\right),
\n
where we perform the coordinate transformation the same as above. Additionally, $w$ is defined as follows:
\m
w^2=&&\frac{|\bm{k-q-q'+p_1+p_2-p_3}|^2}{k^2}\nonumber\\
=&&1+u^2+u'^2+u_1^2+u_2^2+u_3^2-2u\cos\theta-2u'\cos\theta'+2u_1\cos\theta_1+2u_2\cos\theta_2-2u_3\cos\theta_3\nonumber\\
&&+2uu'\[\sin\theta\sin\theta'\cos\varphi_1+\cos\theta\cos\theta'\]-2uu_1\[\sin\theta\sin\theta_1\cos\varphi_2+\cos\theta\cos\theta_1\]-2uu_2\[\sin\theta\sin\theta_2\cos\varphi_3+\cos\theta\cos\theta_2\]\nonumber\\
&&+2uu_3\[\sin\theta\sin\theta_3\cos\varphi_4+\cos\theta\cos\theta_3\]-2u'u_1\[\sin\theta'\sin\theta_1\cos(\varphi_1-\varphi_2)+\cos\theta'\cos\theta_1\]\nonumber\\
&&-2u'u_2\[\sin\theta'\sin\theta_2\cos(\varphi_1-\varphi_3)+\cos\theta'\cos\theta_2\]+2u'u_3\[\sin\theta'\sin\theta_3\cos(\varphi_1-\varphi_4)+\cos\theta'\cos\theta_3\]\nonumber\\
&&+2u_1u_2\[\sin\theta_1\sin\theta_2\cos(\varphi_2-\varphi_3)+\cos\theta_1\cos\theta_2\]-2u_1u_3\[\sin\theta_1\sin\theta_3\cos(\varphi_2-\varphi_4)+\cos\theta_1\cos\theta_3\]\nonumber\\
&&-2u_2u_3\[\sin\theta_2\sin\theta_3\cos(\varphi_3-\varphi_4)+\cos\theta_2\cos\theta_3\].
\n

For the ’2rings’ term, one example of the contraction is shown as follows:
\e
\left\langle\wick{\c1\zeta_g(\bm{p_1}) \c2\zeta_g(\bm{p_2}) \c4\zeta_g(\bm{p_9}) \c1\zeta_g(\bm{p_3}) \c2\zeta_g(\bm{p_4}) \c3\zeta_g(\bm{p_{10}}) \c1\zeta_g(\bm{p_5}) \c2\zeta_g(\bm{p_6}) \c3\zeta_g(\bm{p_{11}}) \c1\zeta_g(\bm{p_7}) \c2\zeta_g(\bm{p_8}) 
\c4\zeta_g(\bm{p_{12}})}\right\rangle,
\q
where $\bm{p_9}$, $\bm{p_{10}}$, $\bm{p_{11}}$ and $\bm{p_{12}}$ are defined the same as above and the symmetry factor in this case is 648. Then we have
\m
\Omega_{\mathrm{GW}}^{2rings}(k)=&&\frac{G_{\mathrm{NL}}^4k^3}{6\pi^2}\int\frac{\mathrm{d}^3q\mathrm{d}^3q'}{(2\pi)^6} \cos2(\phi-\phi') I\left(u,v\right) I\left(u',v'\right) \int\frac{\mathrm{d}^3p_1\mathrm{d}^3p_2\mathrm{d}^3p_3}{(2\pi)^9}\nonumber\\
&&\times P_{g}\left(p_1\right) P_{g}\left(p_2\right) P_{g}\left(p_3\right) P_{g}\left(|\bm{q-p_1-p_2}|\right) P_{g}\left(|\bm{k-q+p_1+p_2}|\right) P_{g}\left(\bm{k-q-q'+p_1+p_2-p_3}\right)\times648\nonumber\\
=&&\frac{27G_{\mathrm{NL}}^4}{64\pi^4}\int_0^{\infty} \mathrm{d}u \int_{|1-u|}^{1+u}\mathrm{d}v \int_0^{\infty} \mathrm{d}u' \int_{|1-u'|}^{1+u'}\mathrm{d}v' \int_0^{\infty} \mathrm{d}u_1 \int_{|1-u_1|}^{1+u_1}\mathrm{d}v_1 \int_0^{\infty} \mathrm{d}u_2 \int_{|1-u_2|}^{1+u_2}\mathrm{d}v_2\int_0^{\infty} \mathrm{d}u_3 \int_{|1-u_3|}^{1+u_3}\mathrm{d}v_3\nonumber\\
&&\times\int_0^{2\pi}\mathrm{d}\varphi_1 \int_0^{2\pi}\mathrm{d}\varphi_2 \int_0^{2\pi}\mathrm{d}\varphi_3 \int_0^{2\pi}\mathrm{d}\varphi_4  \cos2\varphi_1 I(u,v)I(u',v')  \frac{uvu'v'u_1v_1u_2v_2u_3v_3}{(u_1u_2u_3w_{134}w_{0134}w)^3}\nonumber\\
&&\times\mathcal{P}_{g}\left(u_1k\right) \mathcal{P}_{g}\left(u_2k\right)\mathcal{P}_{g}\left(u_3k\right)\mathcal{P}_{g}\left(w_{134}k\right)\mathcal{P}_{g}\left(w_{0134}k\right)\mathcal{P}_{g}\left(wk\right),
\n
where we perform the coordinate transformation the same as above. Additionally, $w_{0134}$ is defined as follows:
\m\label{w0134}
w_{0134}^2=&&\frac{|\bm{k-q+p_1+p_2}|^2}{k^2}\nonumber\\
=&&1+u^2+u_1^2+u_2^2-2u\cos\theta+2u_1\cos\theta_1+2u_2\cos\theta_2-2uu_1\[\sin\theta\sin\theta_1\cos\varphi_2+\cos\theta\cos\theta_1\]\nonumber\\
&&-2uu_2\[\sin\theta\sin\theta_2\cos\varphi_3+\cos\theta\cos\theta_2\]+2u_1u_2\[\sin\theta_1\sin\theta_2\cos(\varphi_2-\varphi_3)+\cos\theta_1\cos\theta_2\].
\n

For the ’net’ term, one example of the contraction is shown as follows:
\e
\left\langle\wick{\c1\zeta_g(\bm{p_1}) \c2\zeta_g(\bm{p_2}) \c3\zeta_g(\bm{p_9}) \c1\zeta_g(\bm{p_3}) \c4\zeta_g(\bm{p_4}) \c5\zeta_g(\bm{p_{10}}) \c2\zeta_g(\bm{p_5}) \c4\zeta_g(\bm{p_6}) \c6\zeta_g(\bm{p_{11}}) \c3\zeta_g(\bm{p_7}) \c5\zeta_g(\bm{p_8}) 
\c6\zeta_g(\bm{p_{12}})}\right\rangle,
\q
where $\bm{p_9}$, $\bm{p_{10}}$, $\bm{p_{11}}$ and $\bm{p_{12}}$ are defined the same as above and the symmetry factor in this case is 1296. Then we have
\m
\Omega_{\mathrm{GW}}^{net}(k)=&&\frac{G_{\mathrm{NL}}^4k^3}{6\pi^2}\int\frac{\mathrm{d}^3q\mathrm{d}^3q'}{(2\pi)^6} \cos2(\phi-\phi') I\left(u,v\right) I\left(u',v'\right) \int\frac{\mathrm{d}^3p_1\mathrm{d}^3p_2\mathrm{d}^3p_3}{(2\pi)^9}\nonumber\\
&&\times P_{g}\left(p_1\right) P_{g}\left(p_2\right) P_{g}\left(p_3\right) P_{g}\left(|\bm{q-p_1-p_2}|\right) P_{g}\left(|\bm{q'-p_2-p_3}|\right) P_{g}\left(\bm{k-q+p_1-p_3}\right)\times1296\nonumber\\
=&&\frac{27G_{\mathrm{NL}}^4}{32\pi^4}\int_0^{\infty} \mathrm{d}u \int_{|1-u|}^{1+u}\mathrm{d}v \int_0^{\infty} \mathrm{d}u' \int_{|1-u'|}^{1+u'}\mathrm{d}v' \int_0^{\infty} \mathrm{d}u_1 \int_{|1-u_1|}^{1+u_1}\mathrm{d}v_1 \int_0^{\infty} \mathrm{d}u_2 \int_{|1-u_2|}^{1+u_2}\mathrm{d}v_2\int_0^{\infty} \mathrm{d}u_3 \int_{|1-u_3|}^{1+u_3}\mathrm{d}v_3\nonumber\\
&&\times\int_0^{2\pi}\mathrm{d}\varphi_1 \int_0^{2\pi}\mathrm{d}\varphi_2 \int_0^{2\pi}\mathrm{d}\varphi_3 \int_0^{2\pi}\mathrm{d}\varphi_4  \cos2\varphi_1 I(u,v)I(u',v')  \frac{uvu'v'u_1v_1u_2v_2u_3v_3}{(u_1u_2u_3w_{134}w_{245}w_{0135})^3}\nonumber\\
&&\times\mathcal{P}_{g}\left(u_1k\right) \mathcal{P}_{g}\left(u_2k\right)\mathcal{P}_{g}\left(u_3k\right)\mathcal{P}_{g}\left(w_{134}k\right)\mathcal{P}_{g}\left(w_{245}k\right)\mathcal{P}_{g}\left(w_{0135}k\right),
\n
where we perform the coordinate transformation the same as above. Additionally, $w_{245}$ and $w_{0135}$ are defined as follows:
\m
w_{245}^2=&&\frac{|\bm{q'-p_2-p_3}|^2}{k^2}\nonumber\\
=&&u'^2+u_2^2+u_3^2-2u'u_2\[\sin\theta'\sin\theta_2\cos(\varphi_1-\varphi_3)+\cos\theta'\cos\theta_2\]-2u'u_3(\sin\theta'\sin\theta_3\cos(\varphi_1-\varphi_4)+\cos\theta'\cos\theta_3)\nonumber\\
&&+2u_2u_3(\sin\theta_2\sin\theta_3\cos(\varphi_3-\varphi_4)+\cos\theta_2\cos\theta_3),\\
w_{0135}^2=&&\frac{|\bm{k-q+p_1-p_3}|^2}{k^2}\nonumber\\
=&&1+u^2+u_1^2+u_3^2-2u\cos\theta+2u_1\cos\theta_1-2u_3\cos\theta_3-2uu_1\[\sin\theta\sin\theta_1\cos\varphi_2+\cos\theta\cos\theta_1\]\nonumber\\
&&+2uu_3\[\sin\theta\sin\theta_3\cos\varphi_4+\cos\theta\cos\theta_3\]-2u_1u_3\[\sin\theta_1\sin\theta_3\cos(\varphi_2-\varphi_4)+\cos\theta_1\cos\theta_3\].
\n

\subsection{$F_{\mathrm{NL}}^2G_{\mathrm{NL}}$ terms}
For terms containing $F_{\mathrm{NL}}^2G_{\mathrm{NL}}$, considering symmetry, the GWs spectrum can be expressed in the following form:
\m
\Omega_{\mathrm{GW}}^{F_{\mathrm{NL}}^2G_{\mathrm{NL}}}(k)=&&\frac{F_{\mathrm{NL}}^2G_{\mathrm{NL}}k^3}{6\pi^2}\int\frac{\mathrm{d}^3q\mathrm{d}^3q'}{(2\pi)^6} \cos2(\phi-\phi') I\left(u,v\right) I\left(u',v'\right) \int\frac{\mathrm{d}^3p_1\mathrm{d}^3p_2\mathrm{d}^3p_3\mathrm{d}^3p_4}{(2\pi)^{12}}\nonumber\\
&&\times\Bigg[8\left\langle\left\langle\zeta_g(\bm{p_1})\zeta_g(\bm{q-p_1})\zeta_g(\bm{p_2})\zeta_g(\bm{p_3})\zeta_g(\bm{k-q-p_2-p_3})\zeta_g(\bm{p_4})\zeta_g(\bm{q'-p_4})\zeta_g(\bm{k'-q'})\right\rangle\right\rangle\nonumber\\
&&+4\left\langle\left\langle\zeta_g(\bm{p_1})\zeta_g(\bm{q-p_1})\zeta_g(\bm{p_2})\zeta_g(\bm{k-q-p_2})\zeta_g(\bm{p_3})\zeta_g(\bm{p_4})\zeta_g(\bm{q'-p_3-p_4})\zeta_g(\bm{k'-q'})\right\rangle\right\rangle\Bigg].
\n
Performing Wick contraction on the eight-point function, there are four distinct non-zero contractions and we name them as the 'loop' term, the '$F^2G(1)$' term, the '$F^2G(2)$' term, and the '$F^2G(3)$' term. Then we have
\e
\Omega_{\mathrm{GW}}^{F_{\mathrm{NL}}^2G_{\mathrm{NL}}}(k)=\Omega_{\mathrm{GW}}^{loop}(k)+\Omega_{\mathrm{GW}}^{F^2G(1)}(k)+\Omega_{\mathrm{GW}}^{F^2G(2)}(k)+\Omega_{\mathrm{GW}}^{F^2G(3)}(k),
\q
and we now demonstrate each of these four parts in detail.

The 'loop' term can be easily obtained as
\m
\Omega_{\mathrm{GW}}^{loop}(k)=6G_{\mathrm{NL}}\int\frac{\mathrm{d}p}{p}\mathcal{P}_g(p) \Omega_{\mathrm{GW}}^{F_{\mathrm{NL}}^2}(k),
\n
and for power spectrum satisfying \Eq{norm}, we have $\Omega_{\mathrm{GW}}^{loop}(k)=6AG_{\mathrm{NL}}\Omega_{\mathrm{GW}}^{F_{\mathrm{NL}}^2}(k)$.

One example of the contraction of the '$F^2G(1)$' term is shown as follows: 
\e
\left\langle\wick{\c1\zeta_g(\bm{p_1}) \c2\zeta_g(\bm{q-p_1}) \c3\zeta_g(\bm{p_2}) \c4\zeta_g(\bm{k-q-p_2}) \c1\zeta_g(\bm{p_3}) \c2\zeta_g(\bm{p_4}) \c3\zeta_g(\bm{q'-p_3-p_4}) \c4\zeta_g(\bm{k'-q'})}\right\rangle,
\q
and the symmetry factor in this case is 24. Expanding the correlation function and using the appearing delta functions to eliminate redundant integrals, we can obtain:
\m
\Omega_{\mathrm{GW}}^{F^2G(1)}(k)=&&\frac{F_{\mathrm{NL}}^2G_{\mathrm{NL}}k^3}{6\pi^2}\int\frac{\mathrm{d}^3q\mathrm{d}^3q'}{(2\pi)^6} \cos2(\phi-\phi') I\left(u,v\right) I\left(u',v'\right) \int\frac{\mathrm{d}^3p}{(2\pi)^3}4P_{g}\left(p\right) P_{g}\left(|\bm{q-p}|\right) P_{g}\left(|\bm{q-q'}|\right) P_{g}\left(\bm{k-q'}\right)\times24 \nonumber\\
=&&\frac{F_{\mathrm{NL}}^2G_{\mathrm{NL}}}{\pi^2} \int_0^{\infty} \mathrm{d}u \int_{|1-u|}^{1+u}\mathrm{d}v \int_0^{\infty} \mathrm{d}u' \int_{|1-u'|}^{1+u'}\mathrm{d}v' \int_0^{\infty} \mathrm{d}u_1 \int_{|1-u_1|}^{1+u_1}\mathrm{d}v_1 \int_0^{2\pi}\mathrm{d}\varphi_1 \int_0^{2\pi}\mathrm{d}\varphi_2 \cos2\varphi_1 I(u,v)I(u',v')\nonumber\\
&&\times \frac{uvu'v'u_1v_1}{(u_1v'w_{13}w_{12})^3} \mathcal{P}_{g}\left(u_1k\right) \mathcal{P}_{g}\left(w_{13}k\right) \mathcal{P}_{g}\left(w_{12}k\right) \mathcal{P}_{g}\left(v'k\right),
\n
where $w_{12}$ and $w_{13}$ are defined the same as \Eq{w12} and \Eq{w13}.

One example of the contraction of the '$F^2G(2)$' term is shown as follows:
\e
\left\langle\wick{\c3\zeta_g(\bm{p_1}) \c1\zeta_g(\bm{q-p_1}) \c1\zeta_g(\bm{p_2}) \c1\zeta_g(\bm{p_3}) \c2\zeta_g(\bm{k-q-p_2-p_3}) \c1\zeta_g(\bm{p_4}) \c2\zeta_g(\bm{q'-p_4}) \c3\zeta_g(\bm{k'-q'})}\right\rangle,
\q
and the symmetry factor in this case is 12. Expanding the correlation function and using the appearing delta functions to eliminate redundant integrals, we can obtain:
\m
\Omega_{\mathrm{GW}}^{F^2G(2)}(k)=&&\frac{F_{\mathrm{NL}}^2G_{\mathrm{NL}}k^3}{6\pi^2}\int\frac{\mathrm{d}^3q\mathrm{d}^3q'}{(2\pi)^6} \cos2(\phi-\phi') I\left(u,v\right) I\left(u',v'\right) \int\frac{\mathrm{d}^3p}{(2\pi)^3}8P_{g}\left(p\right) P_{g}\left(|\bm{q'-p}|\right) P_{g}\left(|\bm{k-q'}|\right)\nonumber\\
&&\times P_{g}\left(\bm{k-q-q'}\right)\times12 \nonumber\\
=&&\frac{F_{\mathrm{NL}}^2G_{\mathrm{NL}}}{\pi^2} \int_0^{\infty} \mathrm{d}u \int_{|1-u|}^{1+u}\mathrm{d}v \int_0^{\infty} \mathrm{d}u' \int_{|1-u'|}^{1+u'}\mathrm{d}v' \int_0^{\infty} \mathrm{d}u_1 \int_{|1-u_1|}^{1+u_1}\mathrm{d}v_1 \int_0^{2\pi}\mathrm{d}\varphi_1 \int_0^{2\pi}\mathrm{d}\varphi_2 \cos2\varphi_1 I(u,v)I(u',v')\nonumber\\
&&\times \frac{uvu'v'u_1v_1}{(u_1v'w_{23}w_{012})^3} \mathcal{P}_{g}\left(u_1k\right) \mathcal{P}_{g}\left(v'k\right) \mathcal{P}_{g}\left(w_{23}k\right) \mathcal{P}_{g}\left(w_{012}k\right),
\n
where $w_{23}$ and $w_{012}$ are defined the same as \Eq{w23} and \Eq{w012}.

One example of the contraction of the '$F^2G(3)$' term is shown as follows:
\e
\left\langle\wick{\c1\zeta_g(\bm{p_1}) \c2\zeta_g(\bm{q-p_1}) \c1\zeta_g(\bm{p_2}) \c1\zeta_g(\bm{p_3}) \c3\zeta_g(\bm{k-q-p_2-p_3}) \c1\zeta_g(\bm{p_4}) \c2\zeta_g(\bm{q'-p_4}) \c3\zeta_g(\bm{k'-q'})}\right\rangle,
\q
and the symmetry factor in this case is 24. Expanding the correlation function and using the appearing delta functions to eliminate redundant integrals, we can obtain:
\m
\Omega_{\mathrm{GW}}^{F^2G(3)}(k)=&&\frac{F_{\mathrm{NL}}^2G_{\mathrm{NL}}k^3}{6\pi^2}\int\frac{\mathrm{d}^3q\mathrm{d}^3q'}{(2\pi)^6} \cos2(\phi-\phi') I\left(u,v\right) I\left(u',v'\right) \int\frac{\mathrm{d}^3p}{(2\pi)^3}8P_{g}\left(p\right) P_{g}\left(|\bm{q-p}|\right) P_{g}\left(|\bm{k-q'}|\right)\nonumber\\
&&\times P_{g}\left(\bm{q-q'-p}\right)\times24 \nonumber\\
=&&\frac{2F_{\mathrm{NL}}^2G_{\mathrm{NL}}}{\pi^2} \int_0^{\infty} \mathrm{d}u \int_{|1-u|}^{1+u}\mathrm{d}v \int_0^{\infty} \mathrm{d}u' \int_{|1-u'|}^{1+u'}\mathrm{d}v' \int_0^{\infty} \mathrm{d}u_1 \int_{|1-u_1|}^{1+u_1}\mathrm{d}v_1 \int_0^{2\pi}\mathrm{d}\varphi_1 \int_0^{2\pi}\mathrm{d}\varphi_2 \cos2\varphi_1 I(u,v)I(u',v')\nonumber\\
&&\times \frac{uvu'v'u_1v_1}{(u_1v'w_{13}w_{123})^3} \mathcal{P}_{g}\left(u_1k\right) \mathcal{P}_{g}\left(v'k\right) \mathcal{P}_{g}\left(w_{13}k\right) \mathcal{P}_{g}\left(w_{123}k\right),
\n
where $w_{13}$ is defined the same as \Eq{w13} and $w_{123}$ is defined as follows:
\m
w_{123}^2=&&\frac{|\bm{q-q'-p}|^2}{k^2}\nonumber\\
=&&u^2+u'^2+u_1^2-2uu'\(\sin\theta\sin\theta'\cos\varphi_1+\cos\theta\cos\theta'\)-2uu_1(\sin\theta\sin\theta_1\cos\varphi_2+\cos\theta\cos\theta_1)\nonumber\\
&&+2u'u_1\[\sin\theta'\sin\theta_1\cos(\varphi_1-\varphi_2)+\cos\theta'\cos\theta_1\].
\n

\subsection{$F_{\mathrm{NL}}^2G_{\mathrm{NL}}^2$ terms}
For terms containing $F_{\mathrm{NL}}^2G_{\mathrm{NL}}^2$, considering symmetry, the GWs spectrum can be expressed in the following form:
\m
\Omega_{\mathrm{GW}}^{F_{\mathrm{NL}}^2G_{\mathrm{NL}}^2}(k)=&&\frac{F_{\mathrm{NL}}^2G_{\mathrm{NL}}^2k^3}{6\pi^2}\int\frac{\mathrm{d}^3q\mathrm{d}^3q'}{(2\pi)^6} \cos2(\phi-\phi') I\left(u,v\right) I\left(u',v'\right) \int\frac{\mathrm{d}^3p_1\mathrm{d}^3p_2\mathrm{d}^3p_3\mathrm{d}^3p_4\mathrm{d}^3p_5\mathrm{d}^3p_6}{(2\pi)^{18}}\nonumber\\
&&\times \Bigg[4\langle\langle\zeta_g(\bm{p_1})\zeta_g(\bm{q-p_1})\zeta_g(\bm{p_2})\zeta_g(\bm{p_3})\zeta_g(\bm{k-q-p_2-p_3})\zeta_g(\bm{p_4})\zeta_g(\bm{q'-p_4})\zeta_g(\bm{p_5})\zeta_g(\bm{p_6})\nonumber\\
&&\quad \times\zeta_g(\bm{k'-q'-p_5-p_6})\rangle\rangle+2\langle\langle\zeta_g(\bm{p_1})\zeta_g(\bm{q-p_1})\zeta_g(\bm{p_2})\zeta_g(\bm{k-q-p_2})\zeta_g(\bm{p_3})\zeta_g(\bm{p_4})\nonumber\\
&&\quad \times\zeta_g(\bm{q'-p_3-p_4})\zeta_g(\bm{p_5})\zeta_g(\bm{p_6})\zeta_g(\bm{k'-q'-p_5-p_6})\rangle\rangle\Bigg].
\n
Performing Wick contraction on the ten-point function, there are 9 distinct non-zero contractions and we name them as the 'loops' term, the '$F^2G^2(1)$' term, the '$F^2G^2(2)$' term, the '$F^2G^2(3)$' term, the '$F^2G^2(4)$' term, the '$F^2G^2(5)$' term, the '$F^2G^2(6)$' term, the '$F^2G^2(7)$' term and the '8$F^2G^2(8)$' term. Then we have
\m
\Omega_{\mathrm{GW}}^{F_{\mathrm{NL}}^2G_{\mathrm{NL}}^2}(k)=&&\Omega_{\mathrm{GW}}^{loops}(k)+\Omega_{\mathrm{GW}}^{F^2G^2(1)}(k)+\Omega_{\mathrm{GW}}^{F^2G^2(2)}(k)+\Omega_{\mathrm{GW}}^{F^2G^2(3)}(k)+\Omega_{\mathrm{GW}}^{F^2G^2(4)}(k)+\Omega_{\mathrm{GW}}^{F^2G^2(5)}(k)+\Omega_{\mathrm{GW}}^{F^2G^2(6)}(k)\nonumber\\
&&+\Omega_{\mathrm{GW}}^{F^2G^2(7)}(k)+\Omega_{\mathrm{GW}}^{F^2G^2(8)}(k),
\n
and we now demonstrate each of these 9 parts in detail.

The 'loops' term can be easily obtained as
\m
\Omega_{\mathrm{GW}}^{loops}(k)=&&9G_{\mathrm{NL}}^2\int\frac{\mathrm{d}p_1\mathrm{d}p_2}{p_1p_2}\mathcal{P}_g(p_1)\mathcal{P}_g(p_2) \Omega_{\mathrm{GW}}^{F_{\mathrm{NL}}^2}(k)+3G_{\mathrm{NL}}\int\frac{\mathrm{d}p_1}{p_1}\mathcal{P}_g(p_1) \(\Omega_{\mathrm{GW}}^{F_{\mathrm{NL}}^2G_{\mathrm{NL}}}(k)-6G_{\mathrm{NL}}\int\frac{\mathrm{d}p_2}{p_2}\mathcal{P}_g(p_2)\Omega_{\mathrm{GW}}^{F_{\mathrm{NL}}^2}(k)\)\nonumber\\
=&&3G_{\mathrm{NL}}\int\frac{\mathrm{d}p_1}{p_1}\mathcal{P}_g(p_1) \(\Omega_{\mathrm{GW}}^{F_{\mathrm{NL}}^2G_{\mathrm{NL}}}(k)-3G_{\mathrm{NL}}\int\frac{\mathrm{d}p_2}{p_2}\mathcal{P}_g(p_2)\Omega_{\mathrm{GW}}^{F_{\mathrm{NL}}^2}(k)\),
\n
and for power spectrum satisfying \Eq{norm}, we have $\Omega_{\mathrm{GW}}^{loops}(k)=3AG_{\mathrm{NL}}\Omega_{\mathrm{GW}}^{F_{\mathrm{NL}}^2G_{\mathrm{NL}}}(k)-9A^2G_{\mathrm{NL}}^2\Omega_{\mathrm{GW}}^{F_{\mathrm{NL}}^2}(k)$.

The '$F^2G^2(1)$' term is a disconnected diagram and one example of the contraction is shown as follows:
\e
\left\langle\wick{\c1\zeta_g(\bm{p_1}) \c2\zeta_g(\bm{q-p_1}) \c3\zeta_g(\bm{p_2}) \c4\zeta_g(\bm{p_3}) \c5\zeta_g(\bm{k-q-p_2-p_3}) \c1\zeta_g(\bm{p_4}) \c2\zeta_g(\bm{q'-p_4}) \c3\zeta_g(\bm{p_5}) \c4\zeta_g(\bm{p_6}) \c5\zeta_g(\bm{k'-q'-p_5-p_6})}\right\rangle.
\q
The symmetry factor in this case is 12. Expanding the correlation function and using the appearing delta functions to eliminate redundant integrals, we can obtain:
\m
\Omega_{\mathrm{GW}}^{F^2G^2(1)}(k)=&&\frac{F_{\mathrm{NL}}^2G_{\mathrm{NL}}^2k^3}{6\pi^2}\int\frac{\mathrm{d}^3q}{(2\pi)^3}  I^2\left(u,v\right) \int\frac{\mathrm{d}^3p_1\mathrm{d}^3p_2\mathrm{d}^3p_3}{(2\pi)^{9}}4P_{g}\left(p_1\right) P_{g}\left(p_2\right)P_{g}\left(p_3\right) P_{g}\left(|\bm{q-p_1}|\right)P_{g}\left(|\bm{k-q-p_2-p_3}|\right)\times12\nonumber\\
=&&F_{\mathrm{NL}}^2G_{\mathrm{NL}}^2 \int_0^{\infty} \mathrm{d}u \int_{|1-u|}^{1+u}\mathrm{d}v \int_0^{\infty} \mathrm{d}u_1 \int_{|1-u_1|}^{1+u_1}\mathrm{d}v_1 \int_0^{\infty} \mathrm{d}u_2 \int_{|1-u_2|}^{1+u_2}\mathrm{d}v_2 \int_0^{\infty} \mathrm{d}u_3 \int_{|1-u_3|}^{1+u_3}\mathrm{d}v_3 I^2(u,v)\nonumber\\
&&\times  \frac{1}{u^2v^2u_1^2v_1^2u_2^2v_2^2u_3^2v_3^2}\mathcal{P} _{g}\left(u_1uk\right) \mathcal{P} _{g}\left(v_1uk\right)\mathcal{P} _{g}\left(u_2vk\right)\mathcal{P} _{g}\left(u_3v_2vk\right) \mathcal{P} _{g}\left(v_3v_2vk\right),
\n
where the second equality in the above equation is obtained by performing the coordinate transformation $u_1 = p_1 / q$, $ v_1=|\bm{q-p_1}| / q$, $u_2 = p_2 / |\bm{k-q}|$, $ v_2=|\bm{k-q-p_2}| /|\bm{k-q}|$, $u_3 = p_3 / |\bm{k-q-p_2}|$, $ v_3=|\bm{k-q-p_2-p_3}| / |\bm{k-q-p_2}|$.

One example of the contraction of the '$F^2G^2(2)$' term is shown as follows:
\e
\left\langle\wick{\c1\zeta_g(\bm{p_1}) \c2\zeta_g(\bm{q-p_1}) \c1\zeta_g(\bm{p_2}) \c3\zeta_g(\bm{p_3}) \c4\zeta_g(\bm{k-q-p_2-p_3}) \c2\zeta_g(\bm{p_4}) \c1\zeta_g(\bm{q'-p_4}) \c1\zeta_g(\bm{p_5}) \c3\zeta_g(\bm{p_6}) \c4\zeta_g(\bm{k'-q'-p_5-p_6})}\right\rangle.
\q
and the symmetry factor in this case is 72. Expanding the correlation function and using the appearing delta functions to eliminate redundant integrals, we can obtain:
\m
\Omega_{\mathrm{GW}}^{F^2G^2(2)}(k)=&&\frac{F_{\mathrm{NL}}^2G_{\mathrm{NL}}^2k^3}{6\pi^2}\int\frac{\mathrm{d}^3q\mathrm{d}^3q'}{(2\pi)^6} \cos2(\phi-\phi') I\left(u,v\right) I\left(u',v'\right) \int\frac{\mathrm{d}^3p_1\mathrm{d}^3p_2}{(2\pi)^6}\nonumber\\
&&\times 4P_{g}\left(p_1\right)P_{g}\left(p_2\right) P_{g}\left(|\bm{q-p_1}|\right) P_{g}\left(|\bm{k+p_1+p_2-q}|\right) P_{g}\left(\bm{q+q'-p_1}\right)\times72 \nonumber\\
=&&\frac{3F_{\mathrm{NL}}^2G_{\mathrm{NL}}^2}{4\pi^3} \int_0^{\infty} \mathrm{d}u \int_{|1-u|}^{1+u}\mathrm{d}v \int_0^{\infty} \mathrm{d}u' \int_{|1-u'|}^{1+u'}\mathrm{d}v' \int_0^{\infty} \mathrm{d}u_1 \int_{|1-u_1|}^{1+u_1}\mathrm{d}v_1 \int_0^{\infty} \mathrm{d}u_2 \int_{|1-u_2|}^{1+u_2}\mathrm{d}v_2\nonumber\\
&&\times \int_0^{2\pi}\mathrm{d}\varphi_1 \int_0^{2\pi}\mathrm{d}\varphi_2 \int_0^{2\pi}\mathrm{d}\varphi_3  \cos2\varphi_1 I(u,v)I(u',v') \frac{uvu'v'u_1v_1u_2v_2}{(u_1u_2w_{13}w_{0134}w_{123})^3} \mathcal{P}_{g}\left(u_1k\right) \mathcal{P}_{g}\left(u_2k\right)\nonumber\\
&&\times \mathcal{P}_{g}\left(w_{13}k\right) \mathcal{P}_{g}\left(w_{0134}k\right) \mathcal{P}_{g}\left(w_{123}k\right),
\n
where $w_{13}$ is defined in \Eq{w13} and $w_{0134}$ and $w_{123}$ are defined as follows:
\m
w_{0134}^2=&&\frac{|\bm{k-q+p_1+p_2}|^2}{k^2}\nonumber\\
=&&1+u^2+u_1^2+u_2^2-2u\cos\theta+2u_1\cos\theta_1+2u_2\cos\theta_2-2uu_1\[\sin\theta\sin\theta_1\cos\varphi_2+\cos\theta\cos\theta_1\]\nonumber\\
&&-2uu_2\[\sin\theta\sin\theta_2\cos\varphi_3+\cos\theta\cos\theta_3\]+2u_1u_2\[\sin\theta_1\sin\theta_2\cos(\varphi_2-\varphi_3)+\cos\theta_1\cos\theta_2\],\\
w_{123}^2=&&\frac{|\bm{q+q'-p_1}|^2}{k^2}\nonumber\\
=&&u^2+u'^2+u_1^2+2uu'\(\sin\theta\sin\theta'\cos\varphi_1+\cos\theta\cos\theta'\)-2uu_1(\sin\theta\sin\theta_1\cos\varphi_2+\cos\theta\cos\theta_1)\nonumber\\
&&-2u'u_1\[\sin\theta'\sin\theta_1\cos(\varphi_1-\varphi_2)+\cos\theta'\cos\theta_1\].
\n

One example of the contraction of the '$F^2G^2(3)$' term is shown as follows:
\e
\left\langle\wick{\c4\zeta_g(\bm{p_1}) \c5\zeta_g(\bm{q-p_1}) \c1\zeta_g(\bm{p_2}) \c2\zeta_g(\bm{p_3}) \c3\zeta_g(\bm{k-q-p_2-p_3}) \c1\zeta_g(\bm{p_4}) \c4\zeta_g(\bm{q'-p_4}) \c2\zeta_g(\bm{p_5}) \c3\zeta_g(\bm{p_6}) \c5\zeta_g(\bm{k'-q'-p_5-p_6})}\right\rangle.
\q
and the symmetry factor in this case is 72. Expanding the correlation function and using the appearing delta functions to eliminate redundant integrals, we can obtain:
\m
\Omega_{\mathrm{GW}}^{F^2G^2(3)}(k)=&&\frac{F_{\mathrm{NL}}^2G_{\mathrm{NL}}^2k^3}{6\pi^2}\int\frac{\mathrm{d}^3q\mathrm{d}^3q'}{(2\pi)^6} \cos2(\phi-\phi') I\left(u,v\right) I\left(u',v'\right) \int\frac{\mathrm{d}^3p_1\mathrm{d}^3p_2}{(2\pi)^6}\nonumber\\
&&\times 4P_{g}\left(p_1\right)P_{g}\left(p_2\right) P_{g}\left(|\bm{q-p_1}|\right) P_{g}\left(|\bm{q'-p_1}|\right) P_{g}\left(|\bm{k-q-q'+p_1-p_2}|\right) \times72 \nonumber\\
=&&\frac{3F_{\mathrm{NL}}^2G_{\mathrm{NL}}^2}{4\pi^3} \int_0^{\infty} \mathrm{d}u \int_{|1-u|}^{1+u}\mathrm{d}v \int_0^{\infty} \mathrm{d}u' \int_{|1-u'|}^{1+u'}\mathrm{d}v' \int_0^{\infty} \mathrm{d}u_1 \int_{|1-u_1|}^{1+u_1}\mathrm{d}v_1 \int_0^{\infty} \mathrm{d}u_2 \int_{|1-u_2|}^{1+u_2}\mathrm{d}v_2\nonumber\\
&&\times \int_0^{2\pi}\mathrm{d}\varphi_1 \int_0^{2\pi}\mathrm{d}\varphi_2 \int_0^{2\pi}\mathrm{d}\varphi_3 \cos2\varphi_1 I(u,v)I(u',v') \frac{uvu'v'u_1v_1u_2v_2}{(u_1u_2w_{13}w_{23}w_{01234})^3} \mathcal{P}_{g}\left(u_1k\right) \mathcal{P}_{g}\left(u_2k\right) \mathcal{P}_{g}\left(w_{13}k\right)\nonumber\\
&&\times \mathcal{P}_{g}\left(w_{23}k\right) \mathcal{P}_{g}\left(w_{01234}k\right),
\n
where $w_{13}$ and $w_{23}$ are defined in \Eq{w13} and \Eq{w23} and $w_{01234}$ are defined as follows:
\m
w_{01234}^2=&&\frac{|\bm{k-q-q'+p_1-p_2}|^2}{k^2}\nonumber\\
=&&1+u^2+u'^2+u_1^2+u_2^2-2u\cos\theta-2u'\cos\theta'+2u_1\cos\theta_1-2u_2\cos\theta_2+2uu'\[\sin\theta\sin\theta'\cos\varphi_1+\cos\theta\cos\theta'\]\nonumber\\
&&-2uu_1\[\sin\theta\sin\theta_1\cos\varphi_2+\cos\theta\cos\theta_1\]+2uu_2\[\sin\theta\sin\theta_2\cos\varphi_3+\cos\theta\cos\theta_2\]\nonumber\\
&&-2u'u_1\[\sin\theta'\sin\theta_1\cos(\varphi_1-\varphi_2)+\cos\theta'\cos\theta_1\]+2u'u_2\[\sin\theta'\sin\theta_2\cos(\varphi_1-\varphi_3)+\cos\theta'\cos\theta_2\]\nonumber\\
&&-2u_1u_2\[\sin\theta_1\sin\theta_2\cos(\varphi_2-\varphi_3)+\cos\theta_1\cos\theta_2\].
\n

One example of the contraction of the '$F^2G^2(4)$' term is shown as follows:
\e
\left\langle\wick{\c1\zeta_g(\bm{p_1}) \c2\zeta_g(\bm{q-p_1}) \c3\zeta_g(\bm{p_2}) \c4\zeta_g(\bm{p_3}) \c5\zeta_g(\bm{k-q-p_2-p_3}) \c3\zeta_g(\bm{p_4}) \c4\zeta_g(\bm{q'-p_4}) \c1\zeta_g(\bm{p_5}) \c2\zeta_g(\bm{p_6}) \c5\zeta_g(\bm{k'-q'-p_5-p_6})}\right\rangle.
\q
and the symmetry factor in this case is 36. Expanding the correlation function and using the appearing delta functions to eliminate redundant integrals, we can obtain:
\m
\Omega_{\mathrm{GW}}^{F^2G^2(4)}(k)=&&\frac{F_{\mathrm{NL}}^2G_{\mathrm{NL}}^2k^3}{6\pi^2}\int\frac{\mathrm{d}^3q\mathrm{d}^3q'}{(2\pi)^6} \cos2(\phi-\phi') I\left(u,v\right) I\left(u',v'\right) \int\frac{\mathrm{d}^3p_1\mathrm{d}^3p_2}{(2\pi)^6}\nonumber\\
&&\times 4P_{g}\left(p_1\right)P_{g}\left(p_2\right) P_{g}\left(|\bm{q-p_1}|\right) P_{g}\left(|\bm{q'-p_2}|\right) P_{g}\left(|\bm{k-q-q'}|\right) \times36 \nonumber\\
=&&\frac{3F_{\mathrm{NL}}^2G_{\mathrm{NL}}^2}{8\pi^3} \int_0^{\infty} \mathrm{d}u \int_{|1-u|}^{1+u}\mathrm{d}v \int_0^{\infty} \mathrm{d}u' \int_{|1-u'|}^{1+u'}\mathrm{d}v' \int_0^{\infty} \mathrm{d}u_1 \int_{|1-u_1|}^{1+u_1}\mathrm{d}v_1 \int_0^{\infty} \mathrm{d}u_2 \int_{|1-u_2|}^{1+u_2}\mathrm{d}v_2\nonumber\\
&&\times \int_0^{2\pi}\mathrm{d}\varphi_1 \int_0^{2\pi}\mathrm{d}\varphi_2 \int_0^{2\pi}\mathrm{d}\varphi_3 \cos2\varphi_1 I(u,v)I(u',v') \frac{uvu'v'u_1v_1u_2v_2}{(u_1u_2w_{13}w_{24}w_{012})^3} \mathcal{P}_{g}\left(u_1k\right) \mathcal{P}_{g}\left(u_2k\right) \mathcal{P}_{g}\left(w_{13}k\right)\nonumber\\
&&\times \mathcal{P}_{g}\left(w_{24}k\right) \mathcal{P}_{g}\left(w_{012}k\right),
\n
where $w_{13}$ and $w_{012}$ are defined in \Eq{w13} and \Eq{w012} and $w_{24}$ is defined as follows:
\m
w_{24}^2&&=\frac{|\bm{q'-p_2}|^2}{k^2}=u'^2+u_2^2-2u'u_2\[\sin\theta'\sin\theta_2\cos(\varphi_1-\varphi_3)+\cos\theta'\cos\theta_2\].
\n

One example of the contraction of the '$F^2G^2(5)$' term is shown as follows:
\e
\left\langle\wick{\c2\zeta_g(\bm{p_1}) \c1\zeta_g(\bm{q-p_1}) \c1\zeta_g(\bm{p_2}) \c3\zeta_g(\bm{k-q-p_2}) \c2\zeta_g(\bm{p_3}) \c2\zeta_g(\bm{p_4}) \c1\zeta_g(\bm{q'-p_3-p_4}) \c1\zeta_g(\bm{p_5}) \c2\zeta_g(\bm{p_6}) \c3\zeta_g(\bm{k'-q'-p_5-p_6})}\right\rangle.
\q
and the symmetry factor in this case is 144. Expanding the correlation function and using the appearing delta functions to eliminate redundant integrals, we can obtain:
\m
\Omega_{\mathrm{GW}}^{F^2G^2(5)}(k)=&&\frac{F_{\mathrm{NL}}^2G_{\mathrm{NL}}^2k^3}{6\pi^2}\int\frac{\mathrm{d}^3q\mathrm{d}^3q'}{(2\pi)^6} \cos2(\phi-\phi') I\left(u,v\right) I\left(u',v'\right) \int\frac{\mathrm{d}^3p_1\mathrm{d}^3p_2}{(2\pi)^6}\nonumber\\
&&\times 2P_{g}\left(p_1\right)P_{g}\left(p_2\right) P_{g}\left(|\bm{q-p_1}|\right) P_{g}\left(|\bm{k-q+p_1}|\right) P_{g}\left(|\bm{q+q'-p_1-p_2}|\right) \times144 \nonumber\\
=&&\frac{3F_{\mathrm{NL}}^2G_{\mathrm{NL}}^2}{4\pi^3} \int_0^{\infty} \mathrm{d}u \int_{|1-u|}^{1+u}\mathrm{d}v \int_0^{\infty} \mathrm{d}u' \int_{|1-u'|}^{1+u'}\mathrm{d}v' \int_0^{\infty} \mathrm{d}u_1 \int_{|1-u_1|}^{1+u_1}\mathrm{d}v_1 \int_0^{\infty} \mathrm{d}u_2 \int_{|1-u_2|}^{1+u_2}\mathrm{d}v_2\nonumber\\
&&\times \int_0^{2\pi}\mathrm{d}\varphi_1 \int_0^{2\pi}\mathrm{d}\varphi_2 \int_0^{2\pi}\mathrm{d}\varphi_3 \cos2\varphi_1 I(u,v)I(u',v') \frac{uvu'v'u_1v_1u_2v_2}{(u_1u_2w_{13}w_{013}w_{1234})^3} \mathcal{P}_{g}\left(u_1k\right) \mathcal{P}_{g}\left(u_2k\right) \mathcal{P}_{g}\left(w_{13}k\right)\nonumber\\
&&\times \mathcal{P}_{g}\left(w_{013}k\right) \mathcal{P}_{g}\left(w_{1234}k\right),
\n
where $w_{013}$ and $w_{1234}$ are defined as follows:
\m
w_{013}^2=&&\frac{|\bm{k-q+p_1}|^2}{k^2}=1+u^2+u_1^2-2u\cos\theta+2u_1\cos\theta_1-2uu_1\[\sin\theta\sin\theta_1\cos\varphi_2+\cos\theta\cos\theta_1\],\\
w_{1234}^2=&&\frac{|\bm{q+q'-p_1-p_2}|^2}{k^2}\nonumber\\
=&&u^2+u'^2+u_1^2+u_2^2+2uu'\[\sin\theta\sin\theta'\cos\varphi_1+\cos\theta\cos\theta'\]-2uu_1\[\sin\theta\sin\theta_1\cos\varphi_2+\cos\theta\cos\theta_1\]\nonumber\\
&&-2uu_2\[\sin\theta\sin\theta_2\cos\varphi_3+\cos\theta\cos\theta_2\]-2u'u_1\[\sin\theta'\sin\theta_1\cos(\varphi_1-\varphi_2)+\cos\theta'\cos\theta_1\]\nonumber\\
&&-2u'u_2\[\sin\theta'\sin\theta_2\cos(\varphi_1-\varphi_3)+\cos\theta'\cos\theta_2\]+2u_1u_2\[\sin\theta_1\sin\theta_2\cos(\varphi_2-\varphi_3)+\cos\theta_1\cos\theta_2\].
\n

One example of the contraction of the '$F^2G^2(6)$' term is shown as follows:
\e
\left\langle\wick{\c1\zeta_g(\bm{p_1}) \c2\zeta_g(\bm{q-p_1}) \c3\zeta_g(\bm{p_2}) \c4\zeta_g(\bm{k-q-p_2}) \c1\zeta_g(\bm{p_3}) \c2\zeta_g(\bm{p_4}) \c1\zeta_g(\bm{q'-p_3-p_4}) \c3\zeta_g(\bm{p_5}) \c4\zeta_g(\bm{p_6}) \c1\zeta_g(\bm{k'-q'-p_5-p_6})}\right\rangle.
\q
and the symmetry factor in this case is 72. Expanding the correlation function and using the appearing delta functions to eliminate redundant integrals, we can obtain: 
\m
\Omega_{\mathrm{GW}}^{F^2G^2(6)}(k)=&&\frac{F_{\mathrm{NL}}^2G_{\mathrm{NL}}^2k^3}{6\pi^2}\int\frac{\mathrm{d}^3q\mathrm{d}^3q'}{(2\pi)^6} \cos2(\phi-\phi') I\left(u,v\right) I\left(u',v'\right) \int\frac{\mathrm{d}^3p_1\mathrm{d}^3p_2}{(2\pi)^6}\nonumber\\
&&\times 2P_{g}\left(p_1\right)P_{g}\left(p_2\right) P_{g}\left(|\bm{q-p_1}|\right) P_{g}\left(|\bm{k-q-p_2}|\right) P_{g}\left(|\bm{q-q'}|\right) \times72 \nonumber\\
=&&\frac{3F_{\mathrm{NL}}^2G_{\mathrm{NL}}^2}{8\pi^3} \int_0^{\infty} \mathrm{d}u \int_{|1-u|}^{1+u}\mathrm{d}v \int_0^{\infty} \mathrm{d}u' \int_{|1-u'|}^{1+u'}\mathrm{d}v' \int_0^{\infty} \mathrm{d}u_1 \int_{|1-u_1|}^{1+u_1}\mathrm{d}v_1 \int_0^{\infty} \mathrm{d}u_2 \int_{|1-u_2|}^{1+u_2}\mathrm{d}v_2\nonumber\\
&&\times \int_0^{2\pi}\mathrm{d}\varphi_1 \int_0^{2\pi}\mathrm{d}\varphi_2 \int_0^{2\pi}\mathrm{d}\varphi_3 \cos2\varphi_1 I(u,v)I(u',v') \frac{uvu'v'u_1v_1u_2v_2}{(u_1u_2w_{13}w_{014}w_{12})^3} \mathcal{P}_{g}\left(u_1k\right) \mathcal{P}_{g}\left(u_2k\right) \mathcal{P}_{g}\left(w_{13}k\right)\nonumber\\
&&\times \mathcal{P}_{g}\left(w_{014}k\right) \mathcal{P}_{g}\left(w_{12}k\right),
\n
where $w_{014}$ is defined as follows:
\m
w_{014}^2=&&\frac{|\bm{k-q-p_2}|^2}{k^2}=1+u^2+u_2^2-2u\cos\theta-2u_2\cos\theta_2+2uu_2\[\sin\theta\sin\theta_2\cos\varphi_3+\cos\theta\cos\theta_2\].
\n

One example of the contraction of the '$F^2G^2(7)$' term is shown as follows:
\e
\left\langle\wick{\c1\zeta_g(\bm{p_1}) \c2\zeta_g(\bm{q-p_1}) \c3\zeta_g(\bm{p_2}) \c4\zeta_g(\bm{k-q-p_2}) \c1\zeta_g(\bm{p_3}) \c3\zeta_g(\bm{p_4}) \c1\zeta_g(\bm{q'-p_3-p_4}) \c2\zeta_g(\bm{p_5}) \c4\zeta_g(\bm{p_6}) \c1\zeta_g(\bm{k'-q'-p_5-p_6})}\right\rangle.
\q
and the symmetry factor in this case is 144. Expanding the correlation function and using the appearing delta functions to eliminate redundant integrals, we can obtain:
\m
\Omega_{\mathrm{GW}}^{F^2G^2(7)}(k)=&&\frac{F_{\mathrm{NL}}^2G_{\mathrm{NL}}^2k^3}{6\pi^2}\int\frac{\mathrm{d}^3q\mathrm{d}^3q'}{(2\pi)^6} \cos2(\phi-\phi') I\left(u,v\right) I\left(u',v'\right) \int\frac{\mathrm{d}^3p_1\mathrm{d}^3p_2}{(2\pi)^6}\nonumber\\
&&\times 2P_{g}\left(p_1\right)P_{g}\left(p_2\right) P_{g}\left(|\bm{q-p_1}|\right) P_{g}\left(|\bm{k-q-p_2}|\right) P_{g}\left(|\bm{q'-p_1-p_2}|\right) \times144 \nonumber\\
=&&\frac{3F_{\mathrm{NL}}^2G_{\mathrm{NL}}^2}{4\pi^3} \int_0^{\infty} \mathrm{d}u \int_{|1-u|}^{1+u}\mathrm{d}v \int_0^{\infty} \mathrm{d}u' \int_{|1-u'|}^{1+u'}\mathrm{d}v' \int_0^{\infty} \mathrm{d}u_1 \int_{|1-u_1|}^{1+u_1}\mathrm{d}v_1 \int_0^{\infty} \mathrm{d}u_2 \int_{|1-u_2|}^{1+u_2}\mathrm{d}v_2\nonumber\\
&&\times \int_0^{2\pi}\mathrm{d}\varphi_1 \int_0^{2\pi}\mathrm{d}\varphi_2 \int_0^{2\pi}\mathrm{d}\varphi_3 \cos2\varphi_1 I(u,v)I(u',v') \frac{uvu'v'u_1v_1u_2v_2}{(u_1u_2w_{13}w_{014}w_{234})^3} \mathcal{P}_{g}\left(u_1k\right) \mathcal{P}_{g}\left(u_2k\right) \mathcal{P}_{g}\left(w_{13}k\right)\nonumber\\
&&\times \mathcal{P}_{g}\left(w_{014}k\right) \mathcal{P}_{g}\left(w_{234}k\right),
\n
where $w_{234}$ is defined as follows:
\m
w_{234}^2=&&\frac{|\bm{q'-p_1-p_2}|^2}{k^2}\nonumber\\
=&&u'^2+u_1^2+u_2^2-2u'u_1\[\sin\theta'\sin\theta_1\cos(\varphi_1-\varphi_2)+\cos\theta'\cos\theta_1\]-2u'u_2\[\sin\theta'\sin\theta_2\cos(\varphi_1-\varphi_3)+\cos\theta'\cos\theta_2\]\nonumber\\
&&+2u_1u_2\[\sin\theta_1\sin\theta_2\cos(\varphi_2-\varphi_3)+\cos\theta_1\cos\theta_2\].
\n

One example of the contraction of the '$F^2G^2(8)$' term is shown as follows:
\e
\left\langle\wick{\c3\zeta_g(\bm{p_1}) \c1\zeta_g(\bm{q-p_1}) \c1\zeta_g(\bm{p_2}) \c2\zeta_g(\bm{p_3}) \c1\zeta_g(\bm{k-q-p_2-p_3}) \c1\zeta_g(\bm{p_4}) \c1\zeta_g(\bm{q'-p_4}) \c1\zeta_g(\bm{p_5}) \c2\zeta_g(\bm{p_6}) \c3\zeta_g(\bm{k'-q'-p_5-p_6})}\right\rangle.
\q
and the symmetry factor in this case is 144. Expanding the correlation function and using the appearing delta functions to eliminate redundant integrals, we can obtain:
\m
\Omega_{\mathrm{GW}}^{F^2G^2(8)}(k)=&&\frac{F_{\mathrm{NL}}^2G_{\mathrm{NL}}^2k^3}{6\pi^2}\int\frac{\mathrm{d}^3q\mathrm{d}^3q'}{(2\pi)^6} \cos2(\phi-\phi') I\left(u,v\right) I\left(u',v'\right) \int\frac{\mathrm{d}^3p_1\mathrm{d}^3p_2}{(2\pi)^6}\nonumber\\
&&\times 4P_{g}\left(p_1\right)P_{g}\left(p_2\right) P_{g}\left(|\bm{q-p_1}|\right) P_{g}\left(|\bm{k-p_1-p_2}|\right) P_{g}\left(|\bm{k+q'-p_1-p_2}|\right) \times144 \nonumber\\
=&&\frac{3F_{\mathrm{NL}}^2G_{\mathrm{NL}}^2}{2\pi^3} \int_0^{\infty} \mathrm{d}u \int_{|1-u|}^{1+u}\mathrm{d}v \int_0^{\infty} \mathrm{d}u' \int_{|1-u'|}^{1+u'}\mathrm{d}v' \int_0^{\infty} \mathrm{d}u_1 \int_{|1-u_1|}^{1+u_1}\mathrm{d}v_1 \int_0^{\infty} \mathrm{d}u_2 \int_{|1-u_2|}^{1+u_2}\mathrm{d}v_2\nonumber\\
&&\times \int_0^{2\pi}\mathrm{d}\varphi_1 \int_0^{2\pi}\mathrm{d}\varphi_2 \int_0^{2\pi}\mathrm{d}\varphi_3 \cos2\varphi_1 I(u,v)I(u',v') \frac{uvu'v'u_1v_1u_2v_2}{(u_1u_2w_{13}w_{034}w_{0234})^3} \mathcal{P}_{g}\left(u_1k\right) \mathcal{P}_{g}\left(u_2k\right) \mathcal{P}_{g}\left(w_{13}k\right)\nonumber\\
&&\times \mathcal{P}_{g}\left(w_{034}k\right) \mathcal{P}_{g}\left(w_{0234}k\right),
\n
where $w_{13}$ is defined in \Eq{w13} and $w_{034}$ and $w_{0234}$ are defined as follows:
\m
w_{034}^2=&&\frac{|\bm{k-p_1-p_2}|^2}{k^2}\nonumber\\
=&&1+u_1^2+u_2^2-2u_1\cos\theta_1-2u_2\cos\theta_2+2u_1u_2\[\sin\theta_1\sin\theta_2\cos(\varphi_2-\varphi_3)+\cos\theta_1\cos\theta_2\],\\
w_{0234}^2=&&\frac{|\bm{k+q'-p_1-p_2}|^2}{k^2}\nonumber\\
=&&1+u'^2+u_1^2+u_2^2+2u'\cos\theta'-2u_1\cos\theta_1-2u_2\cos\theta_2-2u'u_1\[\sin\theta'\sin\theta_1\cos(\varphi_1-\varphi_2)+\cos\theta'\cos\theta_1\]\nonumber\\
&&-2u'u_2\[\sin\theta'\sin\theta_2\cos(\varphi_1-\varphi_3)+\cos\theta'\cos\theta_3\]+2u_1u_2\[\sin\theta_1\sin\theta_2\cos(\varphi_2-\varphi_3)+\cos\theta_1\cos\theta_2\].
\n

\section{Log-dependent behavior in the infrared region}
In this section, we will demonstrate that all the non-Gaussian diagrams have a similar scaling in the infrared region, characterized by the following logarithmic dependence:
\e\label{scaling}
\Omega_{\mathrm{GW}} \propto \(\frac{k}{k_\star} \)^3 \ln^2\(\frac{4k_{\star}^2}{3k^2}\),
\q
where $k_\star$ is a reference scale which we will discuss below and the slope index is given by:
\e\label{ngw}
n_{\mathrm{GW}}\equiv \frac{\mathrm{d}\ln \Omega_{\mathrm{GW}} }{\mathrm{d} \ln k} = 3 - \frac{4}{ \ln\frac{4k_{\star}^2}{3k^2} }.
\q
This logarithmic scaling law was initially investigated in \cite{Yuan:2019wwo} for the Gaussian case, where the authors considered a generic power spectrum with a peak at  $k_*$. More recently, in \cite{Adshead:2021hnm}, the authors also identified logarithmic scaling for $F_{\mathrm{NL}}^2$ terms and $F_{\mathrm{NL}}^4$ terms. In this study, we provide a proof for the ``tri'' term as an example using the methodology outlined in \cite{Yuan:2019wwo}.
First of all, we rewrite the ``tri'' term as follows:
\m
\Omega_{\mathrm{GW}}^{tri}(k)
=&&{G_{\mathrm{NL}}^2 \over 4\pi^2}\int_0^{\infty} \mathrm{d}u \int_{|1-u|}^{1+u}\mathrm{d}v \int_0^{\infty} \mathrm{d}u_1 \int_{|1-u_1|}^{1+u_1}\mathrm{d}v_1 \int_0^{\infty} \mathrm{d}u_2 \int_{|1-u_2|}^{1+u_2}\mathrm{d}v_2 \int_0^{2\pi}\mathrm{d}\varphi_2\int_0^{2\pi}\mathrm{d}\varphi_3\nonumber\\
&&\times I^2(u,v)\frac{uvu_1v_1u_2v_2}{(u_1u_2w_{134}v)^3}\mathcal{P} _{g}\left(u_1k\right) \mathcal{P} _{g}\left(u_2k\right)\mathcal{P} _{g}\left(w_{134}k\right) \mathcal{P} _{g}\left(uk\right).
\n
To effectively analyze the scaling, we consider a generic power spectrum with a peak at $k_*$ and introduce two parameters, $k_-$ and $k_+$, in such a way that the power spectrum is mainly distributed in $k\in [k_-,k_+]$ and we neglect the portion beyond this range. Since the integral involves terms of the form $\mathcal{P}_g(u_1 k)\mathcal{P}_g(u_2 k)\mathcal{P}_g(uk)$, it follows that $u_1 k$, $u_2 k$, and $u k$ are constrained within the range $[k_-, k_+]$. Consequently, this imposes lower and upper limits on the variables $u_1$, $u_2$, and $u$, namely
\m
\Omega_{\mathrm{GW}}^{tri}(k)=&&{G_{\mathrm{NL}}^2 \over 4\pi^2}\int_{k_-/k}^{k_+/k} \mathrm{d}u \int_{|1-u|}^{1+u}\mathrm{d}v \int_{k_-/k}^{k_+/k} \mathrm{d}u_1 \int_{|1-u_1|}^{1+u_1}\mathrm{d}v_1 \int_{k_-/k}^{k_+/k} \mathrm{d}u_2 \int_{|1-u_2|}^{1+u_2}\mathrm{d}v_2 \int_0^{2\pi}\mathrm{d}\varphi_2\int_0^{2\pi}\mathrm{d}\varphi_3\nonumber\\
&&\times I^2(u,v)\frac{uvu_1v_1u_2v_2}{(u_1u_2w_{134}v)^3}\mathcal{P} _{g}\left(u_1k\right) \mathcal{P} _{g}\left(u_2k\right)\mathcal{P} _{g}\left(w_{0134}k\right) \mathcal{P} _{g}\left(uk\right).
\n
Since we are interested in the infrared region where $k\ll k_*$, it follows that $u$, $u_1$, and $u_2$ are much greater than $1$. Consequently, we can simplify the above equation employing the first mean value theorem for definite integrals
\m
\Omega_{\mathrm{GW}}^{tri}(k)=&&{2G_{\mathrm{NL}}^2 \over \pi^2}
\({k_+-k_- \over k}\)^3
I^2(u^*,v^*)\frac{u^*v^*u_{1}^*v_1^*u_2^*v_2^*}{(u_1^*u_2^*w_{134}^*v^*)^3}\mathcal{P} _{g}\left(u_1^*k\right) \mathcal{P} _{g}\left(u_2^*k\right)\mathcal{P} _{g}\left(w_{0134}^*k\right) \mathcal{P} _{g}\left(u^*k\right),
\n
where $u^*,u_1^*,u_2^*\in[k_-/k,k_+/k]$ and $v^*,v_1^*,v_2^*$ are in the range of $[u^*-1,u^*+1]$, $[u_1^*-1,u_1^*+1]$ and $[u_2^*-1,u_2^*+1]$ respectively. $w_{0134}^*$ is defined as replacing $u,v,u_1,v_1,u_2,v_2,\varphi_2,\varphi_3$ in Eq.~(\ref{w0134}) with $u^*,v^*,u_1^*,v_1^*,u_2^*,v_2^*,\varphi_2^*,\varphi_3^*$ and we have $\varphi_2^*,\varphi_3^*\in[0,2\pi]$. By expanding $u^*,u_1^*,u_2^*$ at $k_\star/k$ to leading order where $k_\star \in [k_-,k_+]$ is a reference scale, we obtain
\m
\Omega_{\mathrm{GW}}^{tri}(k)\propto \({k\over k_\star}\)^3I^2\({k_\star\over k},{k_\star \over k}\).
\n
Using the following asymptotic behavior for $u\gg 1$:
\e
I^2(u,u)\simeq{9\over4}\ln^2\({4u^2\over3}\),
\q
we finally get Eq.~(\ref{scaling}). All the scaling of non-Gaussian diagrams in the infrared region can be shown in the similar way.

It has been argued in \cite{Yuan:2019wwo} that this log-dependent scaling could be smoking gun for SIGW. However, the mean value theorm could not give us the exact value of the reference scale, $k_\star$, and one should treat $k_\star$ as free parameter in GW data analysis. 
Moreover, the value of $k_\star$ is different for different power spectrum and different non-Gaussian diagrams. Next, we will show that the scaling of the total energy spectrum also follows Eq.~(\ref{scaling}). 
First of all, we write down the total energy spectrum in the infrared region in a generic form as follows:
\e
\Omega_{\mathrm{GW}}(k) = \sum_{i} A_i \(\frac{k}{k_{\star i}}\)^3
\ln^2\({4k_{\star i}^2 \over 3k^2}\),
\q
where $A_i$ denotes the amplitude of the $i$-th non-Gaussian energy spectrum  and $k_{\star i}$ is the reference scale obtained using the mean value theorem for the $i$-th non-Gaussian energy spectrum. The above equation can be re-written as 
\e\label{scaling1}
\Omega_{\mathrm{GW}}(k) = k^3\sum_{i} 
{A_i \over k_{\star i}^3 } \(
c_i^2+2c_i \ln\({4k_{\star}^2 \over 3k^2}\)
+
\ln^2\({4k_{\star}^2 \over 3k^2}\)
\),
\q
where we introduce $c_i\equiv \ln\({k_{\star i}^2 \over k_\star ^2}\) $ and $k_\star$ is a reference scale. Note that $k_{\star i}$ is obtained using the mean value theorem and it is in the range of $[k_-,k_+]$. On the other hand, we choose $k_\star$ also to be in the range of $[k_-,k_+]$. Then we have
\e
{k_- \over k_+} \lesssim {k_\star \over k_{\star i} }\lesssim {k_+ \over k_-}
\q
In the infrared region where $k\ll k_-$, we obtain the relation
\e
c_i\ll \ln\({4k_{\star}^2 \over 3k^2}\), \mathrm{if} {k \over k_\star } \ll  {k_- \over k_+} 
\q
This indicates that, in the infrared region where $k/k_\star \ll k_-/k_+$, the $c_i$ terms in Eq.~(\ref{scaling1}) are negligible and we can obtain
\e\label{scaling2}
\Omega_{\mathrm{GW}}(k) \simeq k^3\ln^2\({4k_{\star}^2 \over 3k^2}\) \sum_{i} 
{A_i \over k_{\star i}^3 },
\q
This finally lead us to Eq.~(\ref{scaling}) and the slope index is still given by Eq.~(\ref{ngw}).
Note that the region where $k/k_\star \ll k_-/k_+$ depends on the width of the power spectrum through $k_-/k_+$. For narrow spectrum, it reduces to $k_-/ k_+\simeq 1$. While for some wide spectrum where $k_+$ is several orders of magnitude larger than $k_-$, Eq.~(\ref{ngw}) is a approximation only for sufficient small $k$.

Our results show that, the SIGWs from PBH formation will also exhibit a log-dependent scaling in the infrared region, regardless of the specific shape of the power spectrum even in the non-Gaussian case. This log-dependent scaling comes from the oscillating behavior of the evolution of the scalar perturbations during RD, and could be a smoking gun for detecting SIGW from PBHs. 

To have deeper insights into the influence of non-Gaussianities on the energy spectrum SIGWs, we consider a log-normal shape spectrum which is widely used when studying the SIGW spectrum (e.g., \cite{Inomata:2018epa,Pi:2020otn,Yuan:2020iwf,Kohri:2020qqd,Meng:2022ixx,Yuan:2021qgz}), namely
\e\label{lognormal}
\mathcal{P}_g(k) = \frac{A}{\sqrt{2 \pi \sigma_{*}^{2}}} \exp \(-\frac{\ln^2 \(k / k_{*}\)}{2 \sigma_{*}^{2}}\),
\q
where the dimensionless parameter $\sigma_*$ is related to the width of the spectrum and we normalize the power spectrum in such way that $\int \mathcal{P}_g(k) d\ln k = A$.

\begin{figure}
	\centering
	\includegraphics[width=0.8\columnwidth]{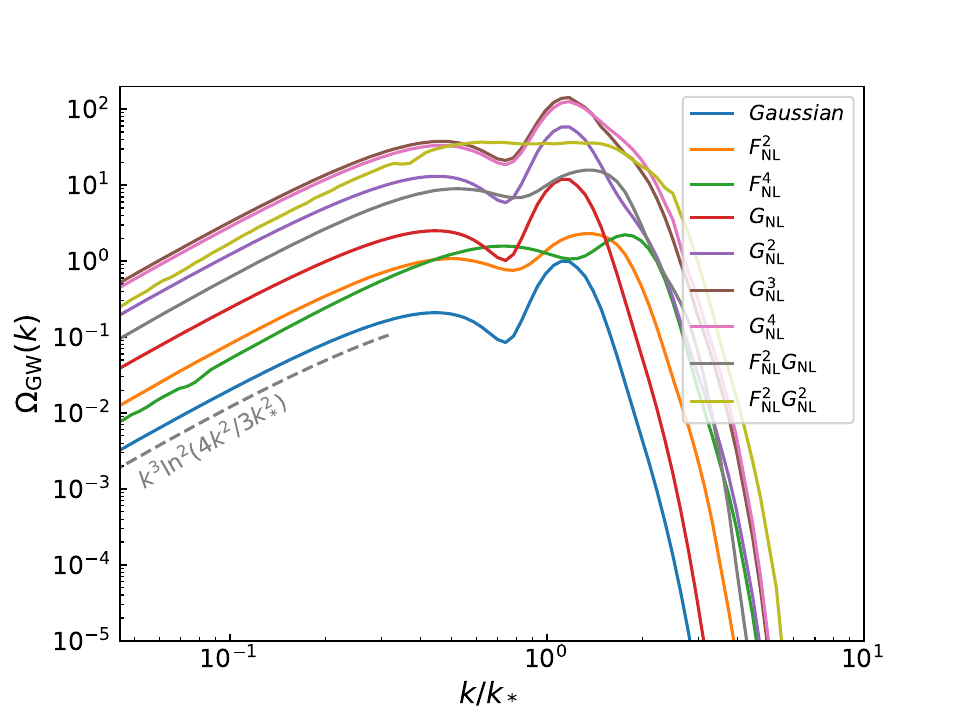}
	\caption{\label{Ogwng} The unscaled (By setting $A=1$, $F_{\mathrm{NL}}=1$ and $G_{\mathrm{NL}}=1$) energy spectrum of SIGW generated by a log-normal power spectrum described by Eq.~(\ref{lognormal}) with $\sigma_*=0.2$.}
\end{figure}

Given that analytical results for the non-Gaussian diagrams are challenging to obtain, in this regard, we use the \texttt{Cuba.jl} package \cite{Hahn:2004fe,Hahn:2014fua} to present the numerical results. Fig.~\ref{Ogwng} illustrates the energy spectrum of SIGW for each non-Gaussian term. As anticipated, the spectra exhibit peaks centered around $k_*$. Furthermore, the spectra sharply decrease for $k/k_* \gtrsim 2$, with the non-Gaussian terms displaying larger drop-off wavelength compared to the Gaussian term due to momentum conservation. Moreover, the non-Gaussian energy spectrum demonstrate a log-dependent scaling described by Eq.~(\ref{scaling}) in the infrared region.

\begin{figure}
	\centering
	\includegraphics[width=0.45\columnwidth]{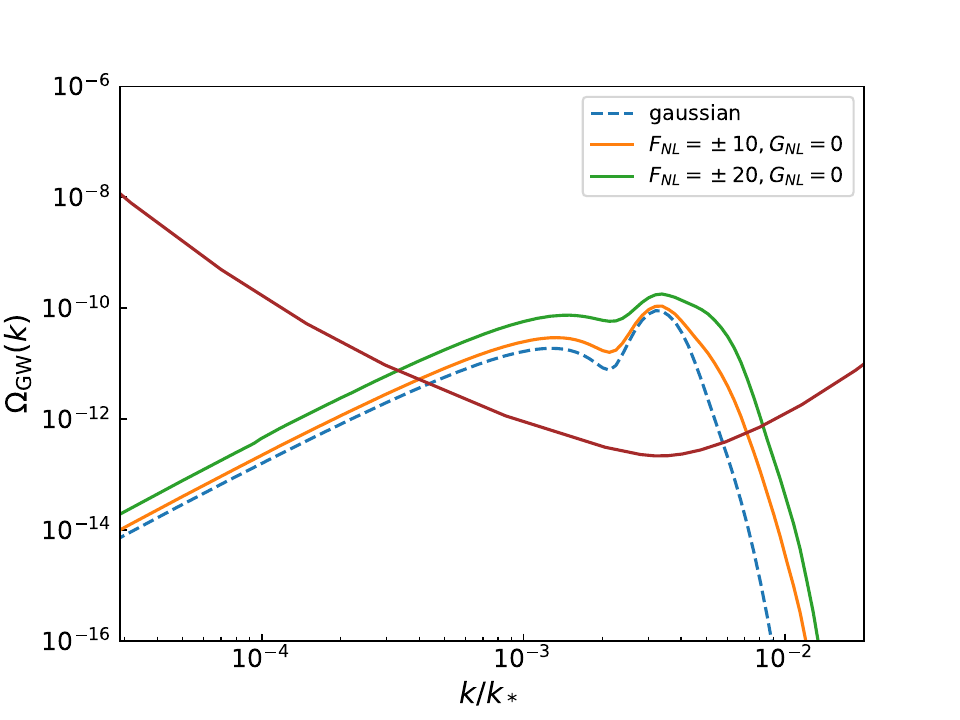}
	\includegraphics[width=0.45\columnwidth]{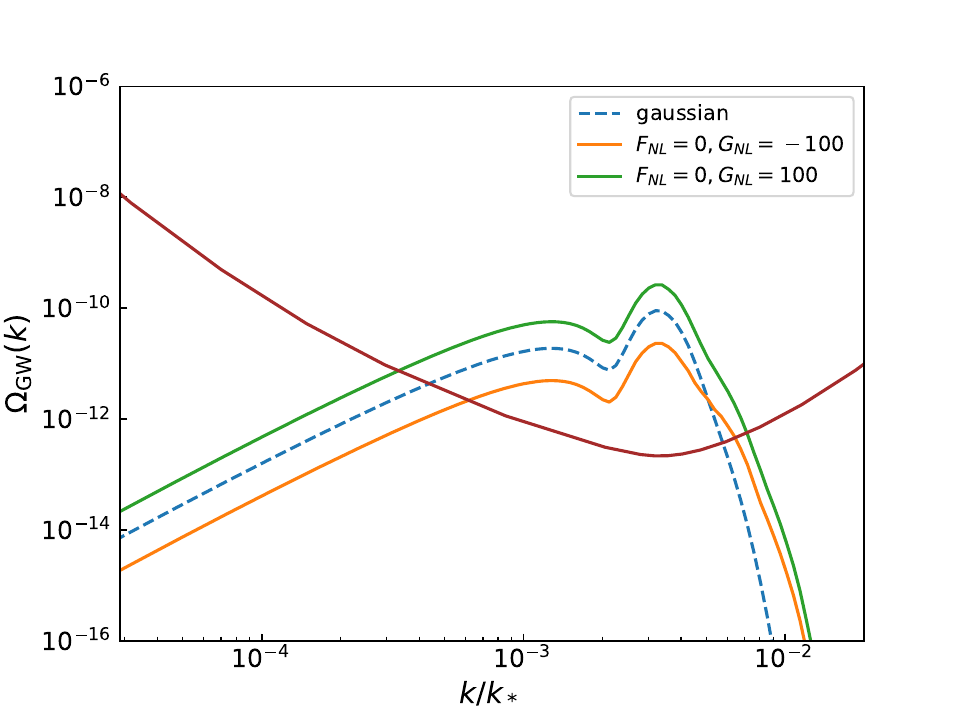}
	\caption{\label{Ogwfnlgnl} The total energy spectrum of SIGW generated by a log-normal power spectrum described by Eq.~(\ref{lognormal}) with $\sigma_*=0.2$. The brown line in each panel denotes the power-law sensitivity curve of LISA, assuming a $4$ year detection time. We set $A=10^{-3}$ for both panels. Left panel: The energy spectrum in the absence of $G_\mathrm{NL}$. Right panel: The energy spectrum in the absence of $F_\mathrm{NL}$.}
\end{figure}

In Fig.~\ref{Ogwfnlgnl}, we present the total energy spectrum of SIGW by today for some representative values of $F_{\mathrm{NL}}$, $G_{\mathrm{NL}}$ and $A$.
As shown in Fig. \ref{Ogwfnlgnl}, the role of non-Gaussian corrections might change the amplitude and the shape of the energy spectrum. The shape of the energy spectrum is modified mainly around $k_*$ but in the infrared region it exhibits a log-dependent scaling given by Eq.~(\ref{ngw}).
It is worth mentioning that we treat $F_\mathrm{NL}$ and $G_\mathrm{NL}$ as free parameters in this paper. Nevertheless, within certain inflationary models, a perturbativity condition could be applicable to the non-Gaussian parameters \cite{Kristiano:2021urj,Meng:2022ixx,Kristiano:2022maq,Garcia-Saenz:2022tzu}. Consequently, the non-Gaussian parameters may not retain.

\section{Conclusion and Discussion}
In this paper, we study the full impacts of a local-type non-Gaussianities up to $G_{\mathrm{NL}}$ order on SIGW and derive semi-analytical results for arbitrary primordial power spectrum. All the non-Gaussian terms to the energy spectrum of SIGW exhibit a log-dependent scaling in the infrared region. This log-dependent scaling distinguishes SIGW from other GW energy spectra generated by currently known physical processes, making it a smoking gun for detecting the SIGW. Recently, NANOGrav collaboration searched the SIGW data for various power spectrum and claimed that the NANOGrav 15-year data is well fit in the low-frequency tail of SIGW \cite{NANOGrav:2023hvm}, indicating the significance of the log-dependent scaling in searching the SIGW signals.

In addition to expanding the curvature perturbation in a non-Gaussian manner, it is feasible to extend the expansion of the source term in the Einstein equations to higher orders, thereby obtaining higher-order corrections to $\Omega_{\mathrm{GW}}$ \cite{Yuan:2019udt,Chang:2022nzu}. For instance, considering the third-order tensor mode $h_{ij}^{(3)}$, the subsequent-order correction to $\Omega_{\mathrm{GW}}$ (assuming $\Gnl=0$ for illustration) would arise from $\left\langle h^{(2)} h^{(3)} \right\rangle\sim \Fnl A_g^3+\Fnl^3A_g^4+\Fnl^5 A_g^5$ or $\left\langle h^{(3)} h^{(3)} \right\rangle\sim A_g^3+\Fnl^2 A_g^4+\Fnl^4A_g^5+\Fnl^6 A_g^6$. However, in scenarios involving the formation of PBHs within non-Gaussian regimes, both $\Fnl A$ and $A_g$ are much smaller than $1$ when the abundance of PBHs in the Universe is fixed (refer to e.g., Fig.~6 in \cite{Cai:2019elf}). Consequently, this leads to the inference that corrections from $\left\langle h^{(2)} h^{(3)} \right\rangle$ and $\left\langle h^{(3)} h^{(3)} \right\rangle$ are at least an order of magnitude smaller, by a factor of $\Fnl A_g$ and $A_g$ respectively, when compared to the leading order of $\Omega_{\mathrm{GW}}$. This conclusion implies that even upon considering the non-Gaussian expansion of higher order source terms, the resultant corrections remain subdominant.

\vspace{5mm}
{\it Acknowledgments. }
D-S.M. would like to thank Guang-shang Chen for his enthusiastic help in computer and programming. The work is supported by the National Key Research and Development Program of China Grant No.2020YFC2201502, grants from NSFC (grant No. 11975019, 11991052, 12047503), Key Research Program of Frontier Sciences, CAS, Grant NO. ZDBS-LY-7009, CAS Project for Young Scientists in Basic Research YSBR-006, the Key Research Program of the Chinese Academy of Sciences (Grant NO. XDPB15). We acknowledge the use of HPC Cluster of ITP-CAS. C.Y. acknowledges financial support provided under the European Union's H2020 ERC Advanced Grant ``Black holes: gravitational engines of discovery'' grant agreement
no.\ Gravitas--101052587. Views and opinions expressed are however those of the author only and do not necessarily reflect those of the European Union or the European Research Council. Neither the European Union nor the granting authority can be held responsible for them.
This project has received funding from the European Union's Horizon 2020 research and innovation programme under the Marie Sklodowska-Curie grant agreement No 101007855.
No 101007855.

\bibliography{./ref}

\begin{thebibliography}{131}%
\makeatletter
\providecommand \@ifxundefined [1]{%
 \@ifx{#1\undefined}
}%
\providecommand \@ifnum [1]{%
 \ifnum #1\expandafter \@firstoftwo
 \else \expandafter \@secondoftwo
 \fi
}%
\providecommand \@ifx [1]{%
 \ifx #1\expandafter \@firstoftwo
 \else \expandafter \@secondoftwo
 \fi
}%
\providecommand \natexlab [1]{#1}%
\providecommand \enquote  [1]{``#1''}%
\providecommand \bibnamefont  [1]{#1}%
\providecommand \bibfnamefont [1]{#1}%
\providecommand \citenamefont [1]{#1}%
\providecommand \href@noop [0]{\@secondoftwo}%
\providecommand \href [0]{\begingroup \@sanitize@url \@href}%
\providecommand \@href[1]{\@@startlink{#1}\@@href}%
\providecommand \@@href[1]{\endgroup#1\@@endlink}%
\providecommand \@sanitize@url [0]{\catcode `\\12\catcode `\$12\catcode
  `\&12\catcode `\#12\catcode `\^12\catcode `\_12\catcode `\%12\relax}%
\providecommand \@@startlink[1]{}%
\providecommand \@@endlink[0]{}%
\providecommand \url  [0]{\begingroup\@sanitize@url \@url }%
\providecommand \@url [1]{\endgroup\@href {#1}{\urlprefix }}%
\providecommand \urlprefix  [0]{URL }%
\providecommand \Eprint [0]{\href }%
\providecommand \doibase [0]{http://dx.doi.org/}%
\providecommand \selectlanguage [0]{\@gobble}%
\providecommand \bibinfo  [0]{\@secondoftwo}%
\providecommand \bibfield  [0]{\@secondoftwo}%
\providecommand \translation [1]{[#1]}%
\providecommand \BibitemOpen [0]{}%
\providecommand \bibitemStop [0]{}%
\providecommand \bibitemNoStop [0]{.\EOS\space}%
\providecommand \EOS [0]{\spacefactor3000\relax}%
\providecommand \BibitemShut  [1]{\csname bibitem#1\endcsname}%
\let\auto@bib@innerbib\@empty
\bibitem [{\citenamefont {Zel'dovich}\ and\ \citenamefont
  {Novikov}(1967)}]{Zeldovich:1967lct}%
  \BibitemOpen
  \bibfield  {author} {\bibinfo {author} {\bibfnamefont {Ya.~B.}\ \bibnamefont
  {Zel'dovich}}\ and\ \bibinfo {author} {\bibfnamefont {I.~D.}\ \bibnamefont
  {Novikov}},\ }\bibfield  {title} {\enquote {\bibinfo {title} {{The Hypothesis
  of Cores Retarded during Expansion and the Hot Cosmological Model}},}\
  }\href@noop {} {\bibfield  {journal} {\bibinfo  {journal} {Soviet Astron. AJ
  (Engl. Transl. ),}\ }\textbf {\bibinfo {volume} {10}},\ \bibinfo {pages}
  {602} (\bibinfo {year} {1967})}\BibitemShut {NoStop}%
\bibitem [{\citenamefont {Hawking}(1971)}]{Hawking:1971ei}%
  \BibitemOpen
  \bibfield  {author} {\bibinfo {author} {\bibfnamefont {Stephen}\ \bibnamefont
  {Hawking}},\ }\bibfield  {title} {\enquote {\bibinfo {title}
  {{Gravitationally collapsed objects of very low mass}},}\ }\href@noop {}
  {\bibfield  {journal} {\bibinfo  {journal} {Mon. Not. Roy. Astron. Soc.}\
  }\textbf {\bibinfo {volume} {152}},\ \bibinfo {pages} {75} (\bibinfo {year}
  {1971})}\BibitemShut {NoStop}%
\bibitem [{\citenamefont {Carr}\ and\ \citenamefont
  {Hawking}(1974)}]{Carr:1974nx}%
  \BibitemOpen
  \bibfield  {author} {\bibinfo {author} {\bibfnamefont {Bernard~J.}\
  \bibnamefont {Carr}}\ and\ \bibinfo {author} {\bibfnamefont {S.~W.}\
  \bibnamefont {Hawking}},\ }\bibfield  {title} {\enquote {\bibinfo {title}
  {{Black holes in the early Universe}},}\ }\href {\doibase
  10.1093/mnras/168.2.399} {\bibfield  {journal} {\bibinfo  {journal} {Mon.
  Not. Roy. Astron. Soc.}\ }\textbf {\bibinfo {volume} {168}},\ \bibinfo
  {pages} {399--415} (\bibinfo {year} {1974})}\BibitemShut {NoStop}%
\bibitem [{\citenamefont {Carr}(1975)}]{Carr:1975qj}%
  \BibitemOpen
  \bibfield  {author} {\bibinfo {author} {\bibfnamefont {Bernard~J.}\
  \bibnamefont {Carr}},\ }\bibfield  {title} {\enquote {\bibinfo {title} {{The
  Primordial black hole mass spectrum}},}\ }\href {\doibase 10.1086/153853}
  {\bibfield  {journal} {\bibinfo  {journal} {Astrophys. J.}\ }\textbf
  {\bibinfo {volume} {201}},\ \bibinfo {pages} {1--19} (\bibinfo {year}
  {1975})}\BibitemShut {NoStop}%
\bibitem [{\citenamefont {Carr}\ \emph {et~al.}(2010)\citenamefont {Carr},
  \citenamefont {Kohri}, \citenamefont {Sendouda},\ and\ \citenamefont
  {Yokoyama}}]{Carr:2009jm}%
  \BibitemOpen
  \bibfield  {author} {\bibinfo {author} {\bibfnamefont {B.~J.}\ \bibnamefont
  {Carr}}, \bibinfo {author} {\bibfnamefont {Kazunori}\ \bibnamefont {Kohri}},
  \bibinfo {author} {\bibfnamefont {Yuuiti}\ \bibnamefont {Sendouda}}, \ and\
  \bibinfo {author} {\bibfnamefont {Jun'ichi}\ \bibnamefont {Yokoyama}},\
  }\bibfield  {title} {\enquote {\bibinfo {title} {{New cosmological
  constraints on primordial black holes}},}\ }\href {\doibase
  10.1103/PhysRevD.81.104019} {\bibfield  {journal} {\bibinfo  {journal} {Phys.
  Rev. D}\ }\textbf {\bibinfo {volume} {81}},\ \bibinfo {pages} {104019}
  (\bibinfo {year} {2010})},\ \Eprint {http://arxiv.org/abs/0912.5297}
  {arXiv:0912.5297 [astro-ph.CO]} \BibitemShut {NoStop}%
\bibitem [{\citenamefont {Graham}\ \emph {et~al.}(2015)\citenamefont {Graham},
  \citenamefont {Rajendran},\ and\ \citenamefont {Varela}}]{Graham:2015apa}%
  \BibitemOpen
  \bibfield  {author} {\bibinfo {author} {\bibfnamefont {Peter~W.}\
  \bibnamefont {Graham}}, \bibinfo {author} {\bibfnamefont {Surjeet}\
  \bibnamefont {Rajendran}}, \ and\ \bibinfo {author} {\bibfnamefont {Jaime}\
  \bibnamefont {Varela}},\ }\bibfield  {title} {\enquote {\bibinfo {title}
  {{Dark Matter Triggers of Supernovae}},}\ }\href {\doibase
  10.1103/PhysRevD.92.063007} {\bibfield  {journal} {\bibinfo  {journal} {Phys.
  Rev.}\ }\textbf {\bibinfo {volume} {D92}},\ \bibinfo {pages} {063007}
  (\bibinfo {year} {2015})},\ \Eprint {http://arxiv.org/abs/1505.04444}
  {arXiv:1505.04444 [hep-ph]} \BibitemShut {NoStop}%
\bibitem [{\citenamefont {Niikura}\ \emph
  {et~al.}(2019{\natexlab{a}})\citenamefont {Niikura} \emph
  {et~al.}}]{Niikura:2017zjd}%
  \BibitemOpen
  \bibfield  {author} {\bibinfo {author} {\bibfnamefont {Hiroko}\ \bibnamefont
  {Niikura}} \emph {et~al.},\ }\bibfield  {title} {\enquote {\bibinfo {title}
  {{Microlensing constraints on primordial black holes with Subaru/HSC
  Andromeda observations}},}\ }\href {\doibase 10.1038/s41550-019-0723-1}
  {\bibfield  {journal} {\bibinfo  {journal} {Nature Astron.}\ }\textbf
  {\bibinfo {volume} {3}},\ \bibinfo {pages} {524--534} (\bibinfo {year}
  {2019}{\natexlab{a}})},\ \Eprint {http://arxiv.org/abs/1701.02151}
  {arXiv:1701.02151 [astro-ph.CO]} \BibitemShut {NoStop}%
\bibitem [{\citenamefont {Tisserand}\ \emph {et~al.}(2007)\citenamefont
  {Tisserand} \emph {et~al.}}]{EROS-2:2006ryy}%
  \BibitemOpen
  \bibfield  {author} {\bibinfo {author} {\bibfnamefont {P.}~\bibnamefont
  {Tisserand}} \emph {et~al.} (\bibinfo {collaboration} {EROS-2}),\ }\bibfield
  {title} {\enquote {\bibinfo {title} {{Limits on the Macho Content of the
  Galactic Halo from the EROS-2 Survey of the Magellanic Clouds}},}\ }\href
  {\doibase 10.1051/0004-6361:20066017} {\bibfield  {journal} {\bibinfo
  {journal} {Astron. Astrophys.}\ }\textbf {\bibinfo {volume} {469}},\ \bibinfo
  {pages} {387--404} (\bibinfo {year} {2007})},\ \Eprint
  {http://arxiv.org/abs/astro-ph/0607207} {arXiv:astro-ph/0607207} \BibitemShut
  {NoStop}%
\bibitem [{\citenamefont {Niikura}\ \emph
  {et~al.}(2019{\natexlab{b}})\citenamefont {Niikura}, \citenamefont {Takada},
  \citenamefont {Yokoyama}, \citenamefont {Sumi},\ and\ \citenamefont
  {Masaki}}]{Niikura:2019kqi}%
  \BibitemOpen
  \bibfield  {author} {\bibinfo {author} {\bibfnamefont {Hiroko}\ \bibnamefont
  {Niikura}}, \bibinfo {author} {\bibfnamefont {Masahiro}\ \bibnamefont
  {Takada}}, \bibinfo {author} {\bibfnamefont {Shuichiro}\ \bibnamefont
  {Yokoyama}}, \bibinfo {author} {\bibfnamefont {Takahiro}\ \bibnamefont
  {Sumi}}, \ and\ \bibinfo {author} {\bibfnamefont {Shogo}\ \bibnamefont
  {Masaki}},\ }\bibfield  {title} {\enquote {\bibinfo {title} {{Constraints on
  Earth-mass primordial black holes from OGLE 5-year microlensing events}},}\
  }\href {\doibase 10.1103/PhysRevD.99.083503} {\bibfield  {journal} {\bibinfo
  {journal} {Phys. Rev. D}\ }\textbf {\bibinfo {volume} {99}},\ \bibinfo
  {pages} {083503} (\bibinfo {year} {2019}{\natexlab{b}})},\ \Eprint
  {http://arxiv.org/abs/1901.07120} {arXiv:1901.07120 [astro-ph.CO]}
  \BibitemShut {NoStop}%
\bibitem [{\citenamefont {Wang}\ \emph {et~al.}(2018)\citenamefont {Wang},
  \citenamefont {Wang}, \citenamefont {Huang},\ and\ \citenamefont
  {Li}}]{Wang:2016ana}%
  \BibitemOpen
  \bibfield  {author} {\bibinfo {author} {\bibfnamefont {Sai}\ \bibnamefont
  {Wang}}, \bibinfo {author} {\bibfnamefont {Yi-Fan}\ \bibnamefont {Wang}},
  \bibinfo {author} {\bibfnamefont {Qing-Guo}\ \bibnamefont {Huang}}, \ and\
  \bibinfo {author} {\bibfnamefont {Tjonnie G.~F.}\ \bibnamefont {Li}},\
  }\bibfield  {title} {\enquote {\bibinfo {title} {{Constraints on the
  Primordial Black Hole Abundance from the First Advanced LIGO Observation Run
  Using the Stochastic Gravitational-Wave Background}},}\ }\href {\doibase
  10.1103/PhysRevLett.120.191102} {\bibfield  {journal} {\bibinfo  {journal}
  {Phys. Rev. Lett.}\ }\textbf {\bibinfo {volume} {120}},\ \bibinfo {pages}
  {191102} (\bibinfo {year} {2018})},\ \Eprint
  {http://arxiv.org/abs/1610.08725} {arXiv:1610.08725 [astro-ph.CO]}
  \BibitemShut {NoStop}%
\bibitem [{\citenamefont {Chen}\ and\ \citenamefont
  {Huang}(2020)}]{Chen:2019irf}%
  \BibitemOpen
  \bibfield  {author} {\bibinfo {author} {\bibfnamefont {Zu-Cheng}\
  \bibnamefont {Chen}}\ and\ \bibinfo {author} {\bibfnamefont {Qing-Guo}\
  \bibnamefont {Huang}},\ }\bibfield  {title} {\enquote {\bibinfo {title}
  {{Distinguishing Primordial Black Holes from Astrophysical Black Holes by
  Einstein Telescope and Cosmic Explorer}},}\ }\href {\doibase
  10.1088/1475-7516/2020/08/039} {\bibfield  {journal} {\bibinfo  {journal}
  {JCAP}\ }\textbf {\bibinfo {volume} {08}},\ \bibinfo {pages} {039} (\bibinfo
  {year} {2020})},\ \Eprint {http://arxiv.org/abs/1904.02396} {arXiv:1904.02396
  [astro-ph.CO]} \BibitemShut {NoStop}%
\bibitem [{\citenamefont {Brandt}(2016)}]{Brandt:2016aco}%
  \BibitemOpen
  \bibfield  {author} {\bibinfo {author} {\bibfnamefont {Timothy~D.}\
  \bibnamefont {Brandt}},\ }\bibfield  {title} {\enquote {\bibinfo {title}
  {{Constraints on MACHO Dark Matter from Compact Stellar Systems in
  Ultra-Faint Dwarf Galaxies}},}\ }\href {\doibase 10.3847/2041-8205/824/2/L31}
  {\bibfield  {journal} {\bibinfo  {journal} {Astrophys. J. Lett.}\ }\textbf
  {\bibinfo {volume} {824}},\ \bibinfo {pages} {L31} (\bibinfo {year}
  {2016})},\ \Eprint {http://arxiv.org/abs/1605.03665} {arXiv:1605.03665
  [astro-ph.GA]} \BibitemShut {NoStop}%
\bibitem [{\citenamefont {Chen}\ \emph {et~al.}(2020)\citenamefont {Chen},
  \citenamefont {Yuan},\ and\ \citenamefont {Huang}}]{Chen:2019xse}%
  \BibitemOpen
  \bibfield  {author} {\bibinfo {author} {\bibfnamefont {Zu-Cheng}\
  \bibnamefont {Chen}}, \bibinfo {author} {\bibfnamefont {Chen}\ \bibnamefont
  {Yuan}}, \ and\ \bibinfo {author} {\bibfnamefont {Qing-Guo}\ \bibnamefont
  {Huang}},\ }\bibfield  {title} {\enquote {\bibinfo {title} {{Pulsar Timing
  Array Constraints on Primordial Black Holes with NANOGrav 11-Year
  Dataset}},}\ }\href {\doibase 10.1103/PhysRevLett.124.251101} {\bibfield
  {journal} {\bibinfo  {journal} {Phys. Rev. Lett.}\ }\textbf {\bibinfo
  {volume} {124}},\ \bibinfo {pages} {251101} (\bibinfo {year} {2020})},\
  \Eprint {http://arxiv.org/abs/1910.12239} {arXiv:1910.12239 [astro-ph.CO]}
  \BibitemShut {NoStop}%
\bibitem [{\citenamefont {Montero-Camacho}\ \emph {et~al.}(2019)\citenamefont
  {Montero-Camacho}, \citenamefont {Fang}, \citenamefont {Vasquez},
  \citenamefont {Silva},\ and\ \citenamefont
  {Hirata}}]{Montero-Camacho:2019jte}%
  \BibitemOpen
  \bibfield  {author} {\bibinfo {author} {\bibfnamefont {Paulo}\ \bibnamefont
  {Montero-Camacho}}, \bibinfo {author} {\bibfnamefont {Xiao}\ \bibnamefont
  {Fang}}, \bibinfo {author} {\bibfnamefont {Gabriel}\ \bibnamefont {Vasquez}},
  \bibinfo {author} {\bibfnamefont {Makana}\ \bibnamefont {Silva}}, \ and\
  \bibinfo {author} {\bibfnamefont {Christopher~M.}\ \bibnamefont {Hirata}},\
  }\bibfield  {title} {\enquote {\bibinfo {title} {{Revisiting constraints on
  asteroid-mass primordial black holes as dark matter candidates}},}\ }\href
  {\doibase 10.1088/1475-7516/2019/08/031} {\bibfield  {journal} {\bibinfo
  {journal} {JCAP}\ }\textbf {\bibinfo {volume} {08}},\ \bibinfo {pages} {031}
  (\bibinfo {year} {2019})},\ \Eprint {http://arxiv.org/abs/1906.05950}
  {arXiv:1906.05950 [astro-ph.CO]} \BibitemShut {NoStop}%
\bibitem [{\citenamefont {Laha}(2019)}]{Laha:2019ssq}%
  \BibitemOpen
  \bibfield  {author} {\bibinfo {author} {\bibfnamefont {Ranjan}\ \bibnamefont
  {Laha}},\ }\bibfield  {title} {\enquote {\bibinfo {title} {{Primordial Black
  Holes as a Dark Matter Candidate Are Severely Constrained by the Galactic
  Center 511 keV $\gamma$ -Ray Line}},}\ }\href {\doibase
  10.1103/PhysRevLett.123.251101} {\bibfield  {journal} {\bibinfo  {journal}
  {Phys. Rev. Lett.}\ }\textbf {\bibinfo {volume} {123}},\ \bibinfo {pages}
  {251101} (\bibinfo {year} {2019})},\ \Eprint
  {http://arxiv.org/abs/1906.09994} {arXiv:1906.09994 [astro-ph.HE]}
  \BibitemShut {NoStop}%
\bibitem [{\citenamefont {Dasgupta}\ \emph {et~al.}(2020)\citenamefont
  {Dasgupta}, \citenamefont {Laha},\ and\ \citenamefont
  {Ray}}]{Dasgupta:2019cae}%
  \BibitemOpen
  \bibfield  {author} {\bibinfo {author} {\bibfnamefont {Basudeb}\ \bibnamefont
  {Dasgupta}}, \bibinfo {author} {\bibfnamefont {Ranjan}\ \bibnamefont {Laha}},
  \ and\ \bibinfo {author} {\bibfnamefont {Anupam}\ \bibnamefont {Ray}},\
  }\bibfield  {title} {\enquote {\bibinfo {title} {{Neutrino and positron
  constraints on spinning primordial black hole dark matter}},}\ }\href
  {\doibase 10.1103/PhysRevLett.125.101101} {\bibfield  {journal} {\bibinfo
  {journal} {Phys. Rev. Lett.}\ }\textbf {\bibinfo {volume} {125}},\ \bibinfo
  {pages} {101101} (\bibinfo {year} {2020})},\ \Eprint
  {http://arxiv.org/abs/1912.01014} {arXiv:1912.01014 [hep-ph]} \BibitemShut
  {NoStop}%
\bibitem [{\citenamefont {Laha}\ \emph {et~al.}(2020)\citenamefont {Laha},
  \citenamefont {Mu\~noz},\ and\ \citenamefont {Slatyer}}]{Laha:2020ivk}%
  \BibitemOpen
  \bibfield  {author} {\bibinfo {author} {\bibfnamefont {Ranjan}\ \bibnamefont
  {Laha}}, \bibinfo {author} {\bibfnamefont {Julian~B.}\ \bibnamefont
  {Mu\~noz}}, \ and\ \bibinfo {author} {\bibfnamefont {Tracy~R.}\ \bibnamefont
  {Slatyer}},\ }\bibfield  {title} {\enquote {\bibinfo {title} {{INTEGRAL
  constraints on primordial black holes and particle dark matter}},}\ }\href
  {\doibase 10.1103/PhysRevD.101.123514} {\bibfield  {journal} {\bibinfo
  {journal} {Phys. Rev. D}\ }\textbf {\bibinfo {volume} {101}},\ \bibinfo
  {pages} {123514} (\bibinfo {year} {2020})},\ \Eprint
  {http://arxiv.org/abs/2004.00627} {arXiv:2004.00627 [astro-ph.CO]}
  \BibitemShut {NoStop}%
\bibitem [{\citenamefont {Saha}\ and\ \citenamefont
  {Laha}(2022)}]{Saha:2021pqf}%
  \BibitemOpen
  \bibfield  {author} {\bibinfo {author} {\bibfnamefont {Akash~Kumar}\
  \bibnamefont {Saha}}\ and\ \bibinfo {author} {\bibfnamefont {Ranjan}\
  \bibnamefont {Laha}},\ }\bibfield  {title} {\enquote {\bibinfo {title}
  {{Sensitivities on nonspinning and spinning primordial black hole dark matter
  with global 21-cm troughs}},}\ }\href {\doibase 10.1103/PhysRevD.105.103026}
  {\bibfield  {journal} {\bibinfo  {journal} {Phys. Rev. D}\ }\textbf {\bibinfo
  {volume} {105}},\ \bibinfo {pages} {103026} (\bibinfo {year} {2022})},\
  \Eprint {http://arxiv.org/abs/2112.10794} {arXiv:2112.10794 [astro-ph.CO]}
  \BibitemShut {NoStop}%
\bibitem [{\citenamefont {Ray}\ \emph {et~al.}(2021)\citenamefont {Ray},
  \citenamefont {Laha}, \citenamefont {Mu\~noz},\ and\ \citenamefont
  {Caputo}}]{Ray:2021mxu}%
  \BibitemOpen
  \bibfield  {author} {\bibinfo {author} {\bibfnamefont {Anupam}\ \bibnamefont
  {Ray}}, \bibinfo {author} {\bibfnamefont {Ranjan}\ \bibnamefont {Laha}},
  \bibinfo {author} {\bibfnamefont {Julian~B.}\ \bibnamefont {Mu\~noz}}, \ and\
  \bibinfo {author} {\bibfnamefont {Regina}\ \bibnamefont {Caputo}},\
  }\bibfield  {title} {\enquote {\bibinfo {title} {{Near future MeV telescopes
  can discover asteroid-mass primordial black hole dark matter}},}\ }\href
  {\doibase 10.1103/PhysRevD.104.023516} {\bibfield  {journal} {\bibinfo
  {journal} {Phys. Rev. D}\ }\textbf {\bibinfo {volume} {104}},\ \bibinfo
  {pages} {023516} (\bibinfo {year} {2021})},\ \Eprint
  {http://arxiv.org/abs/2102.06714} {arXiv:2102.06714 [astro-ph.CO]}
  \BibitemShut {NoStop}%
\bibitem [{\citenamefont {Zheng}\ \emph {et~al.}(2023)\citenamefont {Zheng},
  \citenamefont {Li}, \citenamefont {Chen}, \citenamefont {Zhou},\ and\
  \citenamefont {Zhu}}]{Zheng:2022wqo}%
  \BibitemOpen
  \bibfield  {author} {\bibinfo {author} {\bibfnamefont {Li-Ming}\ \bibnamefont
  {Zheng}}, \bibinfo {author} {\bibfnamefont {Zhengxiang}\ \bibnamefont {Li}},
  \bibinfo {author} {\bibfnamefont {Zu-Cheng}\ \bibnamefont {Chen}}, \bibinfo
  {author} {\bibfnamefont {Huan}\ \bibnamefont {Zhou}}, \ and\ \bibinfo
  {author} {\bibfnamefont {Zong-Hong}\ \bibnamefont {Zhu}},\ }\bibfield
  {title} {\enquote {\bibinfo {title} {{Towards a reliable reconstruction of
  the power spectrum of primordial curvature perturbation on small scales from
  GWTC-3}},}\ }\href {\doibase 10.1016/j.physletb.2023.137720} {\bibfield
  {journal} {\bibinfo  {journal} {Phys. Lett. B}\ }\textbf {\bibinfo {volume}
  {838}},\ \bibinfo {pages} {137720} (\bibinfo {year} {2023})},\ \Eprint
  {http://arxiv.org/abs/2212.05516} {arXiv:2212.05516 [astro-ph.CO]}
  \BibitemShut {NoStop}%
\bibitem [{\citenamefont {Liu}\ \emph {et~al.}(2023{\natexlab{a}})\citenamefont
  {Liu}, \citenamefont {You}, \citenamefont {Wu},\ and\ \citenamefont
  {Chen}}]{Liu:2022iuf}%
  \BibitemOpen
  \bibfield  {author} {\bibinfo {author} {\bibfnamefont {Lang}\ \bibnamefont
  {Liu}}, \bibinfo {author} {\bibfnamefont {Zhi-Qiang}\ \bibnamefont {You}},
  \bibinfo {author} {\bibfnamefont {You}\ \bibnamefont {Wu}}, \ and\ \bibinfo
  {author} {\bibfnamefont {Zu-Cheng}\ \bibnamefont {Chen}},\ }\bibfield
  {title} {\enquote {\bibinfo {title} {{Constraining the merger history of
  primordial-black-hole binaries from GWTC-3}},}\ }\href {\doibase
  10.1103/PhysRevD.107.063035} {\bibfield  {journal} {\bibinfo  {journal}
  {Phys. Rev. D}\ }\textbf {\bibinfo {volume} {107}},\ \bibinfo {pages}
  {063035} (\bibinfo {year} {2023}{\natexlab{a}})},\ \Eprint
  {http://arxiv.org/abs/2210.16094} {arXiv:2210.16094 [astro-ph.CO]}
  \BibitemShut {NoStop}%
\bibitem [{\citenamefont {Chen}\ \emph {et~al.}(2023)\citenamefont {Chen},
  \citenamefont {Du}, \citenamefont {Huang},\ and\ \citenamefont
  {You}}]{Chen:2022fda}%
  \BibitemOpen
  \bibfield  {author} {\bibinfo {author} {\bibfnamefont {Zu-Cheng}\
  \bibnamefont {Chen}}, \bibinfo {author} {\bibfnamefont {Shen-Shi}\
  \bibnamefont {Du}}, \bibinfo {author} {\bibfnamefont {Qing-Guo}\ \bibnamefont
  {Huang}}, \ and\ \bibinfo {author} {\bibfnamefont {Zhi-Qiang}\ \bibnamefont
  {You}},\ }\bibfield  {title} {\enquote {\bibinfo {title} {{Constraints on
  primordial-black-hole population and cosmic expansion history from
  GWTC-3}},}\ }\href {\doibase 10.1088/1475-7516/2023/03/024} {\bibfield
  {journal} {\bibinfo  {journal} {JCAP}\ }\textbf {\bibinfo {volume} {03}},\
  \bibinfo {pages} {024} (\bibinfo {year} {2023})},\ \Eprint
  {http://arxiv.org/abs/2205.11278} {arXiv:2205.11278 [astro-ph.CO]}
  \BibitemShut {NoStop}%
\bibitem [{\citenamefont {Chen}\ \emph {et~al.}(2019)\citenamefont {Chen},
  \citenamefont {Huang},\ and\ \citenamefont {Huang}}]{Chen:2018rzo}%
  \BibitemOpen
  \bibfield  {author} {\bibinfo {author} {\bibfnamefont {Zu-Cheng}\
  \bibnamefont {Chen}}, \bibinfo {author} {\bibfnamefont {Fan}\ \bibnamefont
  {Huang}}, \ and\ \bibinfo {author} {\bibfnamefont {Qing-Guo}\ \bibnamefont
  {Huang}},\ }\bibfield  {title} {\enquote {\bibinfo {title} {{Stochastic
  Gravitational-wave Background from Binary Black Holes and Binary Neutron
  Stars and Implications for LISA}},}\ }\href {\doibase
  10.3847/1538-4357/aaf581} {\bibfield  {journal} {\bibinfo  {journal}
  {Astrophys. J.}\ }\textbf {\bibinfo {volume} {871}},\ \bibinfo {pages} {97}
  (\bibinfo {year} {2019})},\ \Eprint {http://arxiv.org/abs/1809.10360}
  {arXiv:1809.10360 [gr-qc]} \BibitemShut {NoStop}%
\bibitem [{\citenamefont {Mittal}\ \emph {et~al.}(2022)\citenamefont {Mittal},
  \citenamefont {Ray}, \citenamefont {Kulkarni},\ and\ \citenamefont
  {Dasgupta}}]{Mittal:2021egv}%
  \BibitemOpen
  \bibfield  {author} {\bibinfo {author} {\bibfnamefont {Shikhar}\ \bibnamefont
  {Mittal}}, \bibinfo {author} {\bibfnamefont {Anupam}\ \bibnamefont {Ray}},
  \bibinfo {author} {\bibfnamefont {Girish}\ \bibnamefont {Kulkarni}}, \ and\
  \bibinfo {author} {\bibfnamefont {Basudeb}\ \bibnamefont {Dasgupta}},\
  }\bibfield  {title} {\enquote {\bibinfo {title} {{Constraining primordial
  black holes as dark matter using the global 21-cm signal with X-ray heating
  and excess radio background}},}\ }\href {\doibase
  10.1088/1475-7516/2022/03/030} {\bibfield  {journal} {\bibinfo  {journal}
  {JCAP}\ }\textbf {\bibinfo {volume} {03}},\ \bibinfo {pages} {030} (\bibinfo
  {year} {2022})},\ \Eprint {http://arxiv.org/abs/2107.02190} {arXiv:2107.02190
  [astro-ph.CO]} \BibitemShut {NoStop}%
\bibitem [{\citenamefont {Carr}\ \emph {et~al.}(2021)\citenamefont {Carr},
  \citenamefont {Kohri}, \citenamefont {Sendouda},\ and\ \citenamefont
  {Yokoyama}}]{Carr:2020gox}%
  \BibitemOpen
  \bibfield  {author} {\bibinfo {author} {\bibfnamefont {Bernard}\ \bibnamefont
  {Carr}}, \bibinfo {author} {\bibfnamefont {Kazunori}\ \bibnamefont {Kohri}},
  \bibinfo {author} {\bibfnamefont {Yuuiti}\ \bibnamefont {Sendouda}}, \ and\
  \bibinfo {author} {\bibfnamefont {Jun'ichi}\ \bibnamefont {Yokoyama}},\
  }\bibfield  {title} {\enquote {\bibinfo {title} {{Constraints on primordial
  black holes}},}\ }\href {\doibase 10.1088/1361-6633/ac1e31} {\bibfield
  {journal} {\bibinfo  {journal} {Rept. Prog. Phys.}\ }\textbf {\bibinfo
  {volume} {84}},\ \bibinfo {pages} {116902} (\bibinfo {year} {2021})},\
  \Eprint {http://arxiv.org/abs/2002.12778} {arXiv:2002.12778 [astro-ph.CO]}
  \BibitemShut {NoStop}%
\bibitem [{\citenamefont {Franciolini}\ \emph {et~al.}(2018)\citenamefont
  {Franciolini}, \citenamefont {Kehagias}, \citenamefont {Matarrese},\ and\
  \citenamefont {Riotto}}]{Franciolini:2018vbk}%
  \BibitemOpen
  \bibfield  {author} {\bibinfo {author} {\bibfnamefont {G.}~\bibnamefont
  {Franciolini}}, \bibinfo {author} {\bibfnamefont {A.}~\bibnamefont
  {Kehagias}}, \bibinfo {author} {\bibfnamefont {S.}~\bibnamefont {Matarrese}},
  \ and\ \bibinfo {author} {\bibfnamefont {A.}~\bibnamefont {Riotto}},\
  }\bibfield  {title} {\enquote {\bibinfo {title} {{Primordial Black Holes from
  Inflation and non-Gaussianity}},}\ }\href {\doibase
  10.1088/1475-7516/2018/03/016} {\bibfield  {journal} {\bibinfo  {journal}
  {JCAP}\ }\textbf {\bibinfo {volume} {03}},\ \bibinfo {pages} {016} (\bibinfo
  {year} {2018})},\ \Eprint {http://arxiv.org/abs/1801.09415} {arXiv:1801.09415
  [astro-ph.CO]} \BibitemShut {NoStop}%
\bibitem [{\citenamefont {Biagetti}\ \emph {et~al.}(2018)\citenamefont
  {Biagetti}, \citenamefont {Franciolini}, \citenamefont {Kehagias},\ and\
  \citenamefont {Riotto}}]{Biagetti:2018pjj}%
  \BibitemOpen
  \bibfield  {author} {\bibinfo {author} {\bibfnamefont {Matteo}\ \bibnamefont
  {Biagetti}}, \bibinfo {author} {\bibfnamefont {Gabriele}\ \bibnamefont
  {Franciolini}}, \bibinfo {author} {\bibfnamefont {Alex}\ \bibnamefont
  {Kehagias}}, \ and\ \bibinfo {author} {\bibfnamefont {Antonio}\ \bibnamefont
  {Riotto}},\ }\bibfield  {title} {\enquote {\bibinfo {title} {{Primordial
  Black Holes from Inflation and Quantum Diffusion}},}\ }\href {\doibase
  10.1088/1475-7516/2018/07/032} {\bibfield  {journal} {\bibinfo  {journal}
  {JCAP}\ }\textbf {\bibinfo {volume} {07}},\ \bibinfo {pages} {032} (\bibinfo
  {year} {2018})},\ \Eprint {http://arxiv.org/abs/1804.07124} {arXiv:1804.07124
  [astro-ph.CO]} \BibitemShut {NoStop}%
\bibitem [{\citenamefont {Atal}\ and\ \citenamefont
  {Germani}(2019)}]{Atal:2018neu}%
  \BibitemOpen
  \bibfield  {author} {\bibinfo {author} {\bibfnamefont {Vicente}\ \bibnamefont
  {Atal}}\ and\ \bibinfo {author} {\bibfnamefont {Cristiano}\ \bibnamefont
  {Germani}},\ }\bibfield  {title} {\enquote {\bibinfo {title} {{The role of
  non-gaussianities in Primordial Black Hole formation}},}\ }\href {\doibase
  10.1016/j.dark.2019.100275} {\bibfield  {journal} {\bibinfo  {journal} {Phys.
  Dark Univ.}\ }\textbf {\bibinfo {volume} {24}},\ \bibinfo {pages} {100275}
  (\bibinfo {year} {2019})},\ \Eprint {http://arxiv.org/abs/1811.07857}
  {arXiv:1811.07857 [astro-ph.CO]} \BibitemShut {NoStop}%
\bibitem [{\citenamefont {Passaglia}\ \emph {et~al.}(2019)\citenamefont
  {Passaglia}, \citenamefont {Hu},\ and\ \citenamefont
  {Motohashi}}]{Passaglia:2018ixg}%
  \BibitemOpen
  \bibfield  {author} {\bibinfo {author} {\bibfnamefont {Samuel}\ \bibnamefont
  {Passaglia}}, \bibinfo {author} {\bibfnamefont {Wayne}\ \bibnamefont {Hu}}, \
  and\ \bibinfo {author} {\bibfnamefont {Hayato}\ \bibnamefont {Motohashi}},\
  }\bibfield  {title} {\enquote {\bibinfo {title} {{Primordial black holes and
  local non-Gaussianity in canonical inflation}},}\ }\href {\doibase
  10.1103/PhysRevD.99.043536} {\bibfield  {journal} {\bibinfo  {journal} {Phys.
  Rev. D}\ }\textbf {\bibinfo {volume} {99}},\ \bibinfo {pages} {043536}
  (\bibinfo {year} {2019})},\ \Eprint {http://arxiv.org/abs/1812.08243}
  {arXiv:1812.08243 [astro-ph.CO]} \BibitemShut {NoStop}%
\bibitem [{\citenamefont {Atal}\ \emph {et~al.}(2019)\citenamefont {Atal},
  \citenamefont {Garriga},\ and\ \citenamefont
  {Marcos-Caballero}}]{Atal:2019cdz}%
  \BibitemOpen
  \bibfield  {author} {\bibinfo {author} {\bibfnamefont {Vicente}\ \bibnamefont
  {Atal}}, \bibinfo {author} {\bibfnamefont {Jaume}\ \bibnamefont {Garriga}}, \
  and\ \bibinfo {author} {\bibfnamefont {Airam}\ \bibnamefont
  {Marcos-Caballero}},\ }\bibfield  {title} {\enquote {\bibinfo {title}
  {{Primordial black hole formation with non-Gaussian curvature
  perturbations}},}\ }\href {\doibase 10.1088/1475-7516/2019/09/073} {\bibfield
   {journal} {\bibinfo  {journal} {JCAP}\ }\textbf {\bibinfo {volume} {09}},\
  \bibinfo {pages} {073} (\bibinfo {year} {2019})},\ \Eprint
  {http://arxiv.org/abs/1905.13202} {arXiv:1905.13202 [astro-ph.CO]}
  \BibitemShut {NoStop}%
\bibitem [{\citenamefont {Atal}\ \emph {et~al.}(2020)\citenamefont {Atal},
  \citenamefont {Cid}, \citenamefont {Escriv\`a},\ and\ \citenamefont
  {Garriga}}]{Atal:2019erb}%
  \BibitemOpen
  \bibfield  {author} {\bibinfo {author} {\bibfnamefont {Vicente}\ \bibnamefont
  {Atal}}, \bibinfo {author} {\bibfnamefont {Judith}\ \bibnamefont {Cid}},
  \bibinfo {author} {\bibfnamefont {Albert}\ \bibnamefont {Escriv\`a}}, \ and\
  \bibinfo {author} {\bibfnamefont {Jaume}\ \bibnamefont {Garriga}},\
  }\bibfield  {title} {\enquote {\bibinfo {title} {{PBH in single field
  inflation: the effect of shape dispersion and non-Gaussianities}},}\ }\href
  {\doibase 10.1088/1475-7516/2020/05/022} {\bibfield  {journal} {\bibinfo
  {journal} {JCAP}\ }\textbf {\bibinfo {volume} {05}},\ \bibinfo {pages} {022}
  (\bibinfo {year} {2020})},\ \Eprint {http://arxiv.org/abs/1908.11357}
  {arXiv:1908.11357 [astro-ph.CO]} \BibitemShut {NoStop}%
\bibitem [{\citenamefont {Taoso}\ and\ \citenamefont
  {Urbano}(2021)}]{Taoso:2021uvl}%
  \BibitemOpen
  \bibfield  {author} {\bibinfo {author} {\bibfnamefont {Marco}\ \bibnamefont
  {Taoso}}\ and\ \bibinfo {author} {\bibfnamefont {Alfredo}\ \bibnamefont
  {Urbano}},\ }\bibfield  {title} {\enquote {\bibinfo {title}
  {{Non-gaussianities for primordial black hole formation}},}\ }\href {\doibase
  10.1088/1475-7516/2021/08/016} {\bibfield  {journal} {\bibinfo  {journal}
  {JCAP}\ }\textbf {\bibinfo {volume} {08}},\ \bibinfo {pages} {016} (\bibinfo
  {year} {2021})},\ \Eprint {http://arxiv.org/abs/2102.03610} {arXiv:2102.03610
  [astro-ph.CO]} \BibitemShut {NoStop}%
\bibitem [{\citenamefont {Biagetti}\ \emph {et~al.}(2021)\citenamefont
  {Biagetti}, \citenamefont {De~Luca}, \citenamefont {Franciolini},
  \citenamefont {Kehagias},\ and\ \citenamefont {Riotto}}]{Biagetti:2021eep}%
  \BibitemOpen
  \bibfield  {author} {\bibinfo {author} {\bibfnamefont {Matteo}\ \bibnamefont
  {Biagetti}}, \bibinfo {author} {\bibfnamefont {Valerio}\ \bibnamefont
  {De~Luca}}, \bibinfo {author} {\bibfnamefont {Gabriele}\ \bibnamefont
  {Franciolini}}, \bibinfo {author} {\bibfnamefont {Alex}\ \bibnamefont
  {Kehagias}}, \ and\ \bibinfo {author} {\bibfnamefont {Antonio}\ \bibnamefont
  {Riotto}},\ }\bibfield  {title} {\enquote {\bibinfo {title} {{The formation
  probability of primordial black holes}},}\ }\href {\doibase
  10.1016/j.physletb.2021.136602} {\bibfield  {journal} {\bibinfo  {journal}
  {Phys. Lett. B}\ }\textbf {\bibinfo {volume} {820}},\ \bibinfo {pages}
  {136602} (\bibinfo {year} {2021})},\ \Eprint
  {http://arxiv.org/abs/2105.07810} {arXiv:2105.07810 [astro-ph.CO]}
  \BibitemShut {NoStop}%
\bibitem [{\citenamefont {Davies}\ \emph {et~al.}(2022)\citenamefont {Davies},
  \citenamefont {Carrilho},\ and\ \citenamefont {Mulryne}}]{Davies:2021loj}%
  \BibitemOpen
  \bibfield  {author} {\bibinfo {author} {\bibfnamefont {Matthew~W.}\
  \bibnamefont {Davies}}, \bibinfo {author} {\bibfnamefont {Pedro}\
  \bibnamefont {Carrilho}}, \ and\ \bibinfo {author} {\bibfnamefont {David~J.}\
  \bibnamefont {Mulryne}},\ }\bibfield  {title} {\enquote {\bibinfo {title}
  {{Non-Gaussianity in inflationary scenarios for primordial black holes}},}\
  }\href {\doibase 10.1088/1475-7516/2022/06/019} {\bibfield  {journal}
  {\bibinfo  {journal} {JCAP}\ }\textbf {\bibinfo {volume} {06}},\ \bibinfo
  {pages} {019} (\bibinfo {year} {2022})},\ \Eprint
  {http://arxiv.org/abs/2110.08189} {arXiv:2110.08189 [astro-ph.CO]}
  \BibitemShut {NoStop}%
\bibitem [{\citenamefont {Abbott}\ \emph {et~al.}(2016)\citenamefont {Abbott}
  \emph {et~al.}}]{LIGOScientific:2016dsl}%
  \BibitemOpen
  \bibfield  {author} {\bibinfo {author} {\bibfnamefont {B.~P.}\ \bibnamefont
  {Abbott}} \emph {et~al.} (\bibinfo {collaboration} {LIGO Scientific,
  Virgo}),\ }\bibfield  {title} {\enquote {\bibinfo {title} {{Binary Black Hole
  Mergers in the first Advanced LIGO Observing Run}},}\ }\href {\doibase
  10.1103/PhysRevX.6.041015} {\bibfield  {journal} {\bibinfo  {journal} {Phys.
  Rev. X}\ }\textbf {\bibinfo {volume} {6}},\ \bibinfo {pages} {041015}
  (\bibinfo {year} {2016})},\ \bibinfo {note} {[Erratum: Phys.Rev.X 8, 039903
  (2018)]},\ \Eprint {http://arxiv.org/abs/1606.04856} {arXiv:1606.04856
  [gr-qc]} \BibitemShut {NoStop}%
\bibitem [{\citenamefont {Abbott}\ \emph {et~al.}(2021)\citenamefont {Abbott}
  \emph {et~al.}}]{LIGOScientific:2021djp}%
  \BibitemOpen
  \bibfield  {author} {\bibinfo {author} {\bibfnamefont {R.}~\bibnamefont
  {Abbott}} \emph {et~al.} (\bibinfo {collaboration} {LIGO Scientific, VIRGO,
  KAGRA}),\ }\bibfield  {title} {\enquote {\bibinfo {title} {{GWTC-3: Compact
  Binary Coalescences Observed by LIGO and Virgo During the Second Part of the
  Third Observing Run}},}\ }\href@noop {} {\  (\bibinfo {year} {2021})},\
  \Eprint {http://arxiv.org/abs/2111.03606} {arXiv:2111.03606 [gr-qc]}
  \BibitemShut {NoStop}%
\bibitem [{\citenamefont {Sasaki}\ \emph {et~al.}(2016)\citenamefont {Sasaki},
  \citenamefont {Suyama}, \citenamefont {Tanaka},\ and\ \citenamefont
  {Yokoyama}}]{Sasaki:2016jop}%
  \BibitemOpen
  \bibfield  {author} {\bibinfo {author} {\bibfnamefont {Misao}\ \bibnamefont
  {Sasaki}}, \bibinfo {author} {\bibfnamefont {Teruaki}\ \bibnamefont
  {Suyama}}, \bibinfo {author} {\bibfnamefont {Takahiro}\ \bibnamefont
  {Tanaka}}, \ and\ \bibinfo {author} {\bibfnamefont {Shuichiro}\ \bibnamefont
  {Yokoyama}},\ }\bibfield  {title} {\enquote {\bibinfo {title} {{Primordial
  Black Hole Scenario for the Gravitational-Wave Event GW150914}},}\ }\href
  {\doibase 10.1103/PhysRevLett.121.059901, 10.1103/PhysRevLett.117.061101}
  {\bibfield  {journal} {\bibinfo  {journal} {Phys. Rev. Lett.}\ }\textbf
  {\bibinfo {volume} {117}},\ \bibinfo {pages} {061101} (\bibinfo {year}
  {2016})},\ \bibinfo {note} {[erratum: Phys. Rev.
  Lett.121,no.5,059901(2018)]},\ \Eprint {http://arxiv.org/abs/1603.08338}
  {arXiv:1603.08338 [astro-ph.CO]} \BibitemShut {NoStop}%
\bibitem [{\citenamefont {Chen}\ and\ \citenamefont
  {Huang}(2018)}]{Chen:2018czv}%
  \BibitemOpen
  \bibfield  {author} {\bibinfo {author} {\bibfnamefont {Zu-Cheng}\
  \bibnamefont {Chen}}\ and\ \bibinfo {author} {\bibfnamefont {Qing-Guo}\
  \bibnamefont {Huang}},\ }\bibfield  {title} {\enquote {\bibinfo {title}
  {{Merger Rate Distribution of Primordial-Black-Hole Binaries}},}\ }\href
  {\doibase 10.3847/1538-4357/aad6e2} {\bibfield  {journal} {\bibinfo
  {journal} {Astrophys. J.}\ }\textbf {\bibinfo {volume} {864}},\ \bibinfo
  {pages} {61} (\bibinfo {year} {2018})},\ \Eprint
  {http://arxiv.org/abs/1801.10327} {arXiv:1801.10327 [astro-ph.CO]}
  \BibitemShut {NoStop}%
\bibitem [{\citenamefont {Raidal}\ \emph {et~al.}(2019)\citenamefont {Raidal},
  \citenamefont {Spethmann}, \citenamefont {Vaskonen},\ and\ \citenamefont
  {Veermäe}}]{Raidal:2018bbj}%
  \BibitemOpen
  \bibfield  {author} {\bibinfo {author} {\bibfnamefont {Martti}\ \bibnamefont
  {Raidal}}, \bibinfo {author} {\bibfnamefont {Christian}\ \bibnamefont
  {Spethmann}}, \bibinfo {author} {\bibfnamefont {Ville}\ \bibnamefont
  {Vaskonen}}, \ and\ \bibinfo {author} {\bibfnamefont {Hardi}\ \bibnamefont
  {Veermäe}},\ }\bibfield  {title} {\enquote {\bibinfo {title} {{Formation and
  Evolution of Primordial Black Hole Binaries in the Early Universe}},}\ }\href
  {\doibase 10.1088/1475-7516/2019/02/018} {\bibfield  {journal} {\bibinfo
  {journal} {JCAP}\ }\textbf {\bibinfo {volume} {1902}},\ \bibinfo {pages}
  {018} (\bibinfo {year} {2019})},\ \Eprint {http://arxiv.org/abs/1812.01930}
  {arXiv:1812.01930 [astro-ph.CO]} \BibitemShut {NoStop}%
\bibitem [{\citenamefont {De~Luca}\ \emph
  {et~al.}(2020{\natexlab{a}})\citenamefont {De~Luca}, \citenamefont
  {Franciolini}, \citenamefont {Pani},\ and\ \citenamefont
  {Riotto}}]{DeLuca:2020qqa}%
  \BibitemOpen
  \bibfield  {author} {\bibinfo {author} {\bibfnamefont {V.}~\bibnamefont
  {De~Luca}}, \bibinfo {author} {\bibfnamefont {G.}~\bibnamefont
  {Franciolini}}, \bibinfo {author} {\bibfnamefont {P.}~\bibnamefont {Pani}}, \
  and\ \bibinfo {author} {\bibfnamefont {A.}~\bibnamefont {Riotto}},\
  }\bibfield  {title} {\enquote {\bibinfo {title} {{Primordial Black Holes
  Confront LIGO/Virgo data: Current situation}},}\ }\href {\doibase
  10.1088/1475-7516/2020/06/044} {\bibfield  {journal} {\bibinfo  {journal}
  {JCAP}\ }\textbf {\bibinfo {volume} {06}},\ \bibinfo {pages} {044} (\bibinfo
  {year} {2020}{\natexlab{a}})},\ \Eprint {http://arxiv.org/abs/2005.05641}
  {arXiv:2005.05641 [astro-ph.CO]} \BibitemShut {NoStop}%
\bibitem [{\citenamefont {Hall}\ \emph {et~al.}(2020)\citenamefont {Hall},
  \citenamefont {Gow},\ and\ \citenamefont {Byrnes}}]{Hall:2020daa}%
  \BibitemOpen
  \bibfield  {author} {\bibinfo {author} {\bibfnamefont {Alex}\ \bibnamefont
  {Hall}}, \bibinfo {author} {\bibfnamefont {Andrew~D.}\ \bibnamefont {Gow}}, \
  and\ \bibinfo {author} {\bibfnamefont {Christian~T.}\ \bibnamefont
  {Byrnes}},\ }\bibfield  {title} {\enquote {\bibinfo {title} {{Bayesian
  analysis of LIGO-Virgo mergers: Primordial vs. astrophysical black hole
  populations}},}\ }\href {\doibase 10.1103/PhysRevD.102.123524} {\bibfield
  {journal} {\bibinfo  {journal} {Phys. Rev. D}\ }\textbf {\bibinfo {volume}
  {102}},\ \bibinfo {pages} {123524} (\bibinfo {year} {2020})},\ \Eprint
  {http://arxiv.org/abs/2008.13704} {arXiv:2008.13704 [astro-ph.CO]}
  \BibitemShut {NoStop}%
\bibitem [{\citenamefont {Bhagwat}\ \emph {et~al.}(2021)\citenamefont
  {Bhagwat}, \citenamefont {De~Luca}, \citenamefont {Franciolini},
  \citenamefont {Pani},\ and\ \citenamefont {Riotto}}]{Bhagwat:2020bzh}%
  \BibitemOpen
  \bibfield  {author} {\bibinfo {author} {\bibfnamefont {S.}~\bibnamefont
  {Bhagwat}}, \bibinfo {author} {\bibfnamefont {V.}~\bibnamefont {De~Luca}},
  \bibinfo {author} {\bibfnamefont {G.}~\bibnamefont {Franciolini}}, \bibinfo
  {author} {\bibfnamefont {P.}~\bibnamefont {Pani}}, \ and\ \bibinfo {author}
  {\bibfnamefont {A.}~\bibnamefont {Riotto}},\ }\bibfield  {title} {\enquote
  {\bibinfo {title} {{The importance of priors on LIGO-Virgo parameter
  estimation: the case of primordial black holes}},}\ }\href {\doibase
  10.1088/1475-7516/2021/01/037} {\bibfield  {journal} {\bibinfo  {journal}
  {JCAP}\ }\textbf {\bibinfo {volume} {01}},\ \bibinfo {pages} {037} (\bibinfo
  {year} {2021})},\ \Eprint {http://arxiv.org/abs/2008.12320} {arXiv:2008.12320
  [astro-ph.CO]} \BibitemShut {NoStop}%
\bibitem [{\citenamefont {H\"utsi}\ \emph {et~al.}(2020)\citenamefont
  {H\"utsi}, \citenamefont {Raidal}, \citenamefont {Vaskonen},\ and\
  \citenamefont {Veerm\"ae}}]{Hutsi:2020sol}%
  \BibitemOpen
  \bibfield  {author} {\bibinfo {author} {\bibfnamefont {Gert}\ \bibnamefont
  {H\"utsi}}, \bibinfo {author} {\bibfnamefont {Martti}\ \bibnamefont
  {Raidal}}, \bibinfo {author} {\bibfnamefont {Ville}\ \bibnamefont
  {Vaskonen}}, \ and\ \bibinfo {author} {\bibfnamefont {Hardi}\ \bibnamefont
  {Veerm\"ae}},\ }\bibfield  {title} {\enquote {\bibinfo {title} {{Two
  populations of LIGO-Virgo black holes}},}\ }\href@noop {} {\  (\bibinfo
  {year} {2020})},\ \Eprint {http://arxiv.org/abs/2012.02786} {arXiv:2012.02786
  [astro-ph.CO]} \BibitemShut {NoStop}%
\bibitem [{\citenamefont {Wong}\ \emph {et~al.}(2021)\citenamefont {Wong},
  \citenamefont {Franciolini}, \citenamefont {De~Luca}, \citenamefont
  {Baibhav}, \citenamefont {Berti}, \citenamefont {Pani},\ and\ \citenamefont
  {Riotto}}]{Wong:2020yig}%
  \BibitemOpen
  \bibfield  {author} {\bibinfo {author} {\bibfnamefont {Kaze W.~K.}\
  \bibnamefont {Wong}}, \bibinfo {author} {\bibfnamefont {Gabriele}\
  \bibnamefont {Franciolini}}, \bibinfo {author} {\bibfnamefont {Valerio}\
  \bibnamefont {De~Luca}}, \bibinfo {author} {\bibfnamefont {Vishal}\
  \bibnamefont {Baibhav}}, \bibinfo {author} {\bibfnamefont {Emanuele}\
  \bibnamefont {Berti}}, \bibinfo {author} {\bibfnamefont {Paolo}\ \bibnamefont
  {Pani}}, \ and\ \bibinfo {author} {\bibfnamefont {Antonio}\ \bibnamefont
  {Riotto}},\ }\bibfield  {title} {\enquote {\bibinfo {title} {{Constraining
  the primordial black hole scenario with Bayesian inference and machine
  learning: the GWTC-2 gravitational wave catalog}},}\ }\href {\doibase
  10.1103/PhysRevD.103.023026} {\bibfield  {journal} {\bibinfo  {journal}
  {Phys. Rev. D}\ }\textbf {\bibinfo {volume} {103}},\ \bibinfo {pages}
  {023026} (\bibinfo {year} {2021})},\ \Eprint
  {http://arxiv.org/abs/2011.01865} {arXiv:2011.01865 [gr-qc]} \BibitemShut
  {NoStop}%
\bibitem [{\citenamefont {De~Luca}\ \emph {et~al.}(2021)\citenamefont
  {De~Luca}, \citenamefont {Franciolini}, \citenamefont {Pani},\ and\
  \citenamefont {Riotto}}]{DeLuca:2021wjr}%
  \BibitemOpen
  \bibfield  {author} {\bibinfo {author} {\bibfnamefont {V.}~\bibnamefont
  {De~Luca}}, \bibinfo {author} {\bibfnamefont {G.}~\bibnamefont
  {Franciolini}}, \bibinfo {author} {\bibfnamefont {P.}~\bibnamefont {Pani}}, \
  and\ \bibinfo {author} {\bibfnamefont {A.}~\bibnamefont {Riotto}},\
  }\bibfield  {title} {\enquote {\bibinfo {title} {{Bayesian Evidence for Both
  Astrophysical and Primordial Black Holes: Mapping the GWTC-2 Catalog to
  Third-Generation Detectors}},}\ }\href {\doibase
  10.1088/1475-7516/2021/05/003} {\bibfield  {journal} {\bibinfo  {journal}
  {JCAP}\ }\textbf {\bibinfo {volume} {05}},\ \bibinfo {pages} {003} (\bibinfo
  {year} {2021})},\ \Eprint {http://arxiv.org/abs/2102.03809} {arXiv:2102.03809
  [astro-ph.CO]} \BibitemShut {NoStop}%
\bibitem [{\citenamefont {Franciolini}\ \emph {et~al.}(2022)\citenamefont
  {Franciolini}, \citenamefont {Baibhav}, \citenamefont {De~Luca},
  \citenamefont {Ng}, \citenamefont {Wong}, \citenamefont {Berti},
  \citenamefont {Pani}, \citenamefont {Riotto},\ and\ \citenamefont
  {Vitale}}]{Franciolini:2021tla}%
  \BibitemOpen
  \bibfield  {author} {\bibinfo {author} {\bibfnamefont {Gabriele}\
  \bibnamefont {Franciolini}}, \bibinfo {author} {\bibfnamefont {Vishal}\
  \bibnamefont {Baibhav}}, \bibinfo {author} {\bibfnamefont {Valerio}\
  \bibnamefont {De~Luca}}, \bibinfo {author} {\bibfnamefont {Ken K.~Y.}\
  \bibnamefont {Ng}}, \bibinfo {author} {\bibfnamefont {Kaze W.~K.}\
  \bibnamefont {Wong}}, \bibinfo {author} {\bibfnamefont {Emanuele}\
  \bibnamefont {Berti}}, \bibinfo {author} {\bibfnamefont {Paolo}\ \bibnamefont
  {Pani}}, \bibinfo {author} {\bibfnamefont {Antonio}\ \bibnamefont {Riotto}},
  \ and\ \bibinfo {author} {\bibfnamefont {Salvatore}\ \bibnamefont {Vitale}},\
  }\bibfield  {title} {\enquote {\bibinfo {title} {{Searching for a
  subpopulation of primordial black holes in LIGO-Virgo gravitational-wave
  data}},}\ }\href {\doibase 10.1103/PhysRevD.105.083526} {\bibfield  {journal}
  {\bibinfo  {journal} {Phys. Rev. D}\ }\textbf {\bibinfo {volume} {105}},\
  \bibinfo {pages} {083526} (\bibinfo {year} {2022})},\ \Eprint
  {http://arxiv.org/abs/2105.03349} {arXiv:2105.03349 [gr-qc]} \BibitemShut
  {NoStop}%
\bibitem [{\citenamefont {Chen}\ \emph {et~al.}(2022)\citenamefont {Chen},
  \citenamefont {Yuan},\ and\ \citenamefont {Huang}}]{Chen:2021nxo}%
  \BibitemOpen
  \bibfield  {author} {\bibinfo {author} {\bibfnamefont {Zu-Cheng}\
  \bibnamefont {Chen}}, \bibinfo {author} {\bibfnamefont {Chen}\ \bibnamefont
  {Yuan}}, \ and\ \bibinfo {author} {\bibfnamefont {Qing-Guo}\ \bibnamefont
  {Huang}},\ }\bibfield  {title} {\enquote {\bibinfo {title} {{Confronting the
  primordial black hole scenario with the gravitational-wave events detected by
  LIGO-Virgo}},}\ }\href {\doibase 10.1016/j.physletb.2022.137040} {\bibfield
  {journal} {\bibinfo  {journal} {Phys. Lett. B}\ }\textbf {\bibinfo {volume}
  {829}},\ \bibinfo {pages} {137040} (\bibinfo {year} {2022})},\ \Eprint
  {http://arxiv.org/abs/2108.11740} {arXiv:2108.11740 [astro-ph.CO]}
  \BibitemShut {NoStop}%
\bibitem [{\citenamefont {Ananda}\ \emph {et~al.}(2007)\citenamefont {Ananda},
  \citenamefont {Clarkson},\ and\ \citenamefont {Wands}}]{Ananda:2006af}%
  \BibitemOpen
  \bibfield  {author} {\bibinfo {author} {\bibfnamefont {Kishore~N.}\
  \bibnamefont {Ananda}}, \bibinfo {author} {\bibfnamefont {Chris}\
  \bibnamefont {Clarkson}}, \ and\ \bibinfo {author} {\bibfnamefont {David}\
  \bibnamefont {Wands}},\ }\bibfield  {title} {\enquote {\bibinfo {title} {{The
  Cosmological gravitational wave background from primordial density
  perturbations}},}\ }\href {\doibase 10.1103/PhysRevD.75.123518} {\bibfield
  {journal} {\bibinfo  {journal} {Phys. Rev. D}\ }\textbf {\bibinfo {volume}
  {75}},\ \bibinfo {pages} {123518} (\bibinfo {year} {2007})},\ \Eprint
  {http://arxiv.org/abs/gr-qc/0612013} {arXiv:gr-qc/0612013} \BibitemShut
  {NoStop}%
\bibitem [{\citenamefont {Baumann}\ \emph {et~al.}(2007)\citenamefont
  {Baumann}, \citenamefont {Steinhardt}, \citenamefont {Takahashi},\ and\
  \citenamefont {Ichiki}}]{Baumann:2007zm}%
  \BibitemOpen
  \bibfield  {author} {\bibinfo {author} {\bibfnamefont {Daniel}\ \bibnamefont
  {Baumann}}, \bibinfo {author} {\bibfnamefont {Paul~J.}\ \bibnamefont
  {Steinhardt}}, \bibinfo {author} {\bibfnamefont {Keitaro}\ \bibnamefont
  {Takahashi}}, \ and\ \bibinfo {author} {\bibfnamefont {Kiyotomo}\
  \bibnamefont {Ichiki}},\ }\bibfield  {title} {\enquote {\bibinfo {title}
  {{Gravitational Wave Spectrum Induced by Primordial Scalar Perturbations}},}\
  }\href {\doibase 10.1103/PhysRevD.76.084019} {\bibfield  {journal} {\bibinfo
  {journal} {Phys. Rev. D}\ }\textbf {\bibinfo {volume} {76}},\ \bibinfo
  {pages} {084019} (\bibinfo {year} {2007})},\ \Eprint
  {http://arxiv.org/abs/hep-th/0703290} {arXiv:hep-th/0703290} \BibitemShut
  {NoStop}%
\bibitem [{\citenamefont {Saito}\ and\ \citenamefont
  {Yokoyama}(2009)}]{Saito:2008jc}%
  \BibitemOpen
  \bibfield  {author} {\bibinfo {author} {\bibfnamefont {Ryo}\ \bibnamefont
  {Saito}}\ and\ \bibinfo {author} {\bibfnamefont {Jun'ichi}\ \bibnamefont
  {Yokoyama}},\ }\bibfield  {title} {\enquote {\bibinfo {title} {{Gravitational
  wave background as a probe of the primordial black hole abundance}},}\ }\href
  {\doibase 10.1103/PhysRevLett.102.161101, 10.1103/PhysRevLett.107.069901}
  {\bibfield  {journal} {\bibinfo  {journal} {Phys. Rev. Lett.}\ }\textbf
  {\bibinfo {volume} {102}},\ \bibinfo {pages} {161101} (\bibinfo {year}
  {2009})},\ \bibinfo {note} {[Erratum: Phys. Rev. Lett.107,069901(2011)]},\
  \Eprint {http://arxiv.org/abs/0812.4339} {arXiv:0812.4339 [astro-ph]}
  \BibitemShut {NoStop}%
\bibitem [{\citenamefont {Arroja}\ \emph {et~al.}(2009)\citenamefont {Arroja},
  \citenamefont {Assadullahi}, \citenamefont {Koyama},\ and\ \citenamefont
  {Wands}}]{Arroja:2009sh}%
  \BibitemOpen
  \bibfield  {author} {\bibinfo {author} {\bibfnamefont {Frederico}\
  \bibnamefont {Arroja}}, \bibinfo {author} {\bibfnamefont {Hooshyar}\
  \bibnamefont {Assadullahi}}, \bibinfo {author} {\bibfnamefont {Kazuya}\
  \bibnamefont {Koyama}}, \ and\ \bibinfo {author} {\bibfnamefont {David}\
  \bibnamefont {Wands}},\ }\bibfield  {title} {\enquote {\bibinfo {title}
  {{Cosmological matching conditions for gravitational waves at second
  order}},}\ }\href {\doibase 10.1103/PhysRevD.80.123526} {\bibfield  {journal}
  {\bibinfo  {journal} {Phys. Rev. D}\ }\textbf {\bibinfo {volume} {80}},\
  \bibinfo {pages} {123526} (\bibinfo {year} {2009})},\ \Eprint
  {http://arxiv.org/abs/0907.3618} {arXiv:0907.3618 [astro-ph.CO]} \BibitemShut
  {NoStop}%
\bibitem [{\citenamefont {Assadullahi}\ and\ \citenamefont
  {Wands}(2010)}]{Assadullahi:2009jc}%
  \BibitemOpen
  \bibfield  {author} {\bibinfo {author} {\bibfnamefont {Hooshyar}\
  \bibnamefont {Assadullahi}}\ and\ \bibinfo {author} {\bibfnamefont {David}\
  \bibnamefont {Wands}},\ }\bibfield  {title} {\enquote {\bibinfo {title}
  {{Constraints on primordial density perturbations from induced gravitational
  waves}},}\ }\href {\doibase 10.1103/PhysRevD.81.023527} {\bibfield  {journal}
  {\bibinfo  {journal} {Phys. Rev.}\ }\textbf {\bibinfo {volume} {D81}},\
  \bibinfo {pages} {023527} (\bibinfo {year} {2010})},\ \Eprint
  {http://arxiv.org/abs/0907.4073} {arXiv:0907.4073 [astro-ph.CO]} \BibitemShut
  {NoStop}%
\bibitem [{\citenamefont {Bugaev}\ and\ \citenamefont
  {Klimai}(2010{\natexlab{a}})}]{Bugaev:2009kq}%
  \BibitemOpen
  \bibfield  {author} {\bibinfo {author} {\bibfnamefont {E.~V.}\ \bibnamefont
  {Bugaev}}\ and\ \bibinfo {author} {\bibfnamefont {P.~A.}\ \bibnamefont
  {Klimai}},\ }\bibfield  {title} {\enquote {\bibinfo {title} {{Bound on
  induced gravitational wave background from primordial black holes}},}\ }\href
  {\doibase 10.1134/S0021364010010017} {\bibfield  {journal} {\bibinfo
  {journal} {JETP Lett.}\ }\textbf {\bibinfo {volume} {91}},\ \bibinfo {pages}
  {1--5} (\bibinfo {year} {2010}{\natexlab{a}})},\ \Eprint
  {http://arxiv.org/abs/0911.0611} {arXiv:0911.0611 [astro-ph.CO]} \BibitemShut
  {NoStop}%
\bibitem [{\citenamefont {Bugaev}\ and\ \citenamefont
  {Klimai}(2010{\natexlab{b}})}]{Bugaev:2009zh}%
  \BibitemOpen
  \bibfield  {author} {\bibinfo {author} {\bibfnamefont {Edgar}\ \bibnamefont
  {Bugaev}}\ and\ \bibinfo {author} {\bibfnamefont {Peter}\ \bibnamefont
  {Klimai}},\ }\bibfield  {title} {\enquote {\bibinfo {title} {{Induced
  gravitational wave background and primordial black holes}},}\ }\href
  {\doibase 10.1103/PhysRevD.81.023517} {\bibfield  {journal} {\bibinfo
  {journal} {Phys. Rev.}\ }\textbf {\bibinfo {volume} {D81}},\ \bibinfo {pages}
  {023517} (\bibinfo {year} {2010}{\natexlab{b}})},\ \Eprint
  {http://arxiv.org/abs/0908.0664} {arXiv:0908.0664 [astro-ph.CO]} \BibitemShut
  {NoStop}%
\bibitem [{\citenamefont {Saito}\ and\ \citenamefont
  {Yokoyama}(2010)}]{Saito:2009jt}%
  \BibitemOpen
  \bibfield  {author} {\bibinfo {author} {\bibfnamefont {Ryo}\ \bibnamefont
  {Saito}}\ and\ \bibinfo {author} {\bibfnamefont {Jun'ichi}\ \bibnamefont
  {Yokoyama}},\ }\bibfield  {title} {\enquote {\bibinfo {title}
  {{Gravitational-Wave Constraints on the Abundance of Primordial Black
  Holes}},}\ }\href {\doibase 10.1143/PTP.126.351, 10.1143/PTP.123.867}
  {\bibfield  {journal} {\bibinfo  {journal} {Prog. Theor. Phys.}\ }\textbf
  {\bibinfo {volume} {123}},\ \bibinfo {pages} {867--886} (\bibinfo {year}
  {2010})},\ \bibinfo {note} {[Erratum: Prog. Theor. Phys.126,351(2011)]},\
  \Eprint {http://arxiv.org/abs/0912.5317} {arXiv:0912.5317 [astro-ph.CO]}
  \BibitemShut {NoStop}%
\bibitem [{\citenamefont {Bugaev}\ and\ \citenamefont
  {Klimai}(2011)}]{Bugaev:2010bb}%
  \BibitemOpen
  \bibfield  {author} {\bibinfo {author} {\bibfnamefont {Edgar}\ \bibnamefont
  {Bugaev}}\ and\ \bibinfo {author} {\bibfnamefont {Peter}\ \bibnamefont
  {Klimai}},\ }\bibfield  {title} {\enquote {\bibinfo {title} {{Constraints on
  the induced gravitational wave background from primordial black holes}},}\
  }\href {\doibase 10.1103/PhysRevD.83.083521} {\bibfield  {journal} {\bibinfo
  {journal} {Phys. Rev.}\ }\textbf {\bibinfo {volume} {D83}},\ \bibinfo {pages}
  {083521} (\bibinfo {year} {2011})},\ \Eprint {http://arxiv.org/abs/1012.4697}
  {arXiv:1012.4697 [astro-ph.CO]} \BibitemShut {NoStop}%
\bibitem [{\citenamefont {Alabidi}\ \emph {et~al.}(2013)\citenamefont
  {Alabidi}, \citenamefont {Kohri}, \citenamefont {Sasaki},\ and\ \citenamefont
  {Sendouda}}]{Alabidi:2013lya}%
  \BibitemOpen
  \bibfield  {author} {\bibinfo {author} {\bibfnamefont {Laila}\ \bibnamefont
  {Alabidi}}, \bibinfo {author} {\bibfnamefont {Kazunori}\ \bibnamefont
  {Kohri}}, \bibinfo {author} {\bibfnamefont {Misao}\ \bibnamefont {Sasaki}}, \
  and\ \bibinfo {author} {\bibfnamefont {Yuuiti}\ \bibnamefont {Sendouda}},\
  }\bibfield  {title} {\enquote {\bibinfo {title} {{Observable induced
  gravitational waves from an early matter phase}},}\ }\href {\doibase
  10.1088/1475-7516/2013/05/033} {\bibfield  {journal} {\bibinfo  {journal}
  {JCAP}\ }\textbf {\bibinfo {volume} {05}},\ \bibinfo {pages} {033} (\bibinfo
  {year} {2013})},\ \Eprint {http://arxiv.org/abs/1303.4519} {arXiv:1303.4519
  [astro-ph.CO]} \BibitemShut {NoStop}%
\bibitem [{\citenamefont {Nakama}\ and\ \citenamefont
  {Suyama}(2016)}]{Nakama:2016enz}%
  \BibitemOpen
  \bibfield  {author} {\bibinfo {author} {\bibfnamefont {Tomohiro}\
  \bibnamefont {Nakama}}\ and\ \bibinfo {author} {\bibfnamefont {Teruaki}\
  \bibnamefont {Suyama}},\ }\bibfield  {title} {\enquote {\bibinfo {title}
  {{Primordial black holes as a novel probe of primordial gravitational waves.
  II: Detailed analysis}},}\ }\href {\doibase 10.1103/PhysRevD.94.043507}
  {\bibfield  {journal} {\bibinfo  {journal} {Phys. Rev.}\ }\textbf {\bibinfo
  {volume} {D94}},\ \bibinfo {pages} {043507} (\bibinfo {year} {2016})},\
  \Eprint {http://arxiv.org/abs/1605.04482} {arXiv:1605.04482 [gr-qc]}
  \BibitemShut {NoStop}%
\bibitem [{\citenamefont {Nakama}\ \emph {et~al.}(2017)\citenamefont {Nakama},
  \citenamefont {Silk},\ and\ \citenamefont {Kamionkowski}}]{Nakama:2016gzw}%
  \BibitemOpen
  \bibfield  {author} {\bibinfo {author} {\bibfnamefont {Tomohiro}\
  \bibnamefont {Nakama}}, \bibinfo {author} {\bibfnamefont {Joseph}\
  \bibnamefont {Silk}}, \ and\ \bibinfo {author} {\bibfnamefont {Marc}\
  \bibnamefont {Kamionkowski}},\ }\bibfield  {title} {\enquote {\bibinfo
  {title} {{Stochastic gravitational waves associated with the formation of
  primordial black holes}},}\ }\href {\doibase 10.1103/PhysRevD.95.043511}
  {\bibfield  {journal} {\bibinfo  {journal} {Phys. Rev.}\ }\textbf {\bibinfo
  {volume} {D95}},\ \bibinfo {pages} {043511} (\bibinfo {year} {2017})},\
  \Eprint {http://arxiv.org/abs/1612.06264} {arXiv:1612.06264 [astro-ph.CO]}
  \BibitemShut {NoStop}%
\bibitem [{\citenamefont {Inomata}\ \emph {et~al.}(2017)\citenamefont
  {Inomata}, \citenamefont {Kawasaki}, \citenamefont {Mukaida}, \citenamefont
  {Tada},\ and\ \citenamefont {Yanagida}}]{Inomata:2016rbd}%
  \BibitemOpen
  \bibfield  {author} {\bibinfo {author} {\bibfnamefont {Keisuke}\ \bibnamefont
  {Inomata}}, \bibinfo {author} {\bibfnamefont {Masahiro}\ \bibnamefont
  {Kawasaki}}, \bibinfo {author} {\bibfnamefont {Kyohei}\ \bibnamefont
  {Mukaida}}, \bibinfo {author} {\bibfnamefont {Yuichiro}\ \bibnamefont
  {Tada}}, \ and\ \bibinfo {author} {\bibfnamefont {Tsutomu~T.}\ \bibnamefont
  {Yanagida}},\ }\bibfield  {title} {\enquote {\bibinfo {title} {{Inflationary
  primordial black holes for the LIGO gravitational wave events and pulsar
  timing array experiments}},}\ }\href {\doibase 10.1103/PhysRevD.95.123510}
  {\bibfield  {journal} {\bibinfo  {journal} {Phys. Rev. D}\ }\textbf {\bibinfo
  {volume} {95}},\ \bibinfo {pages} {123510} (\bibinfo {year} {2017})},\
  \Eprint {http://arxiv.org/abs/1611.06130} {arXiv:1611.06130 [astro-ph.CO]}
  \BibitemShut {NoStop}%
\bibitem [{\citenamefont {Orlofsky}\ \emph {et~al.}(2017)\citenamefont
  {Orlofsky}, \citenamefont {Pierce},\ and\ \citenamefont
  {Wells}}]{Orlofsky:2016vbd}%
  \BibitemOpen
  \bibfield  {author} {\bibinfo {author} {\bibfnamefont {Nicholas}\
  \bibnamefont {Orlofsky}}, \bibinfo {author} {\bibfnamefont {Aaron}\
  \bibnamefont {Pierce}}, \ and\ \bibinfo {author} {\bibfnamefont {James~D.}\
  \bibnamefont {Wells}},\ }\bibfield  {title} {\enquote {\bibinfo {title}
  {{Inflationary theory and pulsar timing investigations of primordial black
  holes and gravitational waves}},}\ }\href {\doibase
  10.1103/PhysRevD.95.063518} {\bibfield  {journal} {\bibinfo  {journal} {Phys.
  Rev. D}\ }\textbf {\bibinfo {volume} {95}},\ \bibinfo {pages} {063518}
  (\bibinfo {year} {2017})},\ \Eprint {http://arxiv.org/abs/1612.05279}
  {arXiv:1612.05279 [astro-ph.CO]} \BibitemShut {NoStop}%
\bibitem [{\citenamefont {Garcia-Bellido}\ \emph {et~al.}(2017)\citenamefont
  {Garcia-Bellido}, \citenamefont {Peloso},\ and\ \citenamefont
  {Unal}}]{Garcia-Bellido:2017aan}%
  \BibitemOpen
  \bibfield  {author} {\bibinfo {author} {\bibfnamefont {Juan}\ \bibnamefont
  {Garcia-Bellido}}, \bibinfo {author} {\bibfnamefont {Marco}\ \bibnamefont
  {Peloso}}, \ and\ \bibinfo {author} {\bibfnamefont {Caner}\ \bibnamefont
  {Unal}},\ }\bibfield  {title} {\enquote {\bibinfo {title} {{Gravitational
  Wave signatures of inflationary models from Primordial Black Hole Dark
  Matter}},}\ }\href {\doibase 10.1088/1475-7516/2017/09/013} {\bibfield
  {journal} {\bibinfo  {journal} {JCAP}\ }\textbf {\bibinfo {volume} {09}},\
  \bibinfo {pages} {013} (\bibinfo {year} {2017})},\ \Eprint
  {http://arxiv.org/abs/1707.02441} {arXiv:1707.02441 [astro-ph.CO]}
  \BibitemShut {NoStop}%
\bibitem [{\citenamefont {Sasaki}\ \emph {et~al.}(2018)\citenamefont {Sasaki},
  \citenamefont {Suyama}, \citenamefont {Tanaka},\ and\ \citenamefont
  {Yokoyama}}]{Sasaki:2018dmp}%
  \BibitemOpen
  \bibfield  {author} {\bibinfo {author} {\bibfnamefont {Misao}\ \bibnamefont
  {Sasaki}}, \bibinfo {author} {\bibfnamefont {Teruaki}\ \bibnamefont
  {Suyama}}, \bibinfo {author} {\bibfnamefont {Takahiro}\ \bibnamefont
  {Tanaka}}, \ and\ \bibinfo {author} {\bibfnamefont {Shuichiro}\ \bibnamefont
  {Yokoyama}},\ }\bibfield  {title} {\enquote {\bibinfo {title} {{Primordial
  black holes—perspectives in gravitational wave astronomy}},}\ }\href
  {\doibase 10.1088/1361-6382/aaa7b4} {\bibfield  {journal} {\bibinfo
  {journal} {Class. Quant. Grav.}\ }\textbf {\bibinfo {volume} {35}},\ \bibinfo
  {pages} {063001} (\bibinfo {year} {2018})},\ \Eprint
  {http://arxiv.org/abs/1801.05235} {arXiv:1801.05235 [astro-ph.CO]}
  \BibitemShut {NoStop}%
\bibitem [{\citenamefont {Espinosa}\ \emph {et~al.}(2018)\citenamefont
  {Espinosa}, \citenamefont {Racco},\ and\ \citenamefont
  {Riotto}}]{Espinosa:2018eve}%
  \BibitemOpen
  \bibfield  {author} {\bibinfo {author} {\bibfnamefont {José~Ramón}\
  \bibnamefont {Espinosa}}, \bibinfo {author} {\bibfnamefont {Davide}\
  \bibnamefont {Racco}}, \ and\ \bibinfo {author} {\bibfnamefont {Antonio}\
  \bibnamefont {Riotto}},\ }\bibfield  {title} {\enquote {\bibinfo {title} {{A
  Cosmological Signature of the SM Higgs Instability: Gravitational Waves}},}\
  }\href {\doibase 10.1088/1475-7516/2018/09/012} {\bibfield  {journal}
  {\bibinfo  {journal} {JCAP}\ }\textbf {\bibinfo {volume} {1809}},\ \bibinfo
  {pages} {012} (\bibinfo {year} {2018})},\ \Eprint
  {http://arxiv.org/abs/1804.07732} {arXiv:1804.07732 [hep-ph]} \BibitemShut
  {NoStop}%
\bibitem [{\citenamefont {Kohri}\ and\ \citenamefont
  {Terada}(2018)}]{Kohri:2018awv}%
  \BibitemOpen
  \bibfield  {author} {\bibinfo {author} {\bibfnamefont {Kazunori}\
  \bibnamefont {Kohri}}\ and\ \bibinfo {author} {\bibfnamefont {Takahiro}\
  \bibnamefont {Terada}},\ }\bibfield  {title} {\enquote {\bibinfo {title}
  {{Semianalytic calculation of gravitational wave spectrum nonlinearly induced
  from primordial curvature perturbations}},}\ }\href {\doibase
  10.1103/PhysRevD.97.123532} {\bibfield  {journal} {\bibinfo  {journal} {Phys.
  Rev. D}\ }\textbf {\bibinfo {volume} {97}},\ \bibinfo {pages} {123532}
  (\bibinfo {year} {2018})},\ \Eprint {http://arxiv.org/abs/1804.08577}
  {arXiv:1804.08577 [gr-qc]} \BibitemShut {NoStop}%
\bibitem [{\citenamefont {Cai}\ \emph {et~al.}(2019{\natexlab{a}})\citenamefont
  {Cai}, \citenamefont {Pi},\ and\ \citenamefont {Sasaki}}]{Cai:2018dig}%
  \BibitemOpen
  \bibfield  {author} {\bibinfo {author} {\bibfnamefont {Rong-gen}\
  \bibnamefont {Cai}}, \bibinfo {author} {\bibfnamefont {Shi}\ \bibnamefont
  {Pi}}, \ and\ \bibinfo {author} {\bibfnamefont {Misao}\ \bibnamefont
  {Sasaki}},\ }\bibfield  {title} {\enquote {\bibinfo {title} {{Gravitational
  Waves Induced by non-Gaussian Scalar Perturbations}},}\ }\href {\doibase
  10.1103/PhysRevLett.122.201101} {\bibfield  {journal} {\bibinfo  {journal}
  {Phys. Rev. Lett.}\ }\textbf {\bibinfo {volume} {122}},\ \bibinfo {pages}
  {201101} (\bibinfo {year} {2019}{\natexlab{a}})},\ \Eprint
  {http://arxiv.org/abs/1810.11000} {arXiv:1810.11000 [astro-ph.CO]}
  \BibitemShut {NoStop}%
\bibitem [{\citenamefont {Bartolo}\ \emph
  {et~al.}(2019{\natexlab{a}})\citenamefont {Bartolo}, \citenamefont {De~Luca},
  \citenamefont {Franciolini}, \citenamefont {Lewis}, \citenamefont {Peloso},\
  and\ \citenamefont {Riotto}}]{Bartolo:2018evs}%
  \BibitemOpen
  \bibfield  {author} {\bibinfo {author} {\bibfnamefont {N.}~\bibnamefont
  {Bartolo}}, \bibinfo {author} {\bibfnamefont {V.}~\bibnamefont {De~Luca}},
  \bibinfo {author} {\bibfnamefont {G.}~\bibnamefont {Franciolini}}, \bibinfo
  {author} {\bibfnamefont {A.}~\bibnamefont {Lewis}}, \bibinfo {author}
  {\bibfnamefont {M.}~\bibnamefont {Peloso}}, \ and\ \bibinfo {author}
  {\bibfnamefont {A.}~\bibnamefont {Riotto}},\ }\bibfield  {title} {\enquote
  {\bibinfo {title} {{Primordial Black Hole Dark Matter: LISA Serendipity}},}\
  }\href {\doibase 10.1103/PhysRevLett.122.211301} {\bibfield  {journal}
  {\bibinfo  {journal} {Phys. Rev. Lett.}\ }\textbf {\bibinfo {volume} {122}},\
  \bibinfo {pages} {211301} (\bibinfo {year} {2019}{\natexlab{a}})},\ \Eprint
  {http://arxiv.org/abs/1810.12218} {arXiv:1810.12218 [astro-ph.CO]}
  \BibitemShut {NoStop}%
\bibitem [{\citenamefont {Bartolo}\ \emph
  {et~al.}(2019{\natexlab{b}})\citenamefont {Bartolo}, \citenamefont {De~Luca},
  \citenamefont {Franciolini}, \citenamefont {Peloso}, \citenamefont {Racco},\
  and\ \citenamefont {Riotto}}]{Bartolo:2018rku}%
  \BibitemOpen
  \bibfield  {author} {\bibinfo {author} {\bibfnamefont {N.}~\bibnamefont
  {Bartolo}}, \bibinfo {author} {\bibfnamefont {V.}~\bibnamefont {De~Luca}},
  \bibinfo {author} {\bibfnamefont {G.}~\bibnamefont {Franciolini}}, \bibinfo
  {author} {\bibfnamefont {M.}~\bibnamefont {Peloso}}, \bibinfo {author}
  {\bibfnamefont {D.}~\bibnamefont {Racco}}, \ and\ \bibinfo {author}
  {\bibfnamefont {A.}~\bibnamefont {Riotto}},\ }\bibfield  {title} {\enquote
  {\bibinfo {title} {{Testing primordial black holes as dark matter with
  LISA}},}\ }\href {\doibase 10.1103/PhysRevD.99.103521} {\bibfield  {journal}
  {\bibinfo  {journal} {Phys. Rev.}\ }\textbf {\bibinfo {volume} {D99}},\
  \bibinfo {pages} {103521} (\bibinfo {year} {2019}{\natexlab{b}})},\ \Eprint
  {http://arxiv.org/abs/1810.12224} {arXiv:1810.12224 [astro-ph.CO]}
  \BibitemShut {NoStop}%
\bibitem [{\citenamefont {Unal}(2019)}]{Unal:2018yaa}%
  \BibitemOpen
  \bibfield  {author} {\bibinfo {author} {\bibfnamefont {Caner}\ \bibnamefont
  {Unal}},\ }\bibfield  {title} {\enquote {\bibinfo {title} {{Imprints of
  Primordial Non-Gaussianity on Gravitational Wave Spectrum}},}\ }\href
  {\doibase 10.1103/PhysRevD.99.041301} {\bibfield  {journal} {\bibinfo
  {journal} {Phys. Rev. D}\ }\textbf {\bibinfo {volume} {99}},\ \bibinfo
  {pages} {041301} (\bibinfo {year} {2019})},\ \Eprint
  {http://arxiv.org/abs/1811.09151} {arXiv:1811.09151 [astro-ph.CO]}
  \BibitemShut {NoStop}%
\bibitem [{\citenamefont {Byrnes}\ \emph {et~al.}(2019)\citenamefont {Byrnes},
  \citenamefont {Cole},\ and\ \citenamefont {Patil}}]{Byrnes:2018txb}%
  \BibitemOpen
  \bibfield  {author} {\bibinfo {author} {\bibfnamefont {Christian~T.}\
  \bibnamefont {Byrnes}}, \bibinfo {author} {\bibfnamefont {Philippa~S.}\
  \bibnamefont {Cole}}, \ and\ \bibinfo {author} {\bibfnamefont {Subodh~P.}\
  \bibnamefont {Patil}},\ }\bibfield  {title} {\enquote {\bibinfo {title}
  {{Steepest growth of the power spectrum and primordial black holes}},}\
  }\href {\doibase 10.1088/1475-7516/2019/06/028} {\bibfield  {journal}
  {\bibinfo  {journal} {JCAP}\ }\textbf {\bibinfo {volume} {06}},\ \bibinfo
  {pages} {028} (\bibinfo {year} {2019})},\ \Eprint
  {http://arxiv.org/abs/1811.11158} {arXiv:1811.11158 [astro-ph.CO]}
  \BibitemShut {NoStop}%
\bibitem [{\citenamefont {Inomata}\ and\ \citenamefont
  {Nakama}(2019)}]{Inomata:2018epa}%
  \BibitemOpen
  \bibfield  {author} {\bibinfo {author} {\bibfnamefont {Keisuke}\ \bibnamefont
  {Inomata}}\ and\ \bibinfo {author} {\bibfnamefont {Tomohiro}\ \bibnamefont
  {Nakama}},\ }\bibfield  {title} {\enquote {\bibinfo {title} {{Gravitational
  waves induced by scalar perturbations as probes of the small-scale primordial
  spectrum}},}\ }\href {\doibase 10.1103/PhysRevD.99.043511} {\bibfield
  {journal} {\bibinfo  {journal} {Phys. Rev.}\ }\textbf {\bibinfo {volume}
  {D99}},\ \bibinfo {pages} {043511} (\bibinfo {year} {2019})},\ \Eprint
  {http://arxiv.org/abs/1812.00674} {arXiv:1812.00674 [astro-ph.CO]}
  \BibitemShut {NoStop}%
\bibitem [{\citenamefont {Clesse}\ \emph {et~al.}(2018)\citenamefont {Clesse},
  \citenamefont {García-Bellido},\ and\ \citenamefont
  {Orani}}]{Clesse:2018ogk}%
  \BibitemOpen
  \bibfield  {author} {\bibinfo {author} {\bibfnamefont {Sebastien}\
  \bibnamefont {Clesse}}, \bibinfo {author} {\bibfnamefont {Juan}\ \bibnamefont
  {García-Bellido}}, \ and\ \bibinfo {author} {\bibfnamefont {Stefano}\
  \bibnamefont {Orani}},\ }\bibfield  {title} {\enquote {\bibinfo {title}
  {{Detecting the Stochastic Gravitational Wave Background from Primordial
  Black Hole Formation}},}\ }\href@noop {} {\  (\bibinfo {year} {2018})},\
  \Eprint {http://arxiv.org/abs/1812.11011} {arXiv:1812.11011 [astro-ph.CO]}
  \BibitemShut {NoStop}%
\bibitem [{\citenamefont {Cai}\ \emph {et~al.}(2019{\natexlab{b}})\citenamefont
  {Cai}, \citenamefont {Pi}, \citenamefont {Wang},\ and\ \citenamefont
  {Yang}}]{Cai:2019amo}%
  \BibitemOpen
  \bibfield  {author} {\bibinfo {author} {\bibfnamefont {Rong-Gen}\
  \bibnamefont {Cai}}, \bibinfo {author} {\bibfnamefont {Shi}\ \bibnamefont
  {Pi}}, \bibinfo {author} {\bibfnamefont {Shao-Jiang}\ \bibnamefont {Wang}}, \
  and\ \bibinfo {author} {\bibfnamefont {Xing-Yu}\ \bibnamefont {Yang}},\
  }\bibfield  {title} {\enquote {\bibinfo {title} {{Resonant multiple peaks in
  the induced gravitational waves}},}\ }\href {\doibase
  10.1088/1475-7516/2019/05/013} {\bibfield  {journal} {\bibinfo  {journal}
  {JCAP}\ }\textbf {\bibinfo {volume} {1905}},\ \bibinfo {pages} {013}
  (\bibinfo {year} {2019}{\natexlab{b}})},\ \Eprint
  {http://arxiv.org/abs/1901.10152} {arXiv:1901.10152 [astro-ph.CO]}
  \BibitemShut {NoStop}%
\bibitem [{\citenamefont {Inomata}\ \emph {et~al.}(2019)\citenamefont
  {Inomata}, \citenamefont {Kohri}, \citenamefont {Nakama},\ and\ \citenamefont
  {Terada}}]{Inomata:2019ivs}%
  \BibitemOpen
  \bibfield  {author} {\bibinfo {author} {\bibfnamefont {Keisuke}\ \bibnamefont
  {Inomata}}, \bibinfo {author} {\bibfnamefont {Kazunori}\ \bibnamefont
  {Kohri}}, \bibinfo {author} {\bibfnamefont {Tomohiro}\ \bibnamefont
  {Nakama}}, \ and\ \bibinfo {author} {\bibfnamefont {Takahiro}\ \bibnamefont
  {Terada}},\ }\bibfield  {title} {\enquote {\bibinfo {title} {{Enhancement of
  Gravitational Waves Induced by Scalar Perturbations due to a Sudden
  Transition from an Early Matter Era to the Radiation Era}},}\ }\href
  {\doibase 10.1103/PhysRevD.100.043532} {\bibfield  {journal} {\bibinfo
  {journal} {Phys. Rev.}\ }\textbf {\bibinfo {volume} {D100}},\ \bibinfo
  {pages} {043532} (\bibinfo {year} {2019})},\ \Eprint
  {http://arxiv.org/abs/1904.12879} {arXiv:1904.12879 [astro-ph.CO]}
  \BibitemShut {NoStop}%
\bibitem [{\citenamefont {Cai}\ \emph {et~al.}(2019{\natexlab{c}})\citenamefont
  {Cai}, \citenamefont {Pi}, \citenamefont {Wang},\ and\ \citenamefont
  {Yang}}]{Cai:2019elf}%
  \BibitemOpen
  \bibfield  {author} {\bibinfo {author} {\bibfnamefont {Rong-Gen}\
  \bibnamefont {Cai}}, \bibinfo {author} {\bibfnamefont {Shi}\ \bibnamefont
  {Pi}}, \bibinfo {author} {\bibfnamefont {Shao-Jiang}\ \bibnamefont {Wang}}, \
  and\ \bibinfo {author} {\bibfnamefont {Xing-Yu}\ \bibnamefont {Yang}},\
  }\bibfield  {title} {\enquote {\bibinfo {title} {{Pulsar Timing Array
  Constraints on the Induced Gravitational Waves}},}\ }\href {\doibase
  10.1088/1475-7516/2019/10/059} {\bibfield  {journal} {\bibinfo  {journal}
  {JCAP}\ }\textbf {\bibinfo {volume} {10}},\ \bibinfo {pages} {059} (\bibinfo
  {year} {2019}{\natexlab{c}})},\ \Eprint {http://arxiv.org/abs/1907.06372}
  {arXiv:1907.06372 [astro-ph.CO]} \BibitemShut {NoStop}%
\bibitem [{\citenamefont {Yuan}\ \emph {et~al.}(2019)\citenamefont {Yuan},
  \citenamefont {Chen},\ and\ \citenamefont {Huang}}]{Yuan:2019udt}%
  \BibitemOpen
  \bibfield  {author} {\bibinfo {author} {\bibfnamefont {Chen}\ \bibnamefont
  {Yuan}}, \bibinfo {author} {\bibfnamefont {Zu-Cheng}\ \bibnamefont {Chen}}, \
  and\ \bibinfo {author} {\bibfnamefont {Qing-Guo}\ \bibnamefont {Huang}},\
  }\bibfield  {title} {\enquote {\bibinfo {title} {{Probing
  Primordial-Black-Hole Dark Matter with Scalar Induced Gravitational
  Waves}},}\ }\href {\doibase 10.1103/PhysRevD.100.081301} {\bibfield
  {journal} {\bibinfo  {journal} {Phys. Rev.}\ }\textbf {\bibinfo {volume}
  {D100}},\ \bibinfo {pages} {081301} (\bibinfo {year} {2019})},\ \Eprint
  {http://arxiv.org/abs/1906.11549} {arXiv:1906.11549 [astro-ph.CO]}
  \BibitemShut {NoStop}%
\bibitem [{\citenamefont {Cai}\ \emph {et~al.}(2020)\citenamefont {Cai},
  \citenamefont {Pi},\ and\ \citenamefont {Sasaki}}]{Cai:2019cdl}%
  \BibitemOpen
  \bibfield  {author} {\bibinfo {author} {\bibfnamefont {Rong-Gen}\
  \bibnamefont {Cai}}, \bibinfo {author} {\bibfnamefont {Shi}\ \bibnamefont
  {Pi}}, \ and\ \bibinfo {author} {\bibfnamefont {Misao}\ \bibnamefont
  {Sasaki}},\ }\bibfield  {title} {\enquote {\bibinfo {title} {{Universal
  infrared scaling of gravitational wave background spectra}},}\ }\href
  {\doibase 10.1103/PhysRevD.102.083528} {\bibfield  {journal} {\bibinfo
  {journal} {Phys. Rev. D}\ }\textbf {\bibinfo {volume} {102}},\ \bibinfo
  {pages} {083528} (\bibinfo {year} {2020})},\ \Eprint
  {http://arxiv.org/abs/1909.13728} {arXiv:1909.13728 [astro-ph.CO]}
  \BibitemShut {NoStop}%
\bibitem [{\citenamefont {Lu}\ \emph {et~al.}(2019)\citenamefont {Lu},
  \citenamefont {Gong}, \citenamefont {Yi},\ and\ \citenamefont
  {Zhang}}]{Lu:2019sti}%
  \BibitemOpen
  \bibfield  {author} {\bibinfo {author} {\bibfnamefont {Yizhou}\ \bibnamefont
  {Lu}}, \bibinfo {author} {\bibfnamefont {Yungui}\ \bibnamefont {Gong}},
  \bibinfo {author} {\bibfnamefont {Zhu}\ \bibnamefont {Yi}}, \ and\ \bibinfo
  {author} {\bibfnamefont {Fengge}\ \bibnamefont {Zhang}},\ }\bibfield  {title}
  {\enquote {\bibinfo {title} {{Constraints on primordial curvature
  perturbations from primordial black hole dark matter and secondary
  gravitational waves}},}\ }\href {\doibase 10.1088/1475-7516/2019/12/031}
  {\bibfield  {journal} {\bibinfo  {journal} {JCAP}\ }\textbf {\bibinfo
  {volume} {12}},\ \bibinfo {pages} {031} (\bibinfo {year} {2019})},\ \Eprint
  {http://arxiv.org/abs/1907.11896} {arXiv:1907.11896 [gr-qc]} \BibitemShut
  {NoStop}%
\bibitem [{\citenamefont {Yuan}\ \emph
  {et~al.}(2020{\natexlab{a}})\citenamefont {Yuan}, \citenamefont {Chen},\ and\
  \citenamefont {Huang}}]{Yuan:2019wwo}%
  \BibitemOpen
  \bibfield  {author} {\bibinfo {author} {\bibfnamefont {Chen}\ \bibnamefont
  {Yuan}}, \bibinfo {author} {\bibfnamefont {Zu-Cheng}\ \bibnamefont {Chen}}, \
  and\ \bibinfo {author} {\bibfnamefont {Qing-Guo}\ \bibnamefont {Huang}},\
  }\bibfield  {title} {\enquote {\bibinfo {title} {{Log-dependent slope of
  scalar induced gravitational waves in the infrared regions}},}\ }\href
  {\doibase 10.1103/PhysRevD.101.043019} {\bibfield  {journal} {\bibinfo
  {journal} {Phys. Rev. D}\ }\textbf {\bibinfo {volume} {101}},\ \bibinfo
  {pages} {043019} (\bibinfo {year} {2020}{\natexlab{a}})},\ \Eprint
  {http://arxiv.org/abs/1910.09099} {arXiv:1910.09099 [astro-ph.CO]}
  \BibitemShut {NoStop}%
\bibitem [{\citenamefont {Tomikawa}\ and\ \citenamefont
  {Kobayashi}(2020)}]{Tomikawa:2019tvi}%
  \BibitemOpen
  \bibfield  {author} {\bibinfo {author} {\bibfnamefont {Keitaro}\ \bibnamefont
  {Tomikawa}}\ and\ \bibinfo {author} {\bibfnamefont {Tsutomu}\ \bibnamefont
  {Kobayashi}},\ }\bibfield  {title} {\enquote {\bibinfo {title} {{Gauge
  dependence of gravitational waves generated at second order from scalar
  perturbations}},}\ }\href {\doibase 10.1103/PhysRevD.101.083529} {\bibfield
  {journal} {\bibinfo  {journal} {Phys. Rev. D}\ }\textbf {\bibinfo {volume}
  {101}},\ \bibinfo {pages} {083529} (\bibinfo {year} {2020})},\ \Eprint
  {http://arxiv.org/abs/1910.01880} {arXiv:1910.01880 [gr-qc]} \BibitemShut
  {NoStop}%
\bibitem [{\citenamefont {De~Luca}\ \emph
  {et~al.}(2020{\natexlab{b}})\citenamefont {De~Luca}, \citenamefont
  {Franciolini}, \citenamefont {Kehagias},\ and\ \citenamefont
  {Riotto}}]{DeLuca:2019ufz}%
  \BibitemOpen
  \bibfield  {author} {\bibinfo {author} {\bibfnamefont {V.}~\bibnamefont
  {De~Luca}}, \bibinfo {author} {\bibfnamefont {G.}~\bibnamefont
  {Franciolini}}, \bibinfo {author} {\bibfnamefont {A.}~\bibnamefont
  {Kehagias}}, \ and\ \bibinfo {author} {\bibfnamefont {A.}~\bibnamefont
  {Riotto}},\ }\bibfield  {title} {\enquote {\bibinfo {title} {{On the Gauge
  Invariance of Cosmological Gravitational Waves}},}\ }\href {\doibase
  10.1088/1475-7516/2020/03/014} {\bibfield  {journal} {\bibinfo  {journal}
  {JCAP}\ }\textbf {\bibinfo {volume} {03}},\ \bibinfo {pages} {014} (\bibinfo
  {year} {2020}{\natexlab{b}})},\ \Eprint {http://arxiv.org/abs/1911.09689}
  {arXiv:1911.09689 [gr-qc]} \BibitemShut {NoStop}%
\bibitem [{\citenamefont {Yuan}\ \emph
  {et~al.}(2020{\natexlab{b}})\citenamefont {Yuan}, \citenamefont {Chen},\ and\
  \citenamefont {Huang}}]{Yuan:2019fwv}%
  \BibitemOpen
  \bibfield  {author} {\bibinfo {author} {\bibfnamefont {Chen}\ \bibnamefont
  {Yuan}}, \bibinfo {author} {\bibfnamefont {Zu-Cheng}\ \bibnamefont {Chen}}, \
  and\ \bibinfo {author} {\bibfnamefont {Qing-Guo}\ \bibnamefont {Huang}},\
  }\bibfield  {title} {\enquote {\bibinfo {title} {{Scalar induced
  gravitational waves in different gauges}},}\ }\href {\doibase
  10.1103/PhysRevD.101.063018} {\bibfield  {journal} {\bibinfo  {journal}
  {Phys. Rev. D}\ }\textbf {\bibinfo {volume} {101}},\ \bibinfo {pages}
  {063018} (\bibinfo {year} {2020}{\natexlab{b}})},\ \Eprint
  {http://arxiv.org/abs/1912.00885} {arXiv:1912.00885 [astro-ph.CO]}
  \BibitemShut {NoStop}%
\bibitem [{\citenamefont {Inomata}\ \emph {et~al.}(2020)\citenamefont
  {Inomata}, \citenamefont {Kawasaki}, \citenamefont {Mukaida}, \citenamefont
  {Terada},\ and\ \citenamefont {Yanagida}}]{Inomata:2020lmk}%
  \BibitemOpen
  \bibfield  {author} {\bibinfo {author} {\bibfnamefont {Keisuke}\ \bibnamefont
  {Inomata}}, \bibinfo {author} {\bibfnamefont {Masahiro}\ \bibnamefont
  {Kawasaki}}, \bibinfo {author} {\bibfnamefont {Kyohei}\ \bibnamefont
  {Mukaida}}, \bibinfo {author} {\bibfnamefont {Takahiro}\ \bibnamefont
  {Terada}}, \ and\ \bibinfo {author} {\bibfnamefont {Tsutomu~T.}\ \bibnamefont
  {Yanagida}},\ }\bibfield  {title} {\enquote {\bibinfo {title} {{Gravitational
  Wave Production right after a Primordial Black Hole Evaporation}},}\ }\href
  {\doibase 10.1103/PhysRevD.101.123533} {\bibfield  {journal} {\bibinfo
  {journal} {Phys. Rev. D}\ }\textbf {\bibinfo {volume} {101}},\ \bibinfo
  {pages} {123533} (\bibinfo {year} {2020})},\ \Eprint
  {http://arxiv.org/abs/2003.10455} {arXiv:2003.10455 [astro-ph.CO]}
  \BibitemShut {NoStop}%
\bibitem [{\citenamefont {Yuan}\ and\ \citenamefont
  {Huang}(2021{\natexlab{a}})}]{Yuan:2020iwf}%
  \BibitemOpen
  \bibfield  {author} {\bibinfo {author} {\bibfnamefont {Chen}\ \bibnamefont
  {Yuan}}\ and\ \bibinfo {author} {\bibfnamefont {Qing-Guo}\ \bibnamefont
  {Huang}},\ }\bibfield  {title} {\enquote {\bibinfo {title} {{Gravitational
  waves induced by the local-type non-Gaussian curvature perturbations}},}\
  }\href {\doibase 10.1016/j.physletb.2021.136606} {\bibfield  {journal}
  {\bibinfo  {journal} {Phys. Lett. B}\ }\textbf {\bibinfo {volume} {821}},\
  \bibinfo {pages} {136606} (\bibinfo {year} {2021}{\natexlab{a}})},\ \Eprint
  {http://arxiv.org/abs/2007.10686} {arXiv:2007.10686 [astro-ph.CO]}
  \BibitemShut {NoStop}%
\bibitem [{\citenamefont {Papanikolaou}\ \emph {et~al.}(2021)\citenamefont
  {Papanikolaou}, \citenamefont {Vennin},\ and\ \citenamefont
  {Langlois}}]{Papanikolaou:2020qtd}%
  \BibitemOpen
  \bibfield  {author} {\bibinfo {author} {\bibfnamefont {Theodoros}\
  \bibnamefont {Papanikolaou}}, \bibinfo {author} {\bibfnamefont {Vincent}\
  \bibnamefont {Vennin}}, \ and\ \bibinfo {author} {\bibfnamefont {David}\
  \bibnamefont {Langlois}},\ }\bibfield  {title} {\enquote {\bibinfo {title}
  {{Gravitational waves from a universe filled with primordial black holes}},}\
  }\href {\doibase 10.1088/1475-7516/2021/03/053} {\bibfield  {journal}
  {\bibinfo  {journal} {JCAP}\ }\textbf {\bibinfo {volume} {03}},\ \bibinfo
  {pages} {053} (\bibinfo {year} {2021})},\ \Eprint
  {http://arxiv.org/abs/2010.11573} {arXiv:2010.11573 [astro-ph.CO]}
  \BibitemShut {NoStop}%
\bibitem [{\citenamefont {Zhang}\ \emph {et~al.}(2020)\citenamefont {Zhang},
  \citenamefont {Ali}, \citenamefont {Gong}, \citenamefont {Lin},\ and\
  \citenamefont {Lu}}]{Zhang:2020ptw}%
  \BibitemOpen
  \bibfield  {author} {\bibinfo {author} {\bibfnamefont {Fengge}\ \bibnamefont
  {Zhang}}, \bibinfo {author} {\bibfnamefont {Arshad}\ \bibnamefont {Ali}},
  \bibinfo {author} {\bibfnamefont {Yungui}\ \bibnamefont {Gong}}, \bibinfo
  {author} {\bibfnamefont {Jiong}\ \bibnamefont {Lin}}, \ and\ \bibinfo
  {author} {\bibfnamefont {Yizhou}\ \bibnamefont {Lu}},\ }\bibfield  {title}
  {\enquote {\bibinfo {title} {{On the waveform of the scalar induced
  gravitational waves}},}\ }\href@noop {} {\  (\bibinfo {year} {2020})},\
  \Eprint {http://arxiv.org/abs/2008.12961} {arXiv:2008.12961 [gr-qc]}
  \BibitemShut {NoStop}%
\bibitem [{\citenamefont {Kapadia}\ \emph {et~al.}(2021)\citenamefont
  {Kapadia}, \citenamefont {Lal~Pandey}, \citenamefont {Suyama}, \citenamefont
  {Kandhasamy},\ and\ \citenamefont {Ajith}}]{Kapadia:2020pnr}%
  \BibitemOpen
  \bibfield  {author} {\bibinfo {author} {\bibfnamefont {Shasvath~J.}\
  \bibnamefont {Kapadia}}, \bibinfo {author} {\bibfnamefont {Kanhaiya}\
  \bibnamefont {Lal~Pandey}}, \bibinfo {author} {\bibfnamefont {Teruaki}\
  \bibnamefont {Suyama}}, \bibinfo {author} {\bibfnamefont {Shivaraj}\
  \bibnamefont {Kandhasamy}}, \ and\ \bibinfo {author} {\bibfnamefont
  {Parameswaran}\ \bibnamefont {Ajith}},\ }\bibfield  {title} {\enquote
  {\bibinfo {title} {{Search for the Stochastic Gravitational-wave Background
  Induced by Primordial Curvature Perturbations in LIGO\textquoteright{}s
  Second Observing Run}},}\ }\href {\doibase 10.3847/2041-8213/abe86e}
  {\bibfield  {journal} {\bibinfo  {journal} {Astrophys. J. Lett.}\ }\textbf
  {\bibinfo {volume} {910}},\ \bibinfo {pages} {L4} (\bibinfo {year} {2021})},\
  \Eprint {http://arxiv.org/abs/2009.05514} {arXiv:2009.05514 [gr-qc]}
  \BibitemShut {NoStop}%
\bibitem [{\citenamefont {Zhang}\ \emph {et~al.}(2021)\citenamefont {Zhang},
  \citenamefont {Gong}, \citenamefont {Lin}, \citenamefont {Lu},\ and\
  \citenamefont {Yi}}]{Zhang:2020uek}%
  \BibitemOpen
  \bibfield  {author} {\bibinfo {author} {\bibfnamefont {Fengge}\ \bibnamefont
  {Zhang}}, \bibinfo {author} {\bibfnamefont {Yungui}\ \bibnamefont {Gong}},
  \bibinfo {author} {\bibfnamefont {Jiong}\ \bibnamefont {Lin}}, \bibinfo
  {author} {\bibfnamefont {Yizhou}\ \bibnamefont {Lu}}, \ and\ \bibinfo
  {author} {\bibfnamefont {Zhu}\ \bibnamefont {Yi}},\ }\bibfield  {title}
  {\enquote {\bibinfo {title} {{Primordial non-Gaussianity from
  G-inflation}},}\ }\href {\doibase 10.1088/1475-7516/2021/04/045} {\bibfield
  {journal} {\bibinfo  {journal} {JCAP}\ }\textbf {\bibinfo {volume} {04}},\
  \bibinfo {pages} {045} (\bibinfo {year} {2021})},\ \Eprint
  {http://arxiv.org/abs/2012.06960} {arXiv:2012.06960 [astro-ph.CO]}
  \BibitemShut {NoStop}%
\bibitem [{\citenamefont {Dom\`enech}\ \emph {et~al.}(2021)\citenamefont
  {Dom\`enech}, \citenamefont {Lin},\ and\ \citenamefont
  {Sasaki}}]{Domenech:2020ssp}%
  \BibitemOpen
  \bibfield  {author} {\bibinfo {author} {\bibfnamefont {Guillem}\ \bibnamefont
  {Dom\`enech}}, \bibinfo {author} {\bibfnamefont {Chunshan}\ \bibnamefont
  {Lin}}, \ and\ \bibinfo {author} {\bibfnamefont {Misao}\ \bibnamefont
  {Sasaki}},\ }\bibfield  {title} {\enquote {\bibinfo {title} {{Gravitational
  wave constraints on the primordial black hole dominated early universe}},}\
  }\href {\doibase 10.1088/1475-7516/2021/11/E01} {\bibfield  {journal}
  {\bibinfo  {journal} {JCAP}\ }\textbf {\bibinfo {volume} {04}},\ \bibinfo
  {pages} {062} (\bibinfo {year} {2021})},\ \bibinfo {note} {[Erratum: JCAP 11,
  E01 (2021)]},\ \Eprint {http://arxiv.org/abs/2012.08151} {arXiv:2012.08151
  [gr-qc]} \BibitemShut {NoStop}%
\bibitem [{\citenamefont {Dalianis}\ and\ \citenamefont
  {Kouvaris}(2021)}]{Dalianis:2020gup}%
  \BibitemOpen
  \bibfield  {author} {\bibinfo {author} {\bibfnamefont {Ioannis}\ \bibnamefont
  {Dalianis}}\ and\ \bibinfo {author} {\bibfnamefont {Chris}\ \bibnamefont
  {Kouvaris}},\ }\bibfield  {title} {\enquote {\bibinfo {title} {{Gravitational
  waves from density perturbations in an early matter domination era}},}\
  }\href {\doibase 10.1088/1475-7516/2021/07/046} {\bibfield  {journal}
  {\bibinfo  {journal} {JCAP}\ }\textbf {\bibinfo {volume} {07}},\ \bibinfo
  {pages} {046} (\bibinfo {year} {2021})},\ \Eprint
  {http://arxiv.org/abs/2012.09255} {arXiv:2012.09255 [astro-ph.CO]}
  \BibitemShut {NoStop}%
\bibitem [{\citenamefont {Atal}\ and\ \citenamefont
  {Dom\`enech}(2021)}]{Atal:2021jyo}%
  \BibitemOpen
  \bibfield  {author} {\bibinfo {author} {\bibfnamefont {Vicente}\ \bibnamefont
  {Atal}}\ and\ \bibinfo {author} {\bibfnamefont {Guillem}\ \bibnamefont
  {Dom\`enech}},\ }\bibfield  {title} {\enquote {\bibinfo {title} {{Probing
  non-Gaussianities with the high frequency tail of induced gravitational
  waves}},}\ }\href {\doibase 10.1088/1475-7516/2021/06/001} {\bibfield
  {journal} {\bibinfo  {journal} {JCAP}\ }\textbf {\bibinfo {volume} {06}},\
  \bibinfo {pages} {001} (\bibinfo {year} {2021})},\ \Eprint
  {http://arxiv.org/abs/2103.01056} {arXiv:2103.01056 [astro-ph.CO]}
  \BibitemShut {NoStop}%
\bibitem [{\citenamefont {Franciolini}(2021)}]{Franciolini:2021nvv}%
  \BibitemOpen
  \bibfield  {author} {\bibinfo {author} {\bibfnamefont {Gabriele}\
  \bibnamefont {Franciolini}},\ }\emph {\bibinfo {title} {{Primordial Black
  Holes: from Theory to Gravitational Wave Observations}}},\ \href {\doibase
  10.13097/archive-ouverte/unige:156136} {Ph.D. thesis},\ \bibinfo  {school}
  {Geneva U., Dept. Theor. Phys.} (\bibinfo {year} {2021}),\ \Eprint
  {http://arxiv.org/abs/2110.06815} {arXiv:2110.06815 [astro-ph.CO]}
  \BibitemShut {NoStop}%
\bibitem [{\citenamefont {Witkowski}\ \emph {et~al.}(2022)\citenamefont
  {Witkowski}, \citenamefont {Dom\`enech}, \citenamefont {Fumagalli},\ and\
  \citenamefont {Renaux-Petel}}]{Witkowski:2021raz}%
  \BibitemOpen
  \bibfield  {author} {\bibinfo {author} {\bibfnamefont {Lukas~T.}\
  \bibnamefont {Witkowski}}, \bibinfo {author} {\bibfnamefont {Guillem}\
  \bibnamefont {Dom\`enech}}, \bibinfo {author} {\bibfnamefont {Jacopo}\
  \bibnamefont {Fumagalli}}, \ and\ \bibinfo {author} {\bibfnamefont
  {S\'ebastien}\ \bibnamefont {Renaux-Petel}},\ }\bibfield  {title} {\enquote
  {\bibinfo {title} {{Expansion history-dependent oscillations in the
  scalar-induced gravitational wave background}},}\ }\href {\doibase
  10.1088/1475-7516/2022/05/028} {\bibfield  {journal} {\bibinfo  {journal}
  {JCAP}\ }\textbf {\bibinfo {volume} {05}},\ \bibinfo {pages} {028} (\bibinfo
  {year} {2022})},\ \Eprint {http://arxiv.org/abs/2110.09480} {arXiv:2110.09480
  [astro-ph.CO]} \BibitemShut {NoStop}%
\bibitem [{\citenamefont {Balaji}\ \emph {et~al.}(2022)\citenamefont {Balaji},
  \citenamefont {Domenech},\ and\ \citenamefont {Silk}}]{Balaji:2022dbi}%
  \BibitemOpen
  \bibfield  {author} {\bibinfo {author} {\bibfnamefont {Shyam}\ \bibnamefont
  {Balaji}}, \bibinfo {author} {\bibfnamefont {Guillem}\ \bibnamefont
  {Domenech}}, \ and\ \bibinfo {author} {\bibfnamefont {Joseph}\ \bibnamefont
  {Silk}},\ }\bibfield  {title} {\enquote {\bibinfo {title} {{Induced
  gravitational waves from slow-roll inflation after an enhancing phase}},}\
  }\href {\doibase 10.1088/1475-7516/2022/09/016} {\bibfield  {journal}
  {\bibinfo  {journal} {JCAP}\ }\textbf {\bibinfo {volume} {09}},\ \bibinfo
  {pages} {016} (\bibinfo {year} {2022})},\ \Eprint
  {http://arxiv.org/abs/2205.01696} {arXiv:2205.01696 [astro-ph.CO]}
  \BibitemShut {NoStop}%
\bibitem [{\citenamefont {Cang}\ \emph {et~al.}(2022)\citenamefont {Cang},
  \citenamefont {Ma},\ and\ \citenamefont {Gao}}]{Cang:2022oia}%
  \BibitemOpen
  \bibfield  {author} {\bibinfo {author} {\bibfnamefont {Junsong}\ \bibnamefont
  {Cang}}, \bibinfo {author} {\bibfnamefont {Yin-Zhe}\ \bibnamefont {Ma}}, \
  and\ \bibinfo {author} {\bibfnamefont {Yu}~\bibnamefont {Gao}},\ }\bibfield
  {title} {\enquote {\bibinfo {title} {{Constraining primordial black holes
  with relativistic degrees of freedom}},}\ }\href@noop {} {\  (\bibinfo {year}
  {2022})},\ \Eprint {http://arxiv.org/abs/2210.03476} {arXiv:2210.03476
  [astro-ph.CO]} \BibitemShut {NoStop}%
\bibitem [{\citenamefont {Gehrman}\ \emph {et~al.}(2022)\citenamefont
  {Gehrman}, \citenamefont {Shams Es~Haghi}, \citenamefont {Sinha},\ and\
  \citenamefont {Xu}}]{Gehrman:2022imk}%
  \BibitemOpen
  \bibfield  {author} {\bibinfo {author} {\bibfnamefont {Thomas~C.}\
  \bibnamefont {Gehrman}}, \bibinfo {author} {\bibfnamefont {Barmak}\
  \bibnamefont {Shams Es~Haghi}}, \bibinfo {author} {\bibfnamefont {Kuver}\
  \bibnamefont {Sinha}}, \ and\ \bibinfo {author} {\bibfnamefont {Tao}\
  \bibnamefont {Xu}},\ }\bibfield  {title} {\enquote {\bibinfo {title}
  {{Baryogenesis, Primordial Black Holes and MHz-GHz Gravitational Waves}},}\
  }\href@noop {} {\  (\bibinfo {year} {2022})},\ \Eprint
  {http://arxiv.org/abs/2211.08431} {arXiv:2211.08431 [hep-ph]} \BibitemShut
  {NoStop}%
\bibitem [{\citenamefont {Braglia}\ \emph {et~al.}(2021)\citenamefont
  {Braglia}, \citenamefont {Chen},\ and\ \citenamefont
  {Hazra}}]{Braglia:2020taf}%
  \BibitemOpen
  \bibfield  {author} {\bibinfo {author} {\bibfnamefont {Matteo}\ \bibnamefont
  {Braglia}}, \bibinfo {author} {\bibfnamefont {Xingang}\ \bibnamefont {Chen}},
  \ and\ \bibinfo {author} {\bibfnamefont {Dhiraj~Kumar}\ \bibnamefont
  {Hazra}},\ }\bibfield  {title} {\enquote {\bibinfo {title} {{Probing
  Primordial Features with the Stochastic Gravitational Wave Background}},}\
  }\href {\doibase 10.1088/1475-7516/2021/03/005} {\bibfield  {journal}
  {\bibinfo  {journal} {JCAP}\ }\textbf {\bibinfo {volume} {03}},\ \bibinfo
  {pages} {005} (\bibinfo {year} {2021})},\ \Eprint
  {http://arxiv.org/abs/2012.05821} {arXiv:2012.05821 [astro-ph.CO]}
  \BibitemShut {NoStop}%
\bibitem [{\citenamefont {Papanikolaou}(2022)}]{Papanikolaou:2022chm}%
  \BibitemOpen
  \bibfield  {author} {\bibinfo {author} {\bibfnamefont {Theodoros}\
  \bibnamefont {Papanikolaou}},\ }\bibfield  {title} {\enquote {\bibinfo
  {title} {{Gravitational waves induced from primordial black hole
  fluctuations: the~effect of an extended mass function}},}\ }\href {\doibase
  10.1088/1475-7516/2022/10/089} {\bibfield  {journal} {\bibinfo  {journal}
  {JCAP}\ }\textbf {\bibinfo {volume} {10}},\ \bibinfo {pages} {089} (\bibinfo
  {year} {2022})},\ \Eprint {http://arxiv.org/abs/2207.11041} {arXiv:2207.11041
  [astro-ph.CO]} \BibitemShut {NoStop}%
\bibitem [{\citenamefont {Qiu}\ \emph {et~al.}(2023)\citenamefont {Qiu},
  \citenamefont {Wang},\ and\ \citenamefont {Zheng}}]{Qiu:2022klm}%
  \BibitemOpen
  \bibfield  {author} {\bibinfo {author} {\bibfnamefont {Taotao}\ \bibnamefont
  {Qiu}}, \bibinfo {author} {\bibfnamefont {Wenyi}\ \bibnamefont {Wang}}, \
  and\ \bibinfo {author} {\bibfnamefont {Ruifeng}\ \bibnamefont {Zheng}},\
  }\bibfield  {title} {\enquote {\bibinfo {title} {{Generation of primordial
  black holes from an inflation model with modified dispersion relation}},}\
  }\href {\doibase 10.1103/PhysRevD.107.083018} {\bibfield  {journal} {\bibinfo
   {journal} {Phys. Rev. D}\ }\textbf {\bibinfo {volume} {107}},\ \bibinfo
  {pages} {083018} (\bibinfo {year} {2023})},\ \Eprint
  {http://arxiv.org/abs/2212.03403} {arXiv:2212.03403 [astro-ph.CO]}
  \BibitemShut {NoStop}%
\bibitem [{\citenamefont {Escriv\`a}\ \emph {et~al.}(2022)\citenamefont
  {Escriv\`a}, \citenamefont {Kuhnel},\ and\ \citenamefont
  {Tada}}]{Escriva:2022duf}%
  \BibitemOpen
  \bibfield  {author} {\bibinfo {author} {\bibfnamefont {Albert}\ \bibnamefont
  {Escriv\`a}}, \bibinfo {author} {\bibfnamefont {Florian}\ \bibnamefont
  {Kuhnel}}, \ and\ \bibinfo {author} {\bibfnamefont {Yuichiro}\ \bibnamefont
  {Tada}},\ }\bibfield  {title} {\enquote {\bibinfo {title} {{Primordial Black
  Holes}},}\ }\href@noop {} {\  (\bibinfo {year} {2022})},\ \Eprint
  {http://arxiv.org/abs/2211.05767} {arXiv:2211.05767 [astro-ph.CO]}
  \BibitemShut {NoStop}%
\bibitem [{\citenamefont {Meng}\ \emph {et~al.}(2023)\citenamefont {Meng},
  \citenamefont {Yuan},\ and\ \citenamefont {Huang}}]{Meng:2022low}%
  \BibitemOpen
  \bibfield  {author} {\bibinfo {author} {\bibfnamefont {De-Shuang}\
  \bibnamefont {Meng}}, \bibinfo {author} {\bibfnamefont {Chen}\ \bibnamefont
  {Yuan}}, \ and\ \bibinfo {author} {\bibfnamefont {Qing-Guo}\ \bibnamefont
  {Huang}},\ }\bibfield  {title} {\enquote {\bibinfo {title} {{Primordial black
  holes generated by the non-minimal spectator field}},}\ }\href {\doibase
  10.1007/s11433-022-2095-5} {\bibfield  {journal} {\bibinfo  {journal} {Sci.
  China Phys. Mech. Astron.}\ }\textbf {\bibinfo {volume} {66}},\ \bibinfo
  {pages} {280411} (\bibinfo {year} {2023})},\ \Eprint
  {http://arxiv.org/abs/2212.03577} {arXiv:2212.03577 [astro-ph.CO]}
  \BibitemShut {NoStop}%
\bibitem [{\citenamefont {Gehrman}\ \emph {et~al.}(2023)\citenamefont
  {Gehrman}, \citenamefont {Shams Es~Haghi}, \citenamefont {Sinha},\ and\
  \citenamefont {Xu}}]{Gehrman:2023esa}%
  \BibitemOpen
  \bibfield  {author} {\bibinfo {author} {\bibfnamefont {Thomas~C.}\
  \bibnamefont {Gehrman}}, \bibinfo {author} {\bibfnamefont {Barmak}\
  \bibnamefont {Shams Es~Haghi}}, \bibinfo {author} {\bibfnamefont {Kuver}\
  \bibnamefont {Sinha}}, \ and\ \bibinfo {author} {\bibfnamefont {Tao}\
  \bibnamefont {Xu}},\ }\bibfield  {title} {\enquote {\bibinfo {title} {{The
  Primordial Black Holes that Disappeared: Connections to Dark Matter and
  MHz-GHz Gravitational Waves}},}\ }\href@noop {} {\  (\bibinfo {year}
  {2023})},\ \Eprint {http://arxiv.org/abs/2304.09194} {arXiv:2304.09194
  [hep-ph]} \BibitemShut {NoStop}%
\bibitem [{\citenamefont {Dandoy}\ \emph {et~al.}(2023)\citenamefont {Dandoy},
  \citenamefont {Domcke},\ and\ \citenamefont {Rompineve}}]{Dandoy:2023jot}%
  \BibitemOpen
  \bibfield  {author} {\bibinfo {author} {\bibfnamefont {Virgile}\ \bibnamefont
  {Dandoy}}, \bibinfo {author} {\bibfnamefont {Valerie}\ \bibnamefont
  {Domcke}}, \ and\ \bibinfo {author} {\bibfnamefont {Fabrizio}\ \bibnamefont
  {Rompineve}},\ }\bibfield  {title} {\enquote {\bibinfo {title} {{Search for
  scalar induced gravitational waves in the International Pulsar Timing Array
  Data Release 2 and NANOgrav 12.5 years datasets}},}\ }\href@noop {} {\
  (\bibinfo {year} {2023})},\ \Eprint {http://arxiv.org/abs/2302.07901}
  {arXiv:2302.07901 [astro-ph.CO]} \BibitemShut {NoStop}%
\bibitem [{\citenamefont {Ferrante}\ \emph
  {et~al.}(2023{\natexlab{a}})\citenamefont {Ferrante}, \citenamefont
  {Franciolini}, \citenamefont {Iovino},\ and\ \citenamefont
  {Urbano}}]{Ferrante:2023bgz}%
  \BibitemOpen
  \bibfield  {author} {\bibinfo {author} {\bibfnamefont {Giacomo}\ \bibnamefont
  {Ferrante}}, \bibinfo {author} {\bibfnamefont {Gabriele}\ \bibnamefont
  {Franciolini}}, \bibinfo {author} {\bibfnamefont {Antonio}\ \bibnamefont
  {Iovino}, \bibfnamefont {Junior.}}, \ and\ \bibinfo {author} {\bibfnamefont
  {Alfredo}\ \bibnamefont {Urbano}},\ }\bibfield  {title} {\enquote {\bibinfo
  {title} {{Primordial black holes in the curvaton model: possible connections
  to pulsar timing arrays and dark matter}},}\ }\href {\doibase
  10.1088/1475-7516/2023/06/057} {\bibfield  {journal} {\bibinfo  {journal}
  {JCAP}\ }\textbf {\bibinfo {volume} {06}},\ \bibinfo {pages} {057} (\bibinfo
  {year} {2023}{\natexlab{a}})},\ \Eprint {http://arxiv.org/abs/2305.13382}
  {arXiv:2305.13382 [astro-ph.CO]} \BibitemShut {NoStop}%
\bibitem [{\citenamefont {Afzal}\ \emph {et~al.}(2023)\citenamefont {Afzal}
  \emph {et~al.}}]{NANOGrav:2023hvm}%
  \BibitemOpen
  \bibfield  {author} {\bibinfo {author} {\bibfnamefont {Adeela}\ \bibnamefont
  {Afzal}} \emph {et~al.} (\bibinfo {collaboration} {NANOGrav}),\ }\bibfield
  {title} {\enquote {\bibinfo {title} {{The NANOGrav 15 yr Data Set: Search for
  Signals from New Physics}},}\ }\href {\doibase 10.3847/2041-8213/acdc91}
  {\bibfield  {journal} {\bibinfo  {journal} {Astrophys. J. Lett.}\ }\textbf
  {\bibinfo {volume} {951}},\ \bibinfo {pages} {L11} (\bibinfo {year}
  {2023})},\ \Eprint {http://arxiv.org/abs/2306.16219} {arXiv:2306.16219
  [astro-ph.HE]} \BibitemShut {NoStop}%
\bibitem [{\citenamefont {Antoniadis}\ \emph {et~al.}(2023)\citenamefont
  {Antoniadis} \emph {et~al.}}]{Antoniadis:2023zhi}%
  \BibitemOpen
  \bibfield  {author} {\bibinfo {author} {\bibfnamefont {J.}~\bibnamefont
  {Antoniadis}} \emph {et~al.},\ }\bibfield  {title} {\enquote {\bibinfo
  {title} {{The second data release from the European Pulsar Timing Array: V.
  Implications for massive black holes, dark matter and the early Universe}},}\
  }\href@noop {} {\  (\bibinfo {year} {2023})},\ \Eprint
  {http://arxiv.org/abs/2306.16227} {arXiv:2306.16227 [astro-ph.CO]}
  \BibitemShut {NoStop}%
\bibitem [{\citenamefont {Inomata}\ \emph {et~al.}(2023)\citenamefont
  {Inomata}, \citenamefont {Kohri},\ and\ \citenamefont
  {Terada}}]{Inomata:2023zup}%
  \BibitemOpen
  \bibfield  {author} {\bibinfo {author} {\bibfnamefont {Keisuke}\ \bibnamefont
  {Inomata}}, \bibinfo {author} {\bibfnamefont {Kazunori}\ \bibnamefont
  {Kohri}}, \ and\ \bibinfo {author} {\bibfnamefont {Takahiro}\ \bibnamefont
  {Terada}},\ }\bibfield  {title} {\enquote {\bibinfo {title} {{The Detected
  Stochastic Gravitational Waves and Sub-Solar Primordial Black Holes}},}\
  }\href@noop {} {\  (\bibinfo {year} {2023})},\ \Eprint
  {http://arxiv.org/abs/2306.17834} {arXiv:2306.17834 [astro-ph.CO]}
  \BibitemShut {NoStop}%
\bibitem [{\citenamefont {Gu}\ \emph {et~al.}(2023)\citenamefont {Gu},
  \citenamefont {Shu},\ and\ \citenamefont {Yang}}]{Gu:2023mmd}%
  \BibitemOpen
  \bibfield  {author} {\bibinfo {author} {\bibfnamefont {Bao-Min}\ \bibnamefont
  {Gu}}, \bibinfo {author} {\bibfnamefont {Fu-Wen}\ \bibnamefont {Shu}}, \ and\
  \bibinfo {author} {\bibfnamefont {Ke}~\bibnamefont {Yang}},\ }\bibfield
  {title} {\enquote {\bibinfo {title} {{Inflation with shallow dip and
  primordial black holes}},}\ }\href@noop {} {\  (\bibinfo {year} {2023})},\
  \Eprint {http://arxiv.org/abs/2307.00510} {arXiv:2307.00510 [astro-ph.CO]}
  \BibitemShut {NoStop}%
\bibitem [{\citenamefont {Liu}\ \emph {et~al.}(2023{\natexlab{b}})\citenamefont
  {Liu}, \citenamefont {Chen},\ and\ \citenamefont {Huang}}]{Liu:2023ymk}%
  \BibitemOpen
  \bibfield  {author} {\bibinfo {author} {\bibfnamefont {Lang}\ \bibnamefont
  {Liu}}, \bibinfo {author} {\bibfnamefont {Zu-Cheng}\ \bibnamefont {Chen}}, \
  and\ \bibinfo {author} {\bibfnamefont {Qing-Guo}\ \bibnamefont {Huang}},\
  }\bibfield  {title} {\enquote {\bibinfo {title} {{Implications for the
  non-Gaussianity of curvature perturbation from pulsar timing arrays}},}\
  }\href@noop {} {\  (\bibinfo {year} {2023}{\natexlab{b}})},\ \Eprint
  {http://arxiv.org/abs/2307.01102} {arXiv:2307.01102 [astro-ph.CO]}
  \BibitemShut {NoStop}%
\bibitem [{\citenamefont {Yi}\ \emph {et~al.}(2023)\citenamefont {Yi},
  \citenamefont {Gao}, \citenamefont {Gong}, \citenamefont {Wang},\ and\
  \citenamefont {Zhang}}]{Yi:2023mbm}%
  \BibitemOpen
  \bibfield  {author} {\bibinfo {author} {\bibfnamefont {Zhu}\ \bibnamefont
  {Yi}}, \bibinfo {author} {\bibfnamefont {Qing}\ \bibnamefont {Gao}}, \bibinfo
  {author} {\bibfnamefont {Yungui}\ \bibnamefont {Gong}}, \bibinfo {author}
  {\bibfnamefont {Yue}\ \bibnamefont {Wang}}, \ and\ \bibinfo {author}
  {\bibfnamefont {Fengge}\ \bibnamefont {Zhang}},\ }\bibfield  {title}
  {\enquote {\bibinfo {title} {{The waveform of the scalar induced
  gravitational waves in light of Pulsar Timing Array data}},}\ }\href@noop {}
  {\  (\bibinfo {year} {2023})},\ \Eprint {http://arxiv.org/abs/2307.02467}
  {arXiv:2307.02467 [gr-qc]} \BibitemShut {NoStop}%
\bibitem [{\citenamefont {You}\ \emph {et~al.}(2023)\citenamefont {You},
  \citenamefont {Yi},\ and\ \citenamefont {Wu}}]{You:2023rmn}%
  \BibitemOpen
  \bibfield  {author} {\bibinfo {author} {\bibfnamefont {Zhi-Qiang}\
  \bibnamefont {You}}, \bibinfo {author} {\bibfnamefont {Zhu}\ \bibnamefont
  {Yi}}, \ and\ \bibinfo {author} {\bibfnamefont {You}\ \bibnamefont {Wu}},\
  }\bibfield  {title} {\enquote {\bibinfo {title} {{Constraints on primordial
  curvature power spectrum with pulsar timing arrays}},}\ }\href@noop {} {\
  (\bibinfo {year} {2023})},\ \Eprint {http://arxiv.org/abs/2307.04419}
  {arXiv:2307.04419 [gr-qc]} \BibitemShut {NoStop}%
\bibitem [{\citenamefont {Jin}\ \emph {et~al.}(2023)\citenamefont {Jin},
  \citenamefont {Chen}, \citenamefont {Yi}, \citenamefont {You}, \citenamefont
  {Liu},\ and\ \citenamefont {Wu}}]{Jin:2023wri}%
  \BibitemOpen
  \bibfield  {author} {\bibinfo {author} {\bibfnamefont {Jia-Heng}\
  \bibnamefont {Jin}}, \bibinfo {author} {\bibfnamefont {Zu-Cheng}\
  \bibnamefont {Chen}}, \bibinfo {author} {\bibfnamefont {Zhu}\ \bibnamefont
  {Yi}}, \bibinfo {author} {\bibfnamefont {Zhi-Qiang}\ \bibnamefont {You}},
  \bibinfo {author} {\bibfnamefont {Lang}\ \bibnamefont {Liu}}, \ and\ \bibinfo
  {author} {\bibfnamefont {You}\ \bibnamefont {Wu}},\ }\bibfield  {title}
  {\enquote {\bibinfo {title} {{Confronting sound speed resonance with pulsar
  timing arrays}},}\ }\href@noop {} {\  (\bibinfo {year} {2023})},\ \Eprint
  {http://arxiv.org/abs/2307.08687} {arXiv:2307.08687 [astro-ph.CO]}
  \BibitemShut {NoStop}%
\bibitem [{\citenamefont {Balaji}\ \emph {et~al.}(2023)\citenamefont {Balaji},
  \citenamefont {Dom\`enech},\ and\ \citenamefont
  {Franciolini}}]{Balaji:2023ehk}%
  \BibitemOpen
  \bibfield  {author} {\bibinfo {author} {\bibfnamefont {Shyam}\ \bibnamefont
  {Balaji}}, \bibinfo {author} {\bibfnamefont {Guillem}\ \bibnamefont
  {Dom\`enech}}, \ and\ \bibinfo {author} {\bibfnamefont {Gabriele}\
  \bibnamefont {Franciolini}},\ }\bibfield  {title} {\enquote {\bibinfo {title}
  {{Scalar-induced gravitational wave interpretation of PTA data: the role of
  scalar fluctuation propagation speed}},}\ }\href@noop {} {\  (\bibinfo {year}
  {2023})},\ \Eprint {http://arxiv.org/abs/2307.08552} {arXiv:2307.08552
  [gr-qc]} \BibitemShut {NoStop}%
\bibitem [{\citenamefont {Liu}\ \emph {et~al.}(2023{\natexlab{c}})\citenamefont
  {Liu}, \citenamefont {Chen},\ and\ \citenamefont {Huang}}]{Liu:2023pau}%
  \BibitemOpen
  \bibfield  {author} {\bibinfo {author} {\bibfnamefont {Lang}\ \bibnamefont
  {Liu}}, \bibinfo {author} {\bibfnamefont {Zu-Cheng}\ \bibnamefont {Chen}}, \
  and\ \bibinfo {author} {\bibfnamefont {Qing-Guo}\ \bibnamefont {Huang}},\
  }\bibfield  {title} {\enquote {\bibinfo {title} {{Probing the equation of
  state of the early Universe with pulsar timing arrays}},}\ }\href@noop {} {\
  (\bibinfo {year} {2023}{\natexlab{c}})},\ \Eprint
  {http://arxiv.org/abs/2307.14911} {arXiv:2307.14911 [astro-ph.CO]}
  \BibitemShut {NoStop}%
\bibitem [{\citenamefont {Basilakos}\ \emph {et~al.}(2023)\citenamefont
  {Basilakos}, \citenamefont {Nanopoulos}, \citenamefont {Papanikolaou},
  \citenamefont {Saridakis},\ and\ \citenamefont
  {Tzerefos}}]{Basilakos:2023xof}%
  \BibitemOpen
  \bibfield  {author} {\bibinfo {author} {\bibfnamefont {Spyros}\ \bibnamefont
  {Basilakos}}, \bibinfo {author} {\bibfnamefont {Dimitri~V.}\ \bibnamefont
  {Nanopoulos}}, \bibinfo {author} {\bibfnamefont {Theodoros}\ \bibnamefont
  {Papanikolaou}}, \bibinfo {author} {\bibfnamefont {Emmanuel~N.}\ \bibnamefont
  {Saridakis}}, \ and\ \bibinfo {author} {\bibfnamefont {Charalampos}\
  \bibnamefont {Tzerefos}},\ }\bibfield  {title} {\enquote {\bibinfo {title}
  {{Signatures of Superstring theory in NANOGrav}},}\ }\href@noop {} {\
  (\bibinfo {year} {2023})},\ \Eprint {http://arxiv.org/abs/2307.08601}
  {arXiv:2307.08601 [hep-th]} \BibitemShut {NoStop}%
\bibitem [{\citenamefont {Ragavendra}(2022)}]{Ragavendra:2021qdu}%
  \BibitemOpen
  \bibfield  {author} {\bibinfo {author} {\bibfnamefont {H.~V.}\ \bibnamefont
  {Ragavendra}},\ }\bibfield  {title} {\enquote {\bibinfo {title} {{Accounting
  for scalar non-Gaussianity in secondary gravitational waves}},}\ }\href
  {\doibase 10.1103/PhysRevD.105.063533} {\bibfield  {journal} {\bibinfo
  {journal} {Phys. Rev. D}\ }\textbf {\bibinfo {volume} {105}},\ \bibinfo
  {pages} {063533} (\bibinfo {year} {2022})},\ \Eprint
  {http://arxiv.org/abs/2108.04193} {arXiv:2108.04193 [astro-ph.CO]}
  \BibitemShut {NoStop}%
\bibitem [{\citenamefont {Franciolini}\ \emph {et~al.}(2023)\citenamefont
  {Franciolini}, \citenamefont {Iovino}, \citenamefont {Vaskonen},\ and\
  \citenamefont {Veermae}}]{Franciolini:2023pbf}%
  \BibitemOpen
  \bibfield  {author} {\bibinfo {author} {\bibfnamefont {Gabriele}\
  \bibnamefont {Franciolini}}, \bibinfo {author} {\bibfnamefont {Antonio}\
  \bibnamefont {Iovino}, \bibfnamefont {Junior.}}, \bibinfo {author}
  {\bibfnamefont {Ville}\ \bibnamefont {Vaskonen}}, \ and\ \bibinfo {author}
  {\bibfnamefont {Hardi}\ \bibnamefont {Veermae}},\ }\bibfield  {title}
  {\enquote {\bibinfo {title} {{The recent gravitational wave observation by
  pulsar timing arrays and primordial black holes: the importance of
  non-gaussianities}},}\ }\href@noop {} {\  (\bibinfo {year} {2023})},\ \Eprint
  {http://arxiv.org/abs/2306.17149} {arXiv:2306.17149 [astro-ph.CO]}
  \BibitemShut {NoStop}%
\bibitem [{\citenamefont {Ferrante}\ \emph
  {et~al.}(2023{\natexlab{b}})\citenamefont {Ferrante}, \citenamefont
  {Franciolini}, \citenamefont {Iovino},\ and\ \citenamefont
  {Urbano}}]{Ferrante:2022mui}%
  \BibitemOpen
  \bibfield  {author} {\bibinfo {author} {\bibfnamefont {Giacomo}\ \bibnamefont
  {Ferrante}}, \bibinfo {author} {\bibfnamefont {Gabriele}\ \bibnamefont
  {Franciolini}}, \bibinfo {author} {\bibfnamefont {Antonio}\ \bibnamefont
  {Iovino}, \bibfnamefont {Junior.}}, \ and\ \bibinfo {author} {\bibfnamefont
  {Alfredo}\ \bibnamefont {Urbano}},\ }\bibfield  {title} {\enquote {\bibinfo
  {title} {{Primordial non-Gaussianity up to all orders: Theoretical aspects
  and implications for primordial black hole models}},}\ }\href {\doibase
  10.1103/PhysRevD.107.043520} {\bibfield  {journal} {\bibinfo  {journal}
  {Phys. Rev. D}\ }\textbf {\bibinfo {volume} {107}},\ \bibinfo {pages}
  {043520} (\bibinfo {year} {2023}{\natexlab{b}})},\ \Eprint
  {http://arxiv.org/abs/2211.01728} {arXiv:2211.01728 [astro-ph.CO]}
  \BibitemShut {NoStop}%
\bibitem [{\citenamefont {Yuan}\ and\ \citenamefont
  {Huang}(2021{\natexlab{b}})}]{Yuan:2021qgz}%
  \BibitemOpen
  \bibfield  {author} {\bibinfo {author} {\bibfnamefont {Chen}\ \bibnamefont
  {Yuan}}\ and\ \bibinfo {author} {\bibfnamefont {Qing-Guo}\ \bibnamefont
  {Huang}},\ }\bibfield  {title} {\enquote {\bibinfo {title} {{A topic review
  on probing primordial black hole dark matter with scalar induced
  gravitational waves}},}\ }\href@noop {} {\  (\bibinfo {year}
  {2021}{\natexlab{b}})},\ \Eprint {http://arxiv.org/abs/2103.04739}
  {arXiv:2103.04739 [astro-ph.GA]} \BibitemShut {NoStop}%
\bibitem [{\citenamefont {Dom\`enech}(2021)}]{Domenech:2021ztg}%
  \BibitemOpen
  \bibfield  {author} {\bibinfo {author} {\bibfnamefont {Guillem}\ \bibnamefont
  {Dom\`enech}},\ }\bibfield  {title} {\enquote {\bibinfo {title} {{Scalar
  Induced Gravitational Waves Review}},}\ }\href {\doibase
  10.3390/universe7110398} {\bibfield  {journal} {\bibinfo  {journal}
  {Universe}\ }\textbf {\bibinfo {volume} {7}},\ \bibinfo {pages} {398}
  (\bibinfo {year} {2021})},\ \Eprint {http://arxiv.org/abs/2109.01398}
  {arXiv:2109.01398 [gr-qc]} \BibitemShut {NoStop}%
\bibitem [{\citenamefont {Adshead}\ \emph {et~al.}(2021)\citenamefont
  {Adshead}, \citenamefont {Lozanov},\ and\ \citenamefont
  {Weiner}}]{Adshead:2021hnm}%
  \BibitemOpen
  \bibfield  {author} {\bibinfo {author} {\bibfnamefont {Peter}\ \bibnamefont
  {Adshead}}, \bibinfo {author} {\bibfnamefont {Kaloian~D.}\ \bibnamefont
  {Lozanov}}, \ and\ \bibinfo {author} {\bibfnamefont {Zachary~J.}\
  \bibnamefont {Weiner}},\ }\bibfield  {title} {\enquote {\bibinfo {title}
  {{Non-Gaussianity and the induced gravitational wave background}},}\ }\href
  {\doibase 10.1088/1475-7516/2021/10/080} {\bibfield  {journal} {\bibinfo
  {journal} {JCAP}\ }\textbf {\bibinfo {volume} {10}},\ \bibinfo {pages} {080}
  (\bibinfo {year} {2021})},\ \Eprint {http://arxiv.org/abs/2105.01659}
  {arXiv:2105.01659 [astro-ph.CO]} \BibitemShut {NoStop}%
\bibitem [{\citenamefont {Abe}\ \emph {et~al.}(2023)\citenamefont {Abe},
  \citenamefont {Inui}, \citenamefont {Tada},\ and\ \citenamefont
  {Yokoyama}}]{Abe:2022xur}%
  \BibitemOpen
  \bibfield  {author} {\bibinfo {author} {\bibfnamefont {Katsuya~T.}\
  \bibnamefont {Abe}}, \bibinfo {author} {\bibfnamefont {Ryoto}\ \bibnamefont
  {Inui}}, \bibinfo {author} {\bibfnamefont {Yuichiro}\ \bibnamefont {Tada}}, \
  and\ \bibinfo {author} {\bibfnamefont {Shuichiro}\ \bibnamefont {Yokoyama}},\
  }\bibfield  {title} {\enquote {\bibinfo {title} {{Primordial black holes and
  gravitational waves induced by exponential-tailed perturbations}},}\ }\href
  {\doibase 10.1088/1475-7516/2023/05/044} {\bibfield  {journal} {\bibinfo
  {journal} {JCAP}\ }\textbf {\bibinfo {volume} {05}},\ \bibinfo {pages} {044}
  (\bibinfo {year} {2023})},\ \Eprint {http://arxiv.org/abs/2209.13891}
  {arXiv:2209.13891 [astro-ph.CO]} \BibitemShut {NoStop}%
\bibitem [{\citenamefont {Pi}\ and\ \citenamefont {Sasaki}(2020)}]{Pi:2020otn}%
  \BibitemOpen
  \bibfield  {author} {\bibinfo {author} {\bibfnamefont {Shi}\ \bibnamefont
  {Pi}}\ and\ \bibinfo {author} {\bibfnamefont {Misao}\ \bibnamefont
  {Sasaki}},\ }\bibfield  {title} {\enquote {\bibinfo {title} {{Gravitational
  Waves Induced by Scalar Perturbations with a Lognormal Peak}},}\ }\href
  {\doibase 10.1088/1475-7516/2020/09/037} {\bibfield  {journal} {\bibinfo
  {journal} {JCAP}\ }\textbf {\bibinfo {volume} {09}},\ \bibinfo {pages} {037}
  (\bibinfo {year} {2020})},\ \Eprint {http://arxiv.org/abs/2005.12306}
  {arXiv:2005.12306 [gr-qc]} \BibitemShut {NoStop}%
\bibitem [{\citenamefont {Kohri}\ and\ \citenamefont
  {Terada}(2021)}]{Kohri:2020qqd}%
  \BibitemOpen
  \bibfield  {author} {\bibinfo {author} {\bibfnamefont {Kazunori}\
  \bibnamefont {Kohri}}\ and\ \bibinfo {author} {\bibfnamefont {Takahiro}\
  \bibnamefont {Terada}},\ }\bibfield  {title} {\enquote {\bibinfo {title}
  {{Solar-Mass Primordial Black Holes Explain NANOGrav Hint of Gravitational
  Waves}},}\ }\href {\doibase 10.1016/j.physletb.2020.136040} {\bibfield
  {journal} {\bibinfo  {journal} {Phys. Lett. B}\ }\textbf {\bibinfo {volume}
  {813}},\ \bibinfo {pages} {136040} (\bibinfo {year} {2021})},\ \Eprint
  {http://arxiv.org/abs/2009.11853} {arXiv:2009.11853 [astro-ph.CO]}
  \BibitemShut {NoStop}%
\bibitem [{\citenamefont {Meng}\ \emph {et~al.}(2022)\citenamefont {Meng},
  \citenamefont {Yuan},\ and\ \citenamefont {Huang}}]{Meng:2022ixx}%
  \BibitemOpen
  \bibfield  {author} {\bibinfo {author} {\bibfnamefont {De-Shuang}\
  \bibnamefont {Meng}}, \bibinfo {author} {\bibfnamefont {Chen}\ \bibnamefont
  {Yuan}}, \ and\ \bibinfo {author} {\bibfnamefont {Qing-guo}\ \bibnamefont
  {Huang}},\ }\bibfield  {title} {\enquote {\bibinfo {title} {{One-loop
  correction to the enhanced curvature perturbation with local-type
  non-Gaussianity for the formation of primordial black holes}},}\ }\href
  {\doibase 10.1103/PhysRevD.106.063508} {\bibfield  {journal} {\bibinfo
  {journal} {Phys. Rev. D}\ }\textbf {\bibinfo {volume} {106}},\ \bibinfo
  {pages} {063508} (\bibinfo {year} {2022})},\ \Eprint
  {http://arxiv.org/abs/2207.07668} {arXiv:2207.07668 [astro-ph.CO]}
  \BibitemShut {NoStop}%
\bibitem [{\citenamefont {Hahn}(2005)}]{Hahn:2004fe}%
  \BibitemOpen
  \bibfield  {author} {\bibinfo {author} {\bibfnamefont {T.}~\bibnamefont
  {Hahn}},\ }\bibfield  {title} {\enquote {\bibinfo {title} {{CUBA: A Library
  for multidimensional numerical integration}},}\ }\href {\doibase
  10.1016/j.cpc.2005.01.010} {\bibfield  {journal} {\bibinfo  {journal}
  {Comput. Phys. Commun.}\ }\textbf {\bibinfo {volume} {168}},\ \bibinfo
  {pages} {78--95} (\bibinfo {year} {2005})},\ \Eprint
  {http://arxiv.org/abs/hep-ph/0404043} {arXiv:hep-ph/0404043} \BibitemShut
  {NoStop}%
\bibitem [{\citenamefont {Hahn}(2015)}]{Hahn:2014fua}%
  \BibitemOpen
  \bibfield  {author} {\bibinfo {author} {\bibfnamefont {T.}~\bibnamefont
  {Hahn}},\ }\bibfield  {title} {\enquote {\bibinfo {title} {{Concurrent
  Cuba}},}\ }\href {\doibase 10.1088/1742-6596/608/1/012066} {\bibfield
  {journal} {\bibinfo  {journal} {J. Phys. Conf. Ser.}\ }\textbf {\bibinfo
  {volume} {608}},\ \bibinfo {pages} {012066} (\bibinfo {year} {2015})},\
  \Eprint {http://arxiv.org/abs/1408.6373} {arXiv:1408.6373 [physics.comp-ph]}
  \BibitemShut {NoStop}%
\bibitem [{\citenamefont {Kristiano}\ and\ \citenamefont
  {Yokoyama}(2022{\natexlab{a}})}]{Kristiano:2021urj}%
  \BibitemOpen
  \bibfield  {author} {\bibinfo {author} {\bibfnamefont {Jason}\ \bibnamefont
  {Kristiano}}\ and\ \bibinfo {author} {\bibfnamefont {Jun'ichi}\ \bibnamefont
  {Yokoyama}},\ }\bibfield  {title} {\enquote {\bibinfo {title} {{Why Must
  Primordial Non-Gaussianity Be Very Small?}}}\ }\href {\doibase
  10.1103/PhysRevLett.128.061301} {\bibfield  {journal} {\bibinfo  {journal}
  {Phys. Rev. Lett.}\ }\textbf {\bibinfo {volume} {128}},\ \bibinfo {pages}
  {061301} (\bibinfo {year} {2022}{\natexlab{a}})},\ \Eprint
  {http://arxiv.org/abs/2104.01953} {arXiv:2104.01953 [hep-th]} \BibitemShut
  {NoStop}%
\bibitem [{\citenamefont {Kristiano}\ and\ \citenamefont
  {Yokoyama}(2022{\natexlab{b}})}]{Kristiano:2022maq}%
  \BibitemOpen
  \bibfield  {author} {\bibinfo {author} {\bibfnamefont {Jason}\ \bibnamefont
  {Kristiano}}\ and\ \bibinfo {author} {\bibfnamefont {Jun'ichi}\ \bibnamefont
  {Yokoyama}},\ }\bibfield  {title} {\enquote {\bibinfo {title} {{Ruling Out
  Primordial Black Hole Formation From Single-Field Inflation}},}\ }\href@noop
  {} {\  (\bibinfo {year} {2022}{\natexlab{b}})},\ \Eprint
  {http://arxiv.org/abs/2211.03395} {arXiv:2211.03395 [hep-th]} \BibitemShut
  {NoStop}%
\bibitem [{\citenamefont {Garcia-Saenz}\ \emph {et~al.}(2023)\citenamefont
  {Garcia-Saenz}, \citenamefont {Pinol}, \citenamefont {Renaux-Petel},\ and\
  \citenamefont {Werth}}]{Garcia-Saenz:2022tzu}%
  \BibitemOpen
  \bibfield  {author} {\bibinfo {author} {\bibfnamefont {Sebastian}\
  \bibnamefont {Garcia-Saenz}}, \bibinfo {author} {\bibfnamefont {Lucas}\
  \bibnamefont {Pinol}}, \bibinfo {author} {\bibfnamefont {S\'ebastien}\
  \bibnamefont {Renaux-Petel}}, \ and\ \bibinfo {author} {\bibfnamefont
  {Denis}\ \bibnamefont {Werth}},\ }\bibfield  {title} {\enquote {\bibinfo
  {title} {{No-go theorem for scalar-trispectrum-induced gravitational
  waves}},}\ }\href {\doibase 10.1088/1475-7516/2023/03/057} {\bibfield
  {journal} {\bibinfo  {journal} {JCAP}\ }\textbf {\bibinfo {volume} {03}},\
  \bibinfo {pages} {057} (\bibinfo {year} {2023})},\ \Eprint
  {http://arxiv.org/abs/2207.14267} {arXiv:2207.14267 [astro-ph.CO]}
  \BibitemShut {NoStop}%
\bibitem [{\citenamefont {Chang}\ \emph {et~al.}(2023)\citenamefont {Chang},
  \citenamefont {Kuang}, \citenamefont {Zhang},\ and\ \citenamefont
  {Zhou}}]{Chang:2022nzu}%
  \BibitemOpen
  \bibfield  {author} {\bibinfo {author} {\bibfnamefont {Zhe}\ \bibnamefont
  {Chang}}, \bibinfo {author} {\bibfnamefont {Yu-Ting}\ \bibnamefont {Kuang}},
  \bibinfo {author} {\bibfnamefont {Xukun}\ \bibnamefont {Zhang}}, \ and\
  \bibinfo {author} {\bibfnamefont {Jing-Zhi}\ \bibnamefont {Zhou}},\
  }\bibfield  {title} {\enquote {\bibinfo {title} {{Primordial black holes and
  third order scalar induced gravitational waves*}},}\ }\href {\doibase
  10.1088/1674-1137/acc649} {\bibfield  {journal} {\bibinfo  {journal} {Chin.
  Phys. C}\ }\textbf {\bibinfo {volume} {47}},\ \bibinfo {pages} {055104}
  (\bibinfo {year} {2023})},\ \Eprint {http://arxiv.org/abs/2209.12404}
  {arXiv:2209.12404 [astro-ph.CO]} \BibitemShut {NoStop}%
\end{thebibliography}%

\end{document}